\documentclass[fleqn,usenatbib]{mnras}

\usepackage[utf8]{inputenc}
\usepackage[english]{babel}
\usepackage{fontenc,amssymb}
\usepackage{array, makecell}
\usepackage{graphicx}
\usepackage{color}
\usepackage{hyperref}
\usepackage{booktabs}
\usepackage{float}

\graphicspath{{Figures/}}

\newcommand{\galpak}{GalPaK$^{\rm 3D}$}

\newcommand{\FeII}{\hbox{{\rm Fe}{\sc \,ii}}}

\newcommand{\OII}{\hbox{[{\rm O}{\sc \,ii}]}}

\newcommand{\OIII}{\hbox{{\rm O}{\sc \,iii}}}

\newcommand{\MgI}{\hbox{{\rm Mg}{\sc \,i}}}
\newcommand{\MgII}{\hbox{{\rm Mg}{\sc \,ii}}}

\newcommand{\HI}{\hbox{{\rm H}{\sc \,i}}}

\newcommand{\Ha}{\hbox{{\rm H}$\alpha$}}
\newcommand{\Hb}{\hbox{{\rm H}$\beta$}}

\newcommand{\flux}{erg\,s$^{-1}$\,cm$^{-2}$}

 \newcommand{\mpy}{\hbox{M$_{\odot}$~yr$^{-1}$}}

\newcommand{\msun}{\hbox{M$_{\odot}$}}

\newcommand{\ma}{\hbox{$\lambda 2796$}}

\newcommand{\Vmax}{\hbox{$V_{\rm max}$}}
\newcommand{\Vout}{\hbox{$V_{\rm out}$}}
\newcommand{\thetam}{\hbox{$\theta_{\rm max}$}}
\newcommand{\Ms}{\hbox{$M_{\star}$}}
\newcommand{\Mv}{\hbox{$M_{\rm h}$}} 

\newcommand{\kms}{\hbox{km~s$^{-1}$}}
\newcommand{\kpc}{\hbox{kpc}}
\newcommand{\uerglf}{\mathrm{erg}\,\mathrm{s}^{-1}\,\mathrm{cm}^{-2}}

{}
{}
\newcommand{\sext}{\emph{SExtractor}}{}
{}
\newcommand{\camel}{\emph{CAMEL}}{}
{}
{}

\newcommand{\Nfield}{22}
\newcommand{\NabsTot}{79}%% REW>0.3
\newcommand{\Nabs}{59}%% REW>0.3 with one or more galaxy identified
\newcommand{\NgalTot}{165}

\newcommand{\Npairs}{86}
\newcommand{\NpairsfNtwo}{61}

\newcommand{\NabsNtwo}{51}
\newcommand{\NpairsfNtwoFlagsPos}{57}

\newcommand{\NpairsWind}{31}
\newcommand{\NpairsWindFluxthree}{30}
\newcommand{\NpairsFinal}{26}
\newcommand{\NpairsIncl}{28}
\newcommand{\FluxLimit}{{$1.5\times10^{-17}$~\flux}}

\title[Gas outflow in MEGAFLOW]{MusE GAs FLOw and Wind (MEGAFLOW) III: galactic wind properties using background quasars~\thanks{Based on observations made at the ESO telescopes at La Silla Paranal Observatory under programme IDs 094.A-0211(B), 095.A-0365(A), 096.A-0164(A), 097.A-0138(A), 099.A-0059(A), 096.A-0609(A), 097.A-0144(A), 098.A-0310(A), 293.A-5038(A).}}
\author[I. Schroetter et al.]{
        Ilane~Schroetter,$^{1,2}$\thanks{E-mail: ilane.schroetter@obspm.fr},
        Nicolas~F.~Bouch\'e,$^{1,3}$ 
        Johannes~Zabl,$^{1,3}$
        Thierry~Contini,$^{1}$
        \newauthor
        Martin~Wendt,$^{4,5}$
        Joop~Schaye,$^{6}$
        Peter~Mitchell,$^{3}$
        Sowgat~Muzahid,$^{6}$
        Raffaella~A.~Marino,$^{7}$
        \newauthor
        Roland~Bacon,$^{3}$
        Simon~J.~Lilly,$^{7}$
        Johan~Richard,$^{3}$
        Lutz~Wisotzki~$^{5}$
        \\
$^{1}$ Institut de Recherche en Astrophysique et Plan\'etologie (IRAP), Universit\'e de Toulouse, CNRS, UPS, F-31400 Toulouse, France\\
$^{2}$ GEPI, Observatoire de Paris, PSL Université, CNRS,  5 Place Jules Janssen, 92190 Meudon, France\\
$^{3}$ Univ Lyon, Univ Lyon1, Ens de Lyon, CNRS, Centre de Recherche Astrophysique de Lyon UMR5574, F-69230 Saint-Genis-Laval, France\\
$^{4}$ Institut f\"ur Physik und Astronomie, Universit\"at Potsdam, Karl-Liebknecht-Str. 24/25, 14476 Golm, Germany \\
$^{5}$ Leibniz-Institut für Astrophysik Potsdam, An der Sternwarte 16, D-14482 Potsdam, Germany \\
$^{6}$ Leiden Observatory, Leiden University, PO Box 9513, 2300 RA Leiden, The Netherlands \\
$^{7}$ ETH Zurich, Institute of Astronomy, Wolfgang-Pauli-Str. 27, 8093 Zurich, Switzerland \\ }

\begin{document}
\label{firstpage}
\pagerange{\pageref{firstpage}--\pageref{lastpage}}
\maketitle
%----------------------------------------------------------------------------------------
% ABSTRACT
%----------------------------------------------------------------------------------------
\begin{abstract}

We present results from our on-going MusE GAs FLOw and Wind (MEGAFLOW) survey, which consists of 22 quasar lines-of-sight, each observed with the integral field unit (IFU) MUSE and the UVES spectrograph at the ESO Very Large Telescopes (VLT). The goals of this survey are to study the  properties of the circum-galactic medium around $z\sim1$ star-forming galaxies. The absorption-line selected survey consists of \NabsTot\ strong \MgII\ absorbers (with rest-frame equivalent width (REW)$\gtrsim$0.3~\AA) and, currently, \Npairs\  associated galaxies within 100 projected~kpc of the quasar with stellar masses  ($M_\star$) from $10^9$ to $10^{11}$ \msun.
We find that the cool halo gas traced by \MgII\ is not isotropically distributed around these galaxies from the strong bi-modal distribution in the azimuthal angle of the apparent location of the quasar with respect to the galaxy major-axis.
This supports a scenario in which outflows are bi-conical in nature and  co-exist with a coplanar  gaseous structure extending at least up to 60 to 80 kpc.  
Assuming that absorbers near the minor axis probe outflows, the current MEGAFLOW sample allowed us to select \NpairsFinal\ galaxy-quasar pairs suitable for studying winds. 
From this sample, using a simple geometrical model, we find that the outflow velocity only exceeds the escape velocity when $M_{\star}\lesssim 4\times10^9$~\msun, implying the cool material is likely to fall back except in the smallest halos.
Finally, we find  that the mass loading factor $\eta$, the ratio between the ejected mass rate and the star formation rate (SFR), appears to be roughly constant with respect to the galaxy mass. 

\end{abstract}

\begin{keywords}
galaxies: evolution --- galaxies: formation --- galaxies: intergalactic medium  ---  quasars: absorption lines --- 
\end{keywords}

%----------------------------------------------------------------------------------------
% SECTION	1 - Introduction
%----------------------------------------------------------------------------------------
\section{Introduction}
\label{introduction}
\setcounter{footnote}{0}

Galaxies form by the cooling and condensation
of baryons at the centers of  dark
matter halos in an expanding universe   \citep[e.g.][]{ReesM_77a,WhiteS_78a}.
As originally described in \citet{WhiteS_91a}, in halos where the cooling time is shorter than the dynamical time,
galaxies are expected to contain their fair share of baryons, namely $f_B=17$\%, given by the cosmological baryon fraction $\Omega_b/\Omega_m$.
However, galaxies contain, on average, only  10\%\ and at most 20\%\ of their share of baryons \citep[e.g.][]{GuoQ_10a,BehrooziP_13a}. 

This low baryon fraction,
often referred to as the galaxy formation `efficiency' defined as $\Ms/(f_B\;\Mv)$, strongly depends on halo mass
\citep[e.g.][]{GuoQ_10a,BehrooziP_13a}. 
In halos with mass below $10^{12}$\msun, the  decline is directly connected to 
the faint-end slope of the luminosity function, and
galactic (super-)winds from star-forming galaxies are thought to play a major role in causing this decline, as originally proposed by \citet{LarsonR_74b} who noted that the impact of supernovae (SNe) on star formation would be the highest in small halos \citep[see also][]{DekelA_86a}.  
The galactic wind scenario is attractive as it is also thought to play a major role in enriching the inter-galactic medium \citep[e.g.][]{AguirreA_01a,AguirreA_05a,MadauP_01a,TheunsT_02a}.

Theoretically,  the successes of cosmological simulations often rely on the specifics of the feedback implementation \citep[e.g.][]{schaye_10,scannapieco_12,Vogelsberger_13,CrainR_15}.
These implementations depend  on sub-grid prescriptions, 
such as the wind mass loading factor $\eta\equiv \dot M_{\rm out}/$SFR for kinetic implementation of feedback. 
An alternative way to implement the SN-driven outflows relies on a (stochastic) implementation of thermal feedback, where galactic winds develop without 
imposing any input outflow velocity or mass loading factor such as in the EAGLE simulations \citep[e.g.][]{schaye_15}, the FIRE simulations \citep{hopkins_12,hopkins_14, muratov_15}, and the multi-phase scheme of \citet{barai_15}.
For instance, \citet{hopkins_12,HopkinsP_17a} predict that the loading factor is inversely proportional to the galaxy stellar mass, which 
is in agreement with simple momentum conservation expectations but found additional dependencies on star formation rate (SFR) surface density.

Observationally, assumed SN-driven winds are found to be ubiquitous in star-forming galaxies both at low 
\citep[e.g.][]{HeckmanT_90a,HeckmanT_17a,ShopbellP_98a,PettiniM_02c,veilleux_05,MartinC_05a,SatoT_09a,MartinC_09a,arribas_14}
and at high-redshifs 
\citep[e.g.][]{ShapleyA_03a,ForsterSchrieberN_06,WeinerB_09a,ChenY_10a,SteidelC_10a,KorneiK_12a,MartinC_12a,BordoloiR_14a,RubinK_14a,SugaharaY_17,ForsterSchrieberN_18}. 

 Traditionally, galactic winds are found from blue-shifted absorption lines of low-ionization ions such as Na~D 
galaxy spectra \citep[see reviews in ][]{veilleux_05,Bland-HawthornJ_07a,HeckmanT_17b}
or other ions in the rest-frame UV spectra of galaxies \citep[e.g.][]{ChisholmJ_15a,ChisholmJ_16a,SugaharaY_17,ForsterSchrieberN_18}.
Galactic winds can also be studied using various other observational techniques using their emission (X-ray, \Ha\ or CO) properties \citep[e.g.][]{arribas_14,newman_12, bolatto_13,CiconeC_16a,CiconeC_17a,FalgaroneE_17a}, their
UV fluorescent emission \citep[e.g.][]{RubinK_11a,MartinC_13a,TangY_14a,ZhuG_15a,FinleyH_17a}, or
far-infrared spectra \citep[e.g.][]{sturm_11, GonzalezE_17,SpilkerJ_18}. 

There are two main results from these studies. First, galactic outflows appear to be collimated 
 \citep[e.g.][]{ChenY_10a,BordoloiR_11a, BordoloiR_14a, lan_menard_14, RubinK_14a}  consistent with a bi-conical flow with a cone opening angle \thetam\footnote{where \thetam\ is the half-opening angle of a bi-conical flow  underling an area $\Sigma$ of $\pi \cdot \theta_{\rm max}^2$.} that is approximately 30$^\circ$ to 40$^\circ$ from the minor axis of the host galaxy. 
Second,   absorption lines in galaxy spectra give  an accurate measurement of the outflowing gas velocity $V_{\rm out}$,  which is typically 200~\kms\ \citep[depending on the SFR;][]{MartinC_05a},
but this method has a major weakness: it gives a very poor constraint on one key property, namely the mass outflow rate, due to the unknown  location of the absorbing gas, which can be located 0.1, 1 or even 10 kpc away from the host galaxy. 
To illustrate the degree of uncertainty in the assumptions made in the recent literature,  \citet{heckman_15, HeckmanT_17a} assumed a wind launch radius of $2\times$ $R_e$ and spherical symmetry, 
 \citet{ChisholmJ_15a} used a launch radius of 5 kpc,  \citet{arribas_14} assumed a wind launching radius of 0.7~kpc while \citet{ChisholmJ_16a,ChisholmJ_16b,ChisholmJ_17a} puts the wind material at $<100$ pc 
 (inferred from the ionization correction).

This unknown gas location leads to large uncertainties (orders of magnitude) on the ejected mass rate $\dot M_{\rm out}$ , preventing accurate determination of the outflow rate, which increases with the square of the distance. 
Consequently, the loading factor $\eta$  and its dependence on galaxy properties has not been determined unequivocally. 
In order to make further progress and to put strong constraints on models,
 we need to constrain outflow properties using objects for which the gas location can be better determined.

Background quasars naturally provide information on the location of the gas (from the impact parameter $b$), and thus have the potential to lead to  higher accuracy in the wind mass outflow rates and loading factors 
\citep[e.g.][]{bouche_12, KacprzakG_14a, schroetter_15,schroetter_16, muzahid_15, RahmaniH_18}.
Using this background quasar technique, the geometric uncertainty on the mass outflow rate goes from several dex to a factor of two or three. 

This method suffers from the difficulty in finding large numbers of galaxy--quasar pairs, but this can be remedied with appropriate observational strategies.
Over the past few years, the availability of large catalogs of
the common  low-ionization \MgII $\lambda\lambda2796,2803$ absorption  
  in the optical spectra of large samples of quasars from the Sloan Digital Sky Survey \citep{Lan_14,ZhuG_15a} has changed
the situation.  

In \citet[][hereafter paper~I]{schroetter_16},  we presented the first results from this program, the MUSE Gas Flow and Wind  (MEGAFLOW) survey, which aims to collect a statistically significant sample of approximately one 
hundred galaxy-quasar pairs in 22 quasar fields with multiple \MgII\ absorbers.
In \citet[][hereafter paper II]{ZablJ_19}, we analyze the sub-sample of galaxy-quasar pairs suitable for constraining the properties of accreting gas.
In this paper, we present and analyze the pairs suitable to constrain outflow properties.
The full MEGAFLOW survey will be presented in Bouch\'e et al. (in prep.).

This paper is organized as follows. In section~\S~\ref{survey}, we present
the MEGAFLOW observational strategy. The data acquisition is described in section~\S~\ref{observations}. Our sample selection is presented in section~\S~\ref{sample}.
The analysis of our sample is presented in section~\S~\ref{results} while the wind modeling and results are described in section~\S~\ref{wind}.
Finally, we present our conclusions in section~\S~\ref{conclusions}.

Throughout, we use a cosmology of 737 and the \citet{chabrier_03} stellar Initial Mass Function (IMF). 

\section{MEGAFLOW: survey strategy }
\label{survey}

Most of the work on the low-ionization, cool ($T \sim 10^4$~K) component of the circum-galactic medium (CGM) has been focused on the \MgII$\lambda,\lambda$ 2796, 2803 doublet absorption in  quasar spectra \citep{BergeronJ_88a,BergeronJ_91a,BergeronJ_92a,SteidelC_95a,SteidelC_97a,SteidelC_02a}.
However, finding the galaxy counterpart for the \MgII\ absorption is often a complicated process. 
Indeed, it requires deep pre-imaging in order to identify host-galaxy candidates (and to allow the determination of the morphology/inclination)  and multi-object spectroscopy, with the quasar   blocking the view directly along the line-of-sight as an additional problem. 
Furthermore, one must also perform expensive follow-up campaigns to determine the galaxy kinematics. 

Several groups have developed this imaging$+$multi-object spectroscopy technique using ground-based imaging 
\citep[e.g.][]{ChenHW_08a,ChenHW_10a,ChenHW_10b,ZhuB_18,RubinK_18}, 
but usually these lack the spatial resolution to untangle the morphological information, which is crucial to understand the absorption kinematics \citep[e.g.][]{BordoloiR_11a,bouche_12,KacprzakG_12a}. 
Thus, arguably the best sample of \MgII\ based galaxy-quasar pairs with morphological data is the MAGIICAT sample \citep{ChurchillC_13a,NielsenN_13a,NielsenN_13b,NielsenN_15a,NielsenN_16a}, which consists of more than 100 foreground isolated galaxies at $0.3<z<1.0$ imaged with {\it HST}, and with quasar impact parameters ranging from 20 to 110~kpc. 
However, as mentioned, the imaging$+$multi-object spectroscopy technique suffers from several disadvantages: 
(i) it requires pre-imaging and pre-identification of host-galaxy candidates based on the continuum light, thus leading to biases against emission-line galaxies;
(ii) it is nearly impossible close to the line-of-sight (LOS) ;
(iii) it is inefficient, requiring multiple campaigns, for imaging, for redshift identification, and for kinematics determination \citep[e.g.][]{HoS_17a}.

These shortcomings can be bypassed using integral field units (IFUs) data where the galaxy counterpart(s) can be readily identified at once (i.e. without pre-imaging, without knowing its location a priori). 
This identification can be from either emission lines or e.g. H\&K and Balmer absorption lines for passive galaxies. 
In addition, the galaxy kinematics are part of the data, the morphological information can also be determined from 3D data \citep{BoucheN_15a, ContiniT_16a} and the PSF can be more easily subtracted in 3D.
With the  VLT/MUSE instrument \citep{BaconR_06a, BaconR_10a,BaconR_15a} and its exquisite sensitivity, one can now detect galaxies further away ($\approx$250 kpc away at $z=1$) thanks to its field of view of $1\arcmin \times 1 \arcmin$ compared to $8 \arcsec \times8 \arcsec$ for VLT/SINFONI. 
The large wavelength coverage of MUSE (4700\AA\ to 9300\AA) allows us to target quasar fields with multiple \MgII\ $\lambda \lambda 2796, 2803$ absorption lines  having redshifts from 0.4 to 1.5 
for \OII\ $\lambda \lambda 3727, 3729$ identification. 
In the up-coming years, MUSE's Adaptive-Optics (AO) module will increase the quality of data, and the efficiency of such surveys.

The MEGAFLOW survey (papers~I, II) aims at observing a statistical number (80$+$) of galaxy-quasar pairs to allow analysis of the relation between the absorption and the host galaxy properties. 
From the Zhu and M\'enard \MgII\ catalog based on SDSS \citep{zhu_13}, we selected quasars with multiple ($N\geq3$) \MgII$\lambda \lambda 2796, 2802$ absorption lines with redshifts between 0.4 and 1.4
and with a \MgII\ $\lambda 2796$ rest-equivalent width (REW) $W_r^{\ma}$ $\gtrsim$ 0.5\AA. 
The former criteria of having multiple absorbers in one quasar field, ensures that a large number of galaxy-quasar pairs of 80$+$ is reachable with 20$-$25 quasar fields. 
The latter criteria ensures that the host galaxies are within 100~kpc from the quasar LOS (at $z\approx1$), i.e. within the MUSE field-of-view, given
the well known anti-correlation between the impact parameter and $W_r^{\ma}$ \citep{lanzetta_90, steidel_95}. 

Overall, the MEGAFLOW survey is made of \Nfield\ quasar fields, with each quasar spectrum having at least 3  strong ($W_r^{\ma}>$0.5~\AA) \MgII\ absorbers, over the redshift range between 0.4 and 1.5. Including a few serendipitous  systems with $0.3<W_r^{\ma}<0.5$~\AA, the survey containts a total of \NabsTot\  \MgII\ absorbers
with $W_r^{\ma}>$0.3~\AA.

\section{Data}
\label{observations}
\subsection{MUSE Observations and data reduction}

We use the MUSE observations from the MEGAFLOW survey taken from September 2014 to July 2017 during Guaranteed Time Observations (GTO)  runs.
The observations were optimized to cover the inner 20\arcsec\ region uniformly by placing the quasar $\approx5\arcsec$ from the field center, by using small sub-pixel dithers and a rotation of 90$^\circ$ between each exposure. 
The individual exposure time ranges from 900 to 1500s.
The resulting total exposure time per field ranges from two to four hours (See Table~\ref{table:muse_observations}).

The data are reduced as described in paper~II where we used version 1.6 of the MUSE data reduction software \citep[DRS;][]{WeilbacherP_14a,weilbacher_16a} pipeline.  Briefly, the reduction includes an additional step on the pixtables called 'auto-calibration' described in \citet{BaconR_17a}, which removes the slight variations in the background level in each slice of each IFU caused by imperfections in the flat-fielding.  After performing the self-calibration, we resampled the 
pixtables onto datacubes with the sky subtraction, barycentric correction turned on. Finally, we used the
{\it Zurich Atmosphere Purge} ({\it ZAP}) software
\citep{SotoK_16a,SotoK_16b} to remove skyline residuals from each datacube. Finally, we combined the individual cubes weighted by the inverse of the seeing full width half maximum (FWHM) when needed.

\subsection{UVES Observations and data reduction}

Because we are interested in constraining the kinematics of gas surrounding star-forming galaxies, 
we need quasar spectra with a resolution better than MUSE (which has $R\sim2000$ or 150 km/s) in a wavelength range not covered by MUSE (4700-5000\AA). 
We choose high-resolution spectroscopy of the quasars with the VLT$/$UVES instrument.

The \Nfield\ quasar fields were observed with the high-resolution spectrograph UVES \citep{DekkerH_00a} between 2014 and 2016 (Table~\ref{table:uves_observations}).
The settings used in our observation were chosen in order to cover the \MgII$\lambda \lambda 2796,2803$ absorption lines and other elements like \MgI$\lambda2852$, \FeII$\lambda2586$ when possible. 
The details of the observational campaigns are presented in Table \ref{table:uves_observations}. 
A slit width of 1.2 arcsec and a CCD readout with 2x2 binning were used for all the observations, resulting in a spectral resolving power R $\approx 38000$ dispersed on pixels of $\approx$1.3 \kms. 
The Common Pipeline Language (CPL version 6.3) of the UVES pipeline was used to bias correct and flat field the exposures and then to extract the wavelength and flux calibrated spectra. 
After the standard reduction, the custom software UVES Popler \citep[][version 0.66]{MurphyM_16a}  
was used to combine the extracted echelle orders into single 1D spectra. 
The continuum was fitted with low-order polynomial functions. 

\begin{table*}
\centering
\caption{Summary of MUSE observations}
\label{table:muse_observations}
\begin{tabular}{lcrcc}
\hline
Field & Program ID & Exp. time & Obs date & Seeing \\
(1)            & (2)       & (3)                   & (4)  & (5)           \\
\hline
SDSSJ0014m0028 & \makecell[t]{095.A-0365(A),\\ 096.A-0164(A)} & 6300 & \makecell[t]{2015-08-24,\\ 2015-09-11 \& 10-13} & 0.78\\
SDSSJ0014p0912 & 094.A-0211(B) & 10800 & 2014-10-20 10-21 10-25 & 0.85\\
SDSSJ0015m0751  &  \makecell[t]{096.A-0164(A),\\ 097.A-0138(A), \\  099.A-0059(A)} & 9000 & \makecell[t]{2015-10-10\&11,\\ 2016-09-01,\\ 2017-09-22}  & 0.80\\
SDSSJ0058p0111 & \makecell[t]{096.A-0164(A),\\ 097.A-0138(A)} & 7200 & 2015-11-09 2016-08-30  & 0.77\\
SDSSJ0103p1332 & 096.A-0164(A) & 7200 & 2015-11-12 11-13  & 0.84\\
SDSSJ0131p1303 & \makecell[t]{094.A-0211(B),\\ 099.A-0059(A)} & 7200 & 2014-10-28 2017-09-23 09-24  & 0.81\\
SDSSJ0134p0051 & \makecell[t]{096.A-0164(A),\\ 097.A-0138(A)} & 7200 & \makecell[t]{2015-10-15\&16,\\ 2016-09-01,\\ 2017-09-25}  & 0.73\\
SDSSJ0145p1056 & \makecell[t]{096.A-0164(A),\\ 097.A-0138(A)} & 6000 & 2015-11-13 2016-08-30  & 0.85\\
SDSSJ0800p1849 & 094.A-0211(B) & 7200 & 2014-12-25  & 0.56\\
SDSSJ0838p0257 & 096.A-0164(A) & 12000 & 2016-02-02 02-03  & 0.54\\
SDSSJ0937p0656 & 095.A-0365(A) & 7200 & 2015-04-15 04-16 04-18  & 0.67\\
SDSSJ1039p0714 & 097.A-0138(A) & 12000 & 2016-04-07 04-08 04-09  & 0.61\\
SDSSJ1107p1021 & 096.A-0164(A) & 12000 & 2016-03-12  & 0.70\\
SDSSJ1107p1757 & 095.A-0365(A) & 7200 & 2015-04-23 04-24  & 0.88\\
SDSSJ1236p0725 & 096.A-0164(A) & 6000 & 2016-03-13  & 0.91\\
SDSSJ1314p0657 & 097.A-0138(A) & 6000 & 2016-04-07 04-08  & 0.53\\
SDSSJ1352p0614 & 099.A-0059(A) & 6000 & 2017-04-23 04-24  & 0.98\\
SDSSJ1358p1145 & 097.A-0138(A) & 6000 & 2016-04-10  & 0.54\\
SDSSJ1425p1209 & 097.A-0138(A) & 3600 & 2016-05-12  & 0.96\\
SDSSJ1509p1506 & 099.A-0059(A) & 3000 & 2017-04-23  & 0.70\\
SDSSJ2137p0012$^\dagger$ & 094.A-0211(B) & 3600 & 2014-09-20 09-24  & 0.74\\
SDSSJ2152p0625 & 094.A-0211(B) & 7200 & 2014-09-25  & 0.58\\
\hline
\end{tabular}\\
{
(1) Quasar field name;
(2) Program ID; 
(3) Total exposure time (in seconds);
(4) Observation dates of the field;
(5) Seeing FWHM (in $\arcsec$) from a Moffat fit of the QSO at 7000~\AA;
$^\dagger$ 3 hours of this field were rejected due to bad seeing conditions ($>1.2\arcsec$).
}
\end{table*}

\begin{table*}
\centering
\caption{Summary of UVES observations}
\label{table:uves_observations}
\begin{tabular}{lcrccc}
\hline
Field & Program ID & Exp. time & Obs date & Setting & seeing\\
(1)            & (2)       & (3)                   & (4)    &      (5)  & (6) \\
\hline
SDSSJ0014m0028  & 096.A-0609(A) & 9015  & 2015-10-04 & DIC1 390+564 & 0.81 \\
SDSSJ0014p0912  & \makecell[t]{ 096.A-0609(A),\\ 098.A-0310(A)} & 7493  & \makecell[t]{2015-11-10,\\ 2016-10-29} & DIC1 390+564; DIC2 437+760 & 0.66 \\
SDSSJ0015m0751  & 098.A-0310(A) & 12020  & 2016-10-30 12-28 12-29 & DIC1 390+564 & 0.69 \\
SDSSJ0058p0111  & 098.A-0310(A) & 2966  & 2016-12-30 12-31 & DIC1 390+564; DIC2 437+760 & 0.52 \\
SDSSJ0103p1332  & 098.A-0310(A) & 9015  & 2016-10-29 10-30 11-02 & DIC1 390+564 & 0.60 \\
SDSSJ0131p1303  & 096.A-0609(A) & 6010  & 2015-10-15 & DIC1 390+580 & 1.03 \\
SDSSJ0134p0051  & 098.A-0310(A) & 7193  & 2016-10-30 12-04 & DIC1 390+580; DIC2 437+760 & 0.57 \\
SDSSJ0145p1056  & \makecell[t]{ 096.A-0609(A),\\ 097.A-0144(A),\\ 098.A-0310(A)} & 12020  & \makecell[t]{2015-11-12,\\ 2016-09-04, \\ 2016-10-29} & DIC1 390+564 & 0.63 \\
SDSSJ0800p1849  & 096.A-0609(A) & 6010  & 2015-12-11 & RED 520 & 0.90 \\
SDSSJ0838p0257  & \makecell[t]{096.A-0609(A),\\ 098.A-0310(A)} & 2966  & \makecell[t]{2015-11-21,\\ 2016-12-23} & DIC1 390+564; RED 600 & 0.76 \\
SDSSJ0937p0656  & 096.A-0609(A) & 9015  & 2015-12-21 2016-01-12 03-08 & DIC1 390+564 & 0.74 \\
SDSSJ1039p0714  & 097.A-0144(A) & 9015  & 2016-04-04 & DIC1 346+580 & 0.76 \\
SDSSJ1107p1021  & 096.A-0609(A) & 6010  & 2016-02-10 03-08 & DIC1 390+580 & 1.02 \\
SDSSJ1107p1757  & 096.A-0609(A) & 9015  & 2016-01-12 03-07 03-08 & DIC2 437+760 & 0.99 \\
SDSSJ1236p0725  & \makecell[t]{096.A-0609(A),\\ 097.A-0144(A)} & 7493  & \makecell[t]{2016-03-07,\\ 2016-04-07} & DIC2 437+760; RED 600 & 0.61 \\
SDSSJ1314p0657  & 097.A-0144(A) & 1483  & 2016-04-07 & DIC1 390+564 & 0.46 \\
SDSSJ1352p0614  & 097.A-0144(A) & 1483  & 2016-05-31 06-01 & DIC2 437+760 & 0.70 \\
SDSSJ1358p1145  & 097.A-0144(A) & 2966  & 2016-04-07 & DIC1 390+564' 'DIC2 346+860 & 0.51 \\
SDSSJ1425p1209  & 097.A-0144(A) & 2966  & 2016-04-07 06-01 & DIC1 390+564; RED 520 & 0.56 \\
SDSSJ1509p1506  & 097.A-0144(A) & 6010  & 2016-04-04 04-07 & RED 600 & 0.57 \\
SDSSJ2137p0012  & 293.A-5038(A) & 4487  & 2014-10-19 & DIC1 390+564 & 0.99 \\
SDSSJ2152p0625  & 293.A-5038(A) & 9015  & 2014-10-21 10-24 11-18 & DIC1 390+580 & 1.21 \\
\hline
\end{tabular}\\
{
(1) Quasar field name;
(2) Program ID; 
(3) Total exposure time (in seconds);
(4) Observation dates of the quasar;
(5) Instrument setting; 
(6) average seeing FWHM ($\arcsec$)
}
\end{table*}

\section{Sample selection}
\label{sample}
\subsection{Galaxy detection}

In each of the \Nfield\ quasar fields, we search for galaxies (emitters and/or passive) responsible for the \MgII\ absorption lines.
In order to find the potential host galaxy/ies, we run our detection algorithm as described in paper~II.
Briefly, the algorithm is designed to detect galaxies using both emission lines and absorption lines
using  pseudo narrow-band (NB) images made of, depending on the redshift, \OII, \Hb{}, \textsc{{\rm Ca}} H\&K, and/or \OIII{}$\lambda5007$ over a velocity range of 400 \kms. The NB images were created for each absorber, at three different velocity offsets from the absorber redshifts. 
Finally,   galaxy candidates are detected on these pseudo NB images using the source detection algorithm \sext{} \citep{bertin_96}. We optimized \sext{} in order to detect low signal-to-noise ratio (SNR) objects and ensure completeness, leading to a significant fraction of false positives, which had to be removed manually.

Using the wavelength dependent per-pixel noise, we derive a typical $5 \sigma$ detection limit of
$\approx4\times10^{-18}\times(\hbox{FWHM}_{\rm Moffat}/0\arcsec.6)\times(\mbox{T}_{\rm exp}/6\rm ks)^{-0.5}\,\uerglf$ 
(see paper~II) centered at $7000~\AA$. 
This corresponds to an un-obscured SFR limit of $0.07\mpy$ using \OII\ emission line. 

\subsubsection{Redshifts}
For all the detected galaxies, we determined their redshifts using three methods. 
For all three methods we use the MUSE data. 
The first method consists in manually deriving the redshift of each galaxy using the \OII\ emission line position. 
The central position of the line is given by a Gaussian fit. 
A pseudo long slit is also used on each galaxy (along the apparent PA of the galaxy) to obtain a 2D spectrum which provides an additional redshift measurement. 
In the second method, we use a line fitting code which fits the \OII\ doublet automatically using a double Gaussian. 
Using the output of a 3D fitting tool called \galpak\ \citep{bouche_15} is the third method we employ to derive galaxy redshifts. 
Some details on \galpak\ are given below.

Each of those methods gives us with a redshift for each galaxy. 
Those redshifts are consistent with each other and differ by only a few \kms. 
We choose to use redshifts derived by the line fitting method as the standard deviation of the redshift differences (between manual and automatical fits) is lower (15~\kms) than the one using \galpak\ (26~\kms).
Thus, throughout this paper we use the systemic redshifts derived by the line fitting method (i.e. method 2).

\subsection{Absorber-galaxy pairs: Parent sample}

From our \Nfield\ quasar fields, we have found \NgalTot\ galaxies around \NabsTot\ absorbers with $W_r^{\lambda2796}\geq0.3$~\AA. 
Those detected galaxies lie at impact parameters from 0 to 350~\kpc\ from the QSO LOS.
Among these \NgalTot\ galaxies, \Npairs\ have an impact parameter smaller than 100~\kpc\ out of \Nabs\ \MgII\ absorbers. 

\begin{figure}
  \centering
  \includegraphics[width=8.5cm]{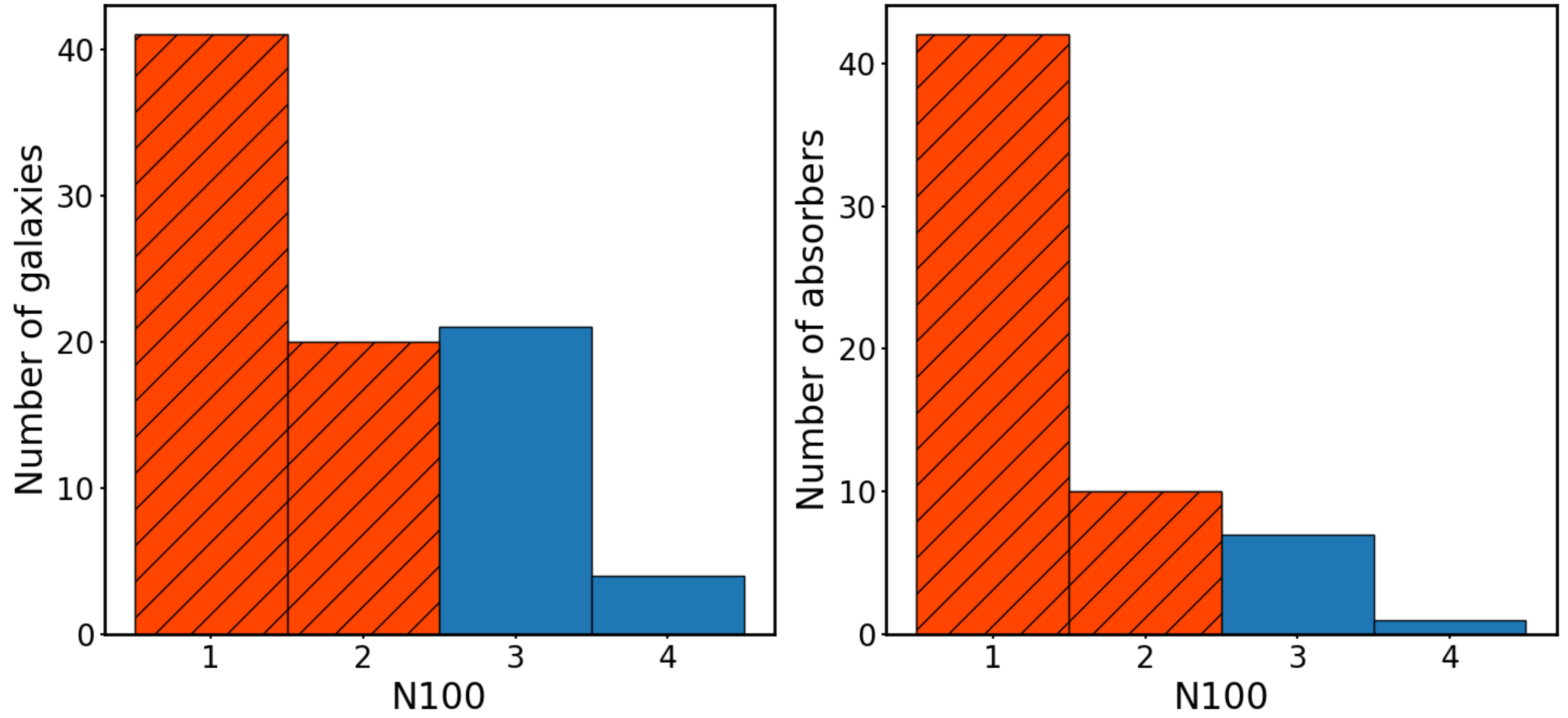}
  \caption{
  Histogram of N100, the number of galaxies at the redshift of an absorber separated by less than 100~\kpc, for the \Npairs\ galaxy-quasar pairs (left) and the 59 absorbers (right).
 Out of the \Npairs\ pairs (59 absorbers), \NpairsfNtwo\ (51) have N100$\leq2$ (hashed), respectively
  }
  \label{fig:N_hist}
\end{figure}

In order to avoid groups of galaxies, we restricted the sample to absorbers with at most two ($\leq2$) galaxies within 100 transverse~kpc from the QSO LOS. 
Among these \Npairs\ galaxies, there are \NpairsfNtwo\ galaxies with N100$\leq2$, where N100 is the number of galaxies within 100~\kpc. 
These \NpairsfNtwo\ galaxies correspond to \NabsNtwo\ \MgII\ absorbers.
The N100 distribution is presented in Figure~\ref{fig:N_hist} where the \NpairsfNtwo\ pairs are represented by hashed regions (on left panel for galaxies and on right panel for absorbers). 
41 of the 61 galaxies are "isolated" (i.e. with N100=1).
For those galaxies we also search for secondary neighbors at $b>100$~\kpc\ and a separation lower than 50~\kpc. We only found two cases of two independent primary galaxies with a secondary companion within approximately 40~\kpc.
Those two primary galaxies are not matching our selection criteria described later in the text (i.e. inclination and azimuthal angle).

\subsection{Absorber-galaxy pairs: Morphology selection}
\label{morpho-kinematics}

From this parent sample of \NpairsfNtwo\ pairs, we wish to select those for which the location of the line of sight to the quasar is favorable for intercepting outflows, assuming that outflows are oriented along the galaxy's minor axis 
\citep[as in][]{bouche_12,schroetter_15,schroetter_16}.
To do so, we   select galaxy-quasar pairs where the apparent quasar location is within $\approx$30$^{\circ}$ of the galaxy's minor-axis. 
Defining $\alpha$ as the azimuthal angle between the galaxy's major axis and the apparent quasar location, we divide the pairs into two classes: ``wind-pair'' and ``inflow-pair'' for pairs with  $55^\circ \leq \alpha \leq 90^\circ$ and $0^\circ \leq \alpha \leq 40^\circ$ respectively.

For each of the \NpairsfNtwo\ galaxies, the orientation is derived using the 3D fitting tool called \galpak\ from \citet{bouche_15}. 
This algorithm uses a parametric disk model with 10 free parameters (such as total line flux, half-light radius, inclination, maximum rotation velocity, velocity dispersion and position angle [PA] of the major-axis) and an MCMC algorithm in order to efficiently probe the parameter space. 
The algorithm also uses a 3-dimensional kernel to account for the instrument PSF and line spread function (LSF).
\galpak\ thus returns the ``intrinsic'' galaxy properties. 
  
Extensive tests presented in \citet{bouche_15} show that the algorithm requires data with a SNR$_{\rm max}>3$ in the brightest pixel.  However, for   SNRs approaching this limit  and for compact galaxies, degeneracies can appear, such as between the  turn-over radius\footnote{which is defined by an arctan function for the rotation curve of the galaxy.} and $V_{\rm max}$.

For of each of the \NpairsfNtwo\ galaxies, we checked manually the morpho-kinematical results  
as well as the \galpak\ MCMC chains. 
We then flagged the results according to the following scheme:
 \begin{itemize}
  \item 0: when neither \Vmax\ nor the morphological parameters (PA, inclination) are constrained. 
  This usually  occurs for galaxies with a very low SNR, e.g. with flux lower than \FluxLimit.
  \item 1: when at least one morphological parameter (at least PA) is constrained. 
  \item 3: when some of the kinematic parameters are either not well constrained or degenerate with other (e.g. \Vmax-inclination, \Vmax-turn-over radius).
  \item 5: when all of the morphological and kinematic parameters are constrained.
 \end{itemize}

Given our $\alpha$ cut,  we select galaxies which have a reliable PA, i.e. with a flag $\geq$ 1. 
From the \NpairsfNtwo\ galaxies, this criterion brings our sample to \NpairsfNtwoFlagsPos\ galaxies. 

\begin{figure}
  \centering
  \includegraphics[width=8cm]{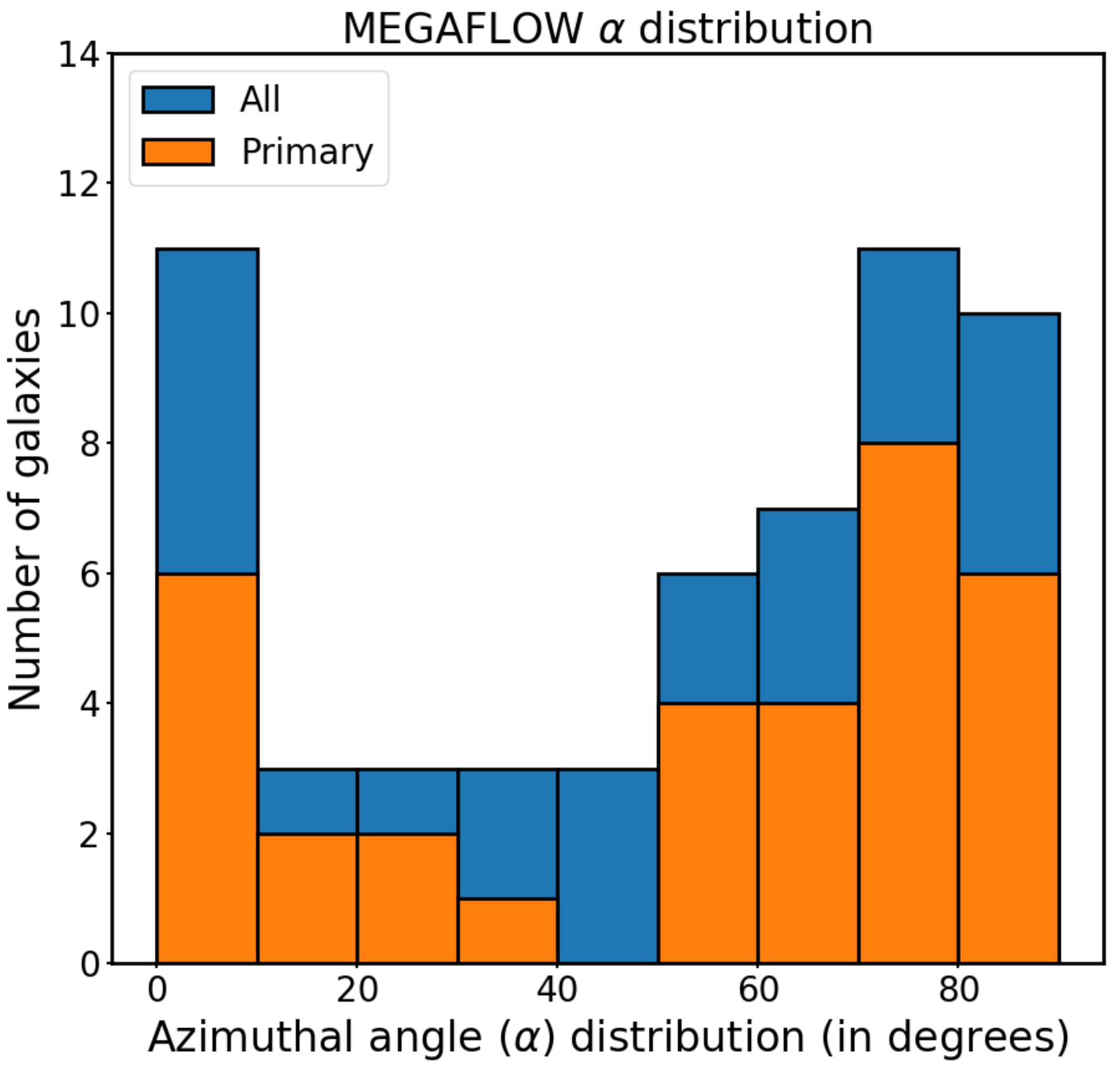}
  \caption{Azimuthal angle distribution of \NpairsfNtwoFlagsPos\ selected galaxies (PA and inclination selected) from the MEGAFLOW survey in blue.
  In orange are the "primary" galaxies (see text).
  We note the bimodal distribution of the whole survey.}
  \label{fig:surveys_alpha}
\end{figure}

Figure~\ref{fig:surveys_alpha} shows the distribution of azimuthal angle for the current MEGAFLOW sample. 
In this Figure, we also show  the subsample of galaxies that are the closest to the QSO LOS as well as being the brightest in \OII\ flux/luminosity (defined as `primary', see Paper II for more details). 
This azimuthal distribution of the primary galaxies (in orange) shows a clear bimodal behavior, confirming previous results \citep[e.g.][]{BordoloiR_11a,bouche_12,KacprzakG_12a}.
This bimodal distribution means that the cool gas traced by \MgII\ is either along the galaxy minor axis or aligned with the disk
and can be interpreted as the simultaneous signature of bi-conical outflows  along the minor axis  and an extended/infalling gaseous disk along the major-axis.

In this paper, we focus on outflowing gas around galaxies, and thus, we restrict ourselves to pairs whose azimuthal angle $\alpha$ is larger than 55$^\circ$ (according to the bimodality of $\alpha$ distribution), bringing  our wind subsample to \NpairsWind\ wind pairs from the \NpairsfNtwoFlagsPos\ galaxy-quasar pairs.

In addition, we impose  a minimum \OII\  flux
of \FluxLimit, 
leaving \NpairsWindFluxthree\ galaxies meeting this criterion.  Finally, we set a minimum inclination of 35$^\circ$ in order to avoid face-on galaxies with inevitably large errors on the galaxy PA (and thus large errors on $\alpha$), bringing our final subsample to \NpairsIncl\ wind pairs.

Out of this subsample, we found two possible major mergers\footnote{Including the galaxy corresponding to J213748+0012G2 in \citet{schroetter_16}.} and chose to not include them in the analysis. Thus, our wind  sub-sample is made of \NpairsFinal\ pairs. 
Out of those \NpairsFinal\ pairs, 21 are "isolated" (N100~$=1$ and and no other galaxy detected with 50~\kpc\ transverse distance and within the searched velocity window from those galaxies).

\subsection{Final subsample selection summary}
 
To summarize, from the \NabsTot\ strong \MgII\ absorbers in our 22  fields, we identified one or more galaxies for \Nabs\ (75\%). 
 A total of \NgalTot\ galaxies were detected, among which \Npairs\ galaxies are found within 100~\kpc\ of the QSO LOS. 
 
 Out of these \Npairs\ galaxies within 100~\kpc\ to the QSO LOS, we selected those with

 \begin{itemize} 
 \item  at most one companion, i.e. N100$\leq2$
 \footnote{we also searched for other companions further away from those galaxies and found none within 50~\kpc.}
 : \NpairsfNtwo;
 \item with a well constrained PA (i.e. a flag $>=1$): \NpairsfNtwoFlagsPos;
 \item that are suitable for studying winds, i.e. with an azimuthal angle $\alpha \geq 55^\circ$: \NpairsWind;
 \item that have a \OII\ flux greater than \FluxLimit: \NpairsWindFluxthree; 
 \item have an inclination $i\geq 35^\circ$: \NpairsIncl;
 \item not be a major merger: \NpairsFinal.
 \end{itemize}

In order to uniquely identify our galaxies, we adopt a specific nomenclature convention for them, like J0103p1332-1048-1-136 in Table~\ref{table:megaflow_wind} for instance.
The first part of a galaxy name corresponds to the quasar field in which it belongs (J0103p1332 for 01:03:32.37 +13:32:36.05).
Then the next numbers correspond to the absorber redshift (1048 for $z=1.048$), the impact parameter (in arcseconds, 1\arcsec\ for this pair) and finally the angle where the galaxy is located with respect to the QSO LOS (defined like the PA of a galaxy, 136$^\circ$).

\begin{table*}
\centering
\tiny
\caption{MEGAFLOW final wind pairs subsample}
\label{table:megaflow_wind}
\begin{tabular}{llcrclccccccc}
\hline
\# & Galaxy name & redshift & $b$ & incl & $V_{\rm max}$& $\sigma$ & $r_{1/2}$ &$\alpha$ & Flux$_{\OII\ }$ & flag &  N100 & comment\\
(1) & (2) & (3) & (4) & (5) & (6) & (7) & (8) &(9) & (10) & (11) & (12)&(13)\\
\hline
1& J0014m0028-0834-1-159 & 0.8340 & 9.7 & 86$\pm 3$ &   8$\pm 5$&  40$\pm 3$ & 3.0$\pm0.1$ & 89$\pm0$ & 0.31$\pm0.01$ & 5 & 1 \\
2& J0014m0028-1052-6-268 & 1.0536 & 52.4 & 65$\pm 5$ &  44$\pm15$&  77$\pm 4$ & 3.5$\pm0.2$ & 80$\pm3$ & 0.35$\pm0.01$ & 5 & 1
 &\\
3& J0015m0751-0500-4-35 & 0.5073 & 24.1 & 87$\pm 2$ & 262$\pm11$&  35$\pm 4$ & 6.6$\pm0.2$ & 71$\pm 1$ & 0.72$\pm0.01$ & 5 & 1
& H\&K \\
4& J0015m0751-0731-5-3 & 0.7305 & 35.5 & 66$\pm 2$ & 266$\pm13$&  33$\pm 6$ & 8.4$\pm0.3$ & 71$\pm 1$ & 0.33$\pm0.01$ & 3 & 1
 & H\&K  \\
5& J0015m0751-0810-3-357 & 0.8160 & 20.7 & 38$\pm 3$ & 284$\pm26$  &  45$\pm 7$ & 2.4$\pm0.2$ & 73$\pm 4$ & 0.35$\pm0.01$ & 3 & 1 
& H\&K \\
6& J0103p1332-1048-1-136 & 1.0483 & 9.1 & 76$\pm 9$ &  45$\pm41$&  87$\pm13$ & 1.9$\pm0.4$ & 89$\pm9$ & 0.33$\pm0.01$ & 3 & 1 \\
7& J0131p1303-1010-3-45 & 1.0103 & 26.4 & 62$\pm 2$ & 108$\pm 4$&  33$\pm 2$ & 3.2$\pm0.1$ & 71$\pm 1$ & 1.24$\pm0.01$ & 5 & 1 \\
8& J0131p1303-1104-9-351 & 1.1049 & 75.5 & 83$\pm 3$ &  94$\pm 8$&  66$\pm 4$ & 3.9$\pm0.1$ & 61$\pm 1$ & 0.86$\pm0.01$ & 5 & 2 
 & \\
9&J0145p1056-0770-1-93 & 0.7699 & 12.9 & 87$\pm 2$ & 103$\pm15$&  21$\pm10$ & 1.9$\pm0.5$ & 89$\pm 5$ & 0.16$\pm0.01$ & 3 & 1 \\
10&J0800p1849-0843-3-254 & 0.8429 & 20.9 & 70$\pm 1$ & 138$\pm 4$&  15$\pm 5$ & 7.0$\pm0.2$ & 79$\pm 1$ & 0.64$\pm0.01$ & 3 & 1 \\
11&J0800p1849-0993-9-282 & 0.9936 & 78.0 & 71$\pm 1$ &  79$\pm 2$&  46$\pm 1$ & 8.3$\pm0.1$ & 65$\pm 1$ & 1.48$\pm0.02$ & 5 & 1 \\
12&J0937p0656-0702-10-197 & 0.7019 & 69.0 & 50$\pm 1$ & 136$\pm 5$&  39$\pm 2$ & 3.0$\pm0.1$ & 58$\pm 1$ & 1.14$\pm0.01$ & 5 & 2
& 2nd: 13\\
 13&J0937p0656-0702-6-209 & 0.7020 & 38.7 & 55$\pm 1$ & 215$\pm11$&  51$\pm 2$ & 4.3$\pm0.1$ & 87$\pm 1$ & 1.81$\pm0.01$ & 3 & 2
& H\&K;  2nd: 12\\
14& J0937p0656-0933-5-6 & 0.9337 & 41.4 & 77$\pm 1$ & 107$\pm 7$&  44$\pm 2$ & 4.6$\pm0.1$ & 75$\pm 1$ & 0.91$\pm0.01$ & 5 & 1 \\
15&J1039p0714-0819-3-124 & 0.8192 & 24.5 & 73$\pm 1$ & 243$\pm 6$&  30$\pm 5$ & 6.7$\pm0.2$ & 63$\pm 1$ & 0.23$\pm0.01$ & 5 & 1 & H\&K\\
16&J1039p0714-0949-9-344 & 0.9492 & 72.2 & 61$\pm 2$ & 129$\pm10$&  50$\pm 6$ & 1.5$\pm0.1$ & 68$\pm 3$ & 0.34$\pm0.01$ & 3 & 2 & 2nd: in paper~II\\
 17&J1039p0714-1359-1-123 & 1.3589 & 8.6 & 70$\pm 1$ &  34$\pm12$ &  46$\pm 2$& 6.1$\pm0.2$ & 80$\pm 1$ & 0.37$\pm0.01$ & 3 & 1 & \\
 18&J1107p1021-1015-10-272 & 1.0150 & 80.9 & 54$\pm 3$ & 373$\pm18$ &  10$\pm 9$ & 7.2$\pm0.4$ & 75$\pm 3$ & 0.26$\pm0.01$ & 5 & 1
 & H\&K\\
 19&J1107p1757-1063-3-140 & 1.0637 & 22.1 & 77$\pm 7$ &  75$\pm14$&  45$\pm 6$ & 2.2$\pm0.3$ & 78$\pm 7$ & 0.39$\pm0.01$ & 5 & 1 \\
 20&J1107p1757-1163-6-166 & 1.1618 & 44.4 & 57$\pm 5$ & 113$\pm13$&  44$\pm 4$ & 4.5$\pm0.3$ & 88$\pm 4$ & 0.78$\pm0.03$ & 5 & 2 & 2nd: low $i$, accr \\
21& J1236p0725-0639-10-256 & 0.6382 & 66.9 & 68$\pm 1$ & 230$\pm10$&  24$\pm 6$ & 6.7$\pm0.2$ & 70$\pm 1$ & 0.85$\pm0.01$ & 5 & 2
& H\&K, 2nd closer\\
 22&J1352p0614-0604-2-260 & 0.6039 & 14.0 & 80$\pm 7$ &  35$\pm11$&  24$\pm 8$ & 3.8$\pm0.5$ & 79$\pm1$ & 0.29$\pm0.02$ & 5 & 1 \\
 23&J1358p1145-0809-2-202 & 0.8093 & 12.7 & 65$\pm 2$ &  61$\pm 5$&  47$\pm 2$ & 2.5$\pm0.1$ & 80$\pm 2$ & 0.55$\pm0.01$ & 5 & 1\\
 24&J1425p1209-0597-1-87 & 0.5968 & 9.6 & 54$\pm 1$ &  56$\pm 3$&   6$\pm 4$ & 0.9$\pm0.1$ & 64$\pm 2$ & 1.09$\pm0.02$ & 5 & 1
& compact \\
 25&J1425p1209-0865-8-353 & 0.8657 & 60.8 & 43$\pm 2$ & 101$\pm 5$&  18$\pm 2$ & 3.4$\pm0.1$ & 59$\pm 2$ & 1.02$\pm0.01$ & 5 & 1 \\
 26&J2152p0625-1319-4-187 & 1.3181 & 32.5 & 71$\pm 6$ &  82$\pm16$&  37$\pm 6$ & 4.3$\pm0.4$ & 88$\pm 4$ & 0.19$\pm0.01$ & 3 & 1
& has companion\\

\hline
\end{tabular}\\
{  
(1) Galaxy number;
(2) Extended name;
(3) Redshift;
(4) Impact parameter $b$ (kpc) (with typical errors of $\approx$0.2 kpc);
(5) inclination (degrees);
(6) Maximum rotational velocity $V_{\rm max}$ (\kms);
(7) Dispersion velocity $\sigma$ (\kms);
(8) Half-light radius (kpc); 
(9) Azimuthal angle $\alpha$ (degrees);
(10) \OII\  flux ($\times 10^{-16}$ \flux);
(11) \galpak\ flag;
(12) N100, the number of galaxies for one absorber within 100~\kpc;
(13) Comments if needed.
}

{\it Note.} Errors are $1\sigma$.
\end{table*}

\section{Results}
\label{results}

\subsection{Radial dependence: How far do winds propagate?}
\label{section:samples_how_far}

\begin{figure}
  \centering
  \includegraphics[width=8cm]{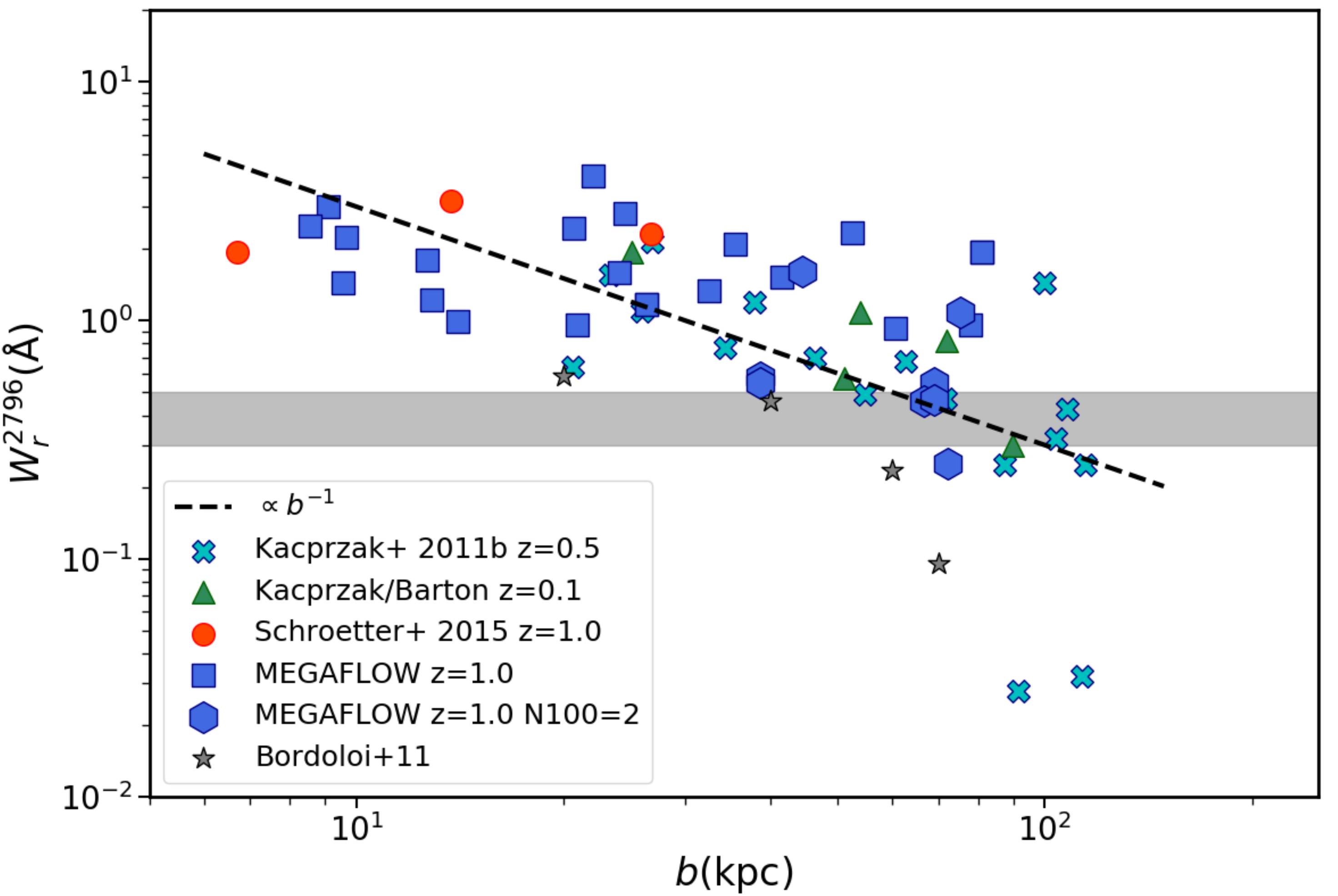}
  \caption{\MgII\ ($\lambda 2796$) rest equivalent width  as a function of impact parameter $b$ for galaxy-quasar pairs classified as wind-pairs.
  The REWs were measured from the UVES spectra.
  The gray area represents the REW selection criterion (see text).
  The thick black dashed line represents the expected $W_r^{\lambda 2796} \propto b^{-1}$. 
   The blue square and hexagon below the threshold appears because we plot the UVES derived REWs whereas the survey threshold was for the SDSS spectra. 
   The blue hexagons are the cases with 2 galaxies detected within 100~\kpc\ from the QSO LOS.
  }
  \label{fig:rew_vs_b_all}
\end{figure}

For the \NpairsFinal\ wind-pairs in our sample,  we investigate the radial dependence of  $W_r^{\lambda 2796}$ as a function of impact parameter $b$. 
Figure~\ref{fig:rew_vs_b_all} shows the \MgII\ REW\footnote{\MgII\ REW are from SDSS catalog and also derived from our UVES data for cross checking.} as a function $b$ for each of the \NpairsFinal\ galaxies.
The blue squares are the MEGAFLOW wind-pairs whereas orange circles are from the SINFONI-based SIMPLE sample \citep{bouche_07,schroetter_15}. 
Hexagons correspond to the MEGAFLOW pairs for which N100=2. 
Dark stars and cyan crosses are wind-pairs from \citet{BordoloiR_11a}\footnote{as in their paper, due to low spectral resolution, they only have EWs for both \MgII\ components, we divided their values by a factor 2 in this Figure.} and \citet{KacprzakG_11b} respectively. 
We choose not to include the fit from MAGIICAT \citep{NielsenN_13b} since we are only showing the wind-selected galaxies and they do not make such selection.
The dashed line shows the relation $W_r^{\lambda 2796}\propto b^{-1}$ expected for a bi-conical geometry and mass conservation from \citet{bouche_12}. 
It is evident from this figure that an anti-correlation between $W_r^{\lambda 2796}$ and $b$ appears to be consistent with the $b^{-1}$ expectation.
In other words, it seems that galactic outflows (as traced by strong \MgII\ absorbers) are able to travel at least 80-100~\kpc\ away from their host galaxy. In section~\S~\ref{escape}, we will address the question whether the clouds escape the potential well of the host.

\subsection{Galaxy properties}
\subsubsection{Stellar Mass}
\label{mass_estimates}

Our sample of galaxy-absorber pairs is \MgII\ absorption selected sample. 
We therefore investigate whether the host galaxies are normal star-forming galaxies, i.e. whether they lie  ont the SFR$-M_\star$  main sequence (MS).  Any deviations from the MS could shed light on the connection between outflow properties and star-formation activity.
 
We first  estimate the galaxy stellar masses from the tight correlation between stellar mass and the dynamical estimator $S_{05}=\sqrt{0.5\times V_{\rm max}^2 + \sigma^2}$ \citep[e.g.][]{weiner_06, kassin_07,price_16,StraatmanC_17,AlcornL_18a,AquinoE_18}, 
which combines the galaxy dispersion velocity, $\sigma$\footnote{Derived by \galpak}, and its rotational velocity $V_{\rm max}$.
Then, we use the following relation  from \citet{AlcornL_18a}:
\begin{equation}
\log(S_{0.5}) = A\log(M_\star/\rm M_\odot -10) +B \label{equation:alcorn}
\end{equation}
where the slope $A=0.34$ and the zero-point $B=2.05$, appropriate for  a \citet{chabrier_03} IMF. 

For self-consistency, we checked that this relation (obtained from 2D spectra) agrees when the kinematics are determined with IFU 3D data, such as in our case using  the kinematic 3D data-set obtained with MUSE at  $\approx30$hr depth. There are two such data sets. The first one is from
\citet{continiT_16a}  who presented the kinematic analysis of the Hubble-Deep-Field-South \citep[HDFS][]{BaconR_15a}, extending the Tully-Fisher (TF) relation to the low mass regime, $M_\star=10^8$-$10^9 \rm M_\odot$ for $\approx$30 galaxies. The second data set consists of $\approx$ 300 galaxies from Contini et al. (in prep.), who used the 3'$\times$3' MUSE mosaic of the Hubble-Ultra-Deep-Field \citep[HUDF][]{BaconR_17a}. The S05-$M_\star$ relation of  Eq.~\ref{equation:alcorn}  is found to be consistent with the MUSE results of Contini et al. (in prep.).

\subsubsection{Star Formation Rate (SFR)}
\label{subsection:data-analysis-sfr}

To estimate SFRs from \OII\ fluxes, we proceed as in Paper~II, namely, we use the $M_\star - E(B-V)$ relation obtained by \citet{GarnBest_10} due to the lack of multiple lines and direct constraints on the amount of extinction, $E(B-V)$.
The \citet{GarnBest_10} relation corrected from a \citet{KroupaP_01} to a \citet{chabrier_03} IMF is:
\begin{equation}
E(B-V) = (0.93+0.77X+0.11X^2-0.09X^3)/k_{H_\alpha}
\label{garnbest_equ}
\end{equation}
where $X=\log(M_\star/\rm M_\odot)-10$ and $k_{H_\alpha}=3.326$ for the \citet{CalzettiD_00a} extinction law. 
Errors on $E(B-V)$ are calculated from the 0.3 mag scatter of this relation combined with the 0.15 dex error from the $M_\star$ estimation.

We correct the observed \OII\ luminosities $L_{\rm o}$ with these extinctions  using  a \citet{CalzettiD_00a} extinction curve.
From these intrinsic luminosities $L_{\rm i}$, we estimate the SFRs using the \citet{kewley_04} calibration
\begin{equation}
SFR(\OII\ ) = 4.1\times 10^{-42}(L_{\rm i}\OII\  \rm erg \rm ~s^{-1}) \rm M_\odot \rm yr^{-1} \label{eq:kewley_sfr}
\end{equation}
adjusted from a \citet{salpeter_55} to a \citet{chabrier_03} IMF.

 \subsection{Main-Sequence}
 
\begin{figure}
  \centering
  \includegraphics[width=8cm]{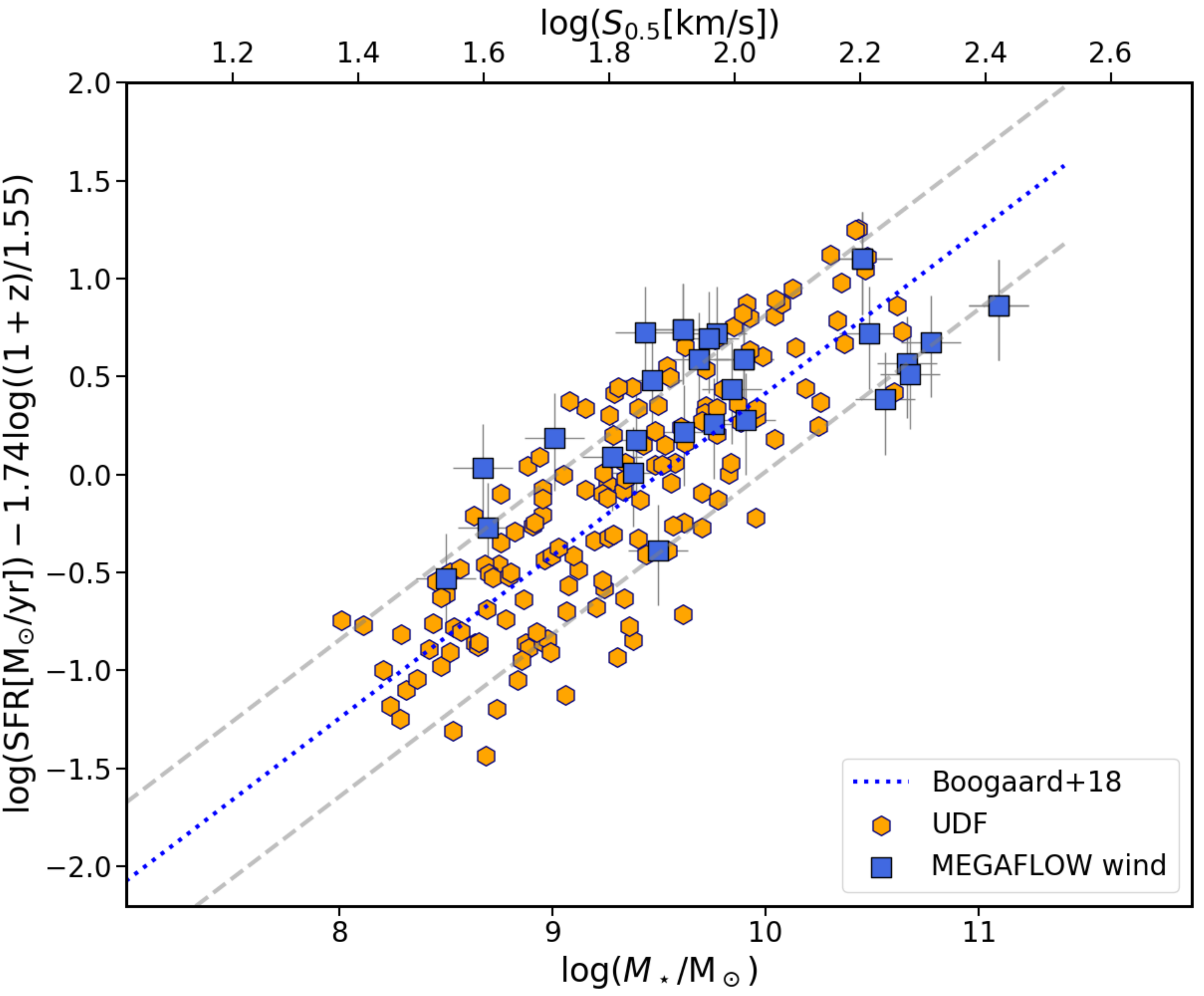}
  \caption{Star formation rate  as a function of galaxy stellar mass (bottom x-axis) and dynamical estimator $\log(S_{0.5})$ (top x-axis).
  The blue squares represents the MEGAFLOW wind subsample, while the orange points represent
  the   MUSE-HUDF data from Contini et al. (in prep.).
The data are corrected to  redshift $z=0.55$ using the  \citet{BoogaardL_18} redshift evolution of the MS.
  The blue dashed line represents the \citet{BoogaardL_18} fit to the MS and the grey dashed lines the 0.4~dex intrinsic scatter of this relation.
  }
  \label{fig:sfr_vs_mstar_vmax}
\end{figure}

Having estimated stellar masses and  star-formation rates, we can place our galaxies on the SFR-$M_\star$ diagram.
Figure~\ref{fig:sfr_vs_mstar_vmax} shows the SFR-$M_\star$ diagram for the HUDF (orange hexagons) and the MEGAFLOW (blue squares) wind subsample. In this Figure, the SFRs and stellar masses were derived using the method described before and the MS is presented at a common redshift ($z=0.55$) using the redshift evolution from \citet{BoogaardL_18}. The MS relation obtained from the HUDF using different $M_\star$ and SFR derivations is shown by the blue dashed line \citep{BoogaardL_18}.

 Figure~\ref{fig:sfr_vs_mstar_vmax} shows that  the wind subsample from the MEGAFLOW survey tends to follow the galaxy MS. However, the data
suggest that below $\log(M_\star/\rm M_\odot)\approx 10$, our wind galaxies could be  slightly above the galaxy MS, while galaxies
above this mass tend to be preferentially   below the MS, suggesting that these are   in the process of quenching their SF.
See \citet{RhodinN_18} for a similar result for HI-selected hosts. 

\section{Wind modeling}
\label{wind}

Having measured the morpho-kinematic properties of our galaxies, we focus on deriving outflow properties. 
For the \NpairsFinal\ wind-pairs, we attempt to constrain the wind kinematics using the same method as used in \citet{bouche_12,schroetter_15, schroetter_16}. 

\subsection{Classic wind model}
We use a bi-conical wind model filled with randomly distributed particles\footnote{These particles represent cold gas clouds being pushed away by the hot medium or radiation pressure.}.
We assume mass conservation throughout the outflowing cone (thus, density evolves like $1/r^2$, $r$ being the distance to the galaxy center). 
The clouds are also assumed to be accelerated with respect to their terminal velocity \Vout\ in a few \kpc\ ($<10$ \kpc), i.e. the wind speed is assumed to be constant in the observed impact parameter (range from 10 to 100 \kpc).  

The particle observed velocities are then projected onto the quasar LOS at the impact parameter. 
This projection gives an optical depth $\tau_\nu$ which we turn into a simulated absorption profile (flux $\propto \exp(-\tau_\nu)$). 

The geometrical configuration, namely the wind direction, is determined from the galaxy's orientation (inclination and PA), assuming a wind flowing radially from the host galaxy. 
The wind model thus has two free parameters: the wind speed $V_{\rm out}$ and the cone opening angle $\thetam$.
They can both be adjusted to match the absorption profile seen in the data\footnote{We also model a disk contribution for each wind model,  as in \citet{schroetter_15, ZablJ_19} but found the disk contribution too low due to the large galactic-radius resulting from the high-inclinations of our galaxies.}.  

In order to facilitate comparison with the data, we add Poisson noise (corresponding to instrumental noise) to the simulated absorption profile. 
We thus derive an outflow velocity as well as a cone opening angle for each individual wind-pair. 
This is achieved by visually matching\footnote{The EW, taking into account the depth of the profile, cannot be estimated as the normalization of the particles in $\tau$ in our model is arbitrary.} the absorption profile edges, shape and asymmetry. 

\subsection{Empty inner cone}

While we use a filled cone by default, in some cases, the data require us to use a hollow (within $\theta_{\rm in}$) cone. 
This hollow inner cone produces a gap in absorption velocities in our simulated profiles. 
These gaps in absorption velocities can occur in the data when $\alpha$ is close to 90$^\circ$, i.e. when the quasar LOS intercepts the middle of the outflowing cone. 
This is the case for the galaxy-quasar pairs \#1, 7, 9, 17, 19, 20, 22, 24 and 26.
9 out of 26 galaxies with $\alpha \geq 65^\circ$ require a hollow inner cone.
 
As mentioned in Paper I, this empty inner cone could be the signature of a highly ionized gas component filling the inner cone.
Thus, the low-ionized gas which we are tracing is entrained along the outskirts of the outflowing cone, in a manner similar to \citet{fox_15} for the MilkyWay as well as observations from \citet{veilleux_02,veilleux_03} and \citet{bland_hawthorn_07}.

\begin{figure*} 
   \centering 
   \includegraphics[width=12.0cm]{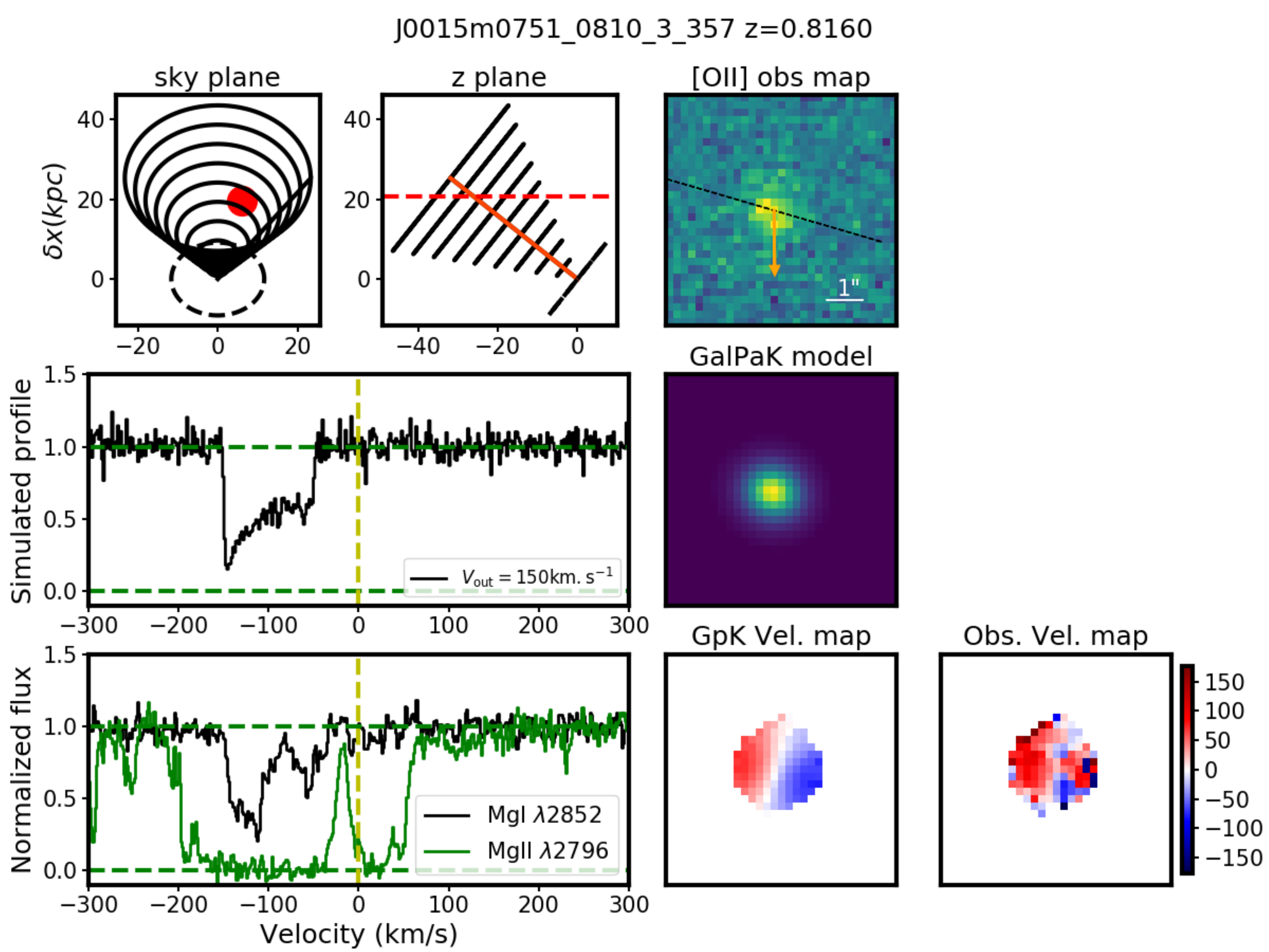} 
   \caption{Simulated profile and quasar spectrum associated with the J0015m0751-0810-3-357 (\#5) galaxy. 
  The first top-left two panels represent the geometrical configuration of the galaxy-quasar system with the quasar line of sight in red (dot for sky plane representation in the top left panel and dashed horizontal line for the z plane), 
  the outflowing cone as black circles and the host galaxy in a dashed black circle.   
  The middle left panel shows the best fit simulated wind profile corresponding to the observed \MgI\ absorption profile (centered at $z=0.8160$) from UVES shown in the bottom left panel (the green \MgII\ absorption profile is present to show that this line is saturated). 
  This outflow has \Vout\ = $150\pm10$ \kms\ and opening angle \thetam\ = $35\pm2^\circ$.
  The top right panel shows the observed \OII\ flux map of the galaxy from the MUSE cube including a representation of the galaxy PA (black dashed line) as well as an orange arrow showing the direction to the associated quasar. 
  The middle right panel shows the \galpak\ model (convolved with the PSF), the panel below shows the model velocity map of the galaxy obtained with \galpak\ (convolved with the PSF), and the bottom right panel shows the observed velocity map obtained with \camel.} 
   \label{fig:J0014m0028_0834_1} 
\end{figure*}

For four wind-pairs (\#7, 12, 13 and 18), the \MgII\ absorption seen in the UVES quasar spectrum is too complex to determine which component is actually the signature of outflows.  
Therefore, we create a wind model for each component when possible. 
The results of these models are listed in Table~\ref{table:megaflow_outflow}. 

Figure~\ref{fig:J0014m0028_0834_1} shows the best-fit wind model for the galaxy J0015m0751-0810-3-357 (\#5). 
The top two left panels represent the geometrical configuration of the system. 
The top left panel represents the sky view of this galaxy-quasar pair.
The QSO LOS is represented by the red dot and the galaxy by the dashed black circle. 
The outflowing cone is represented by the black circles. 
The middle top panel of shows a side view of the same system. 
The quasar LOS is the horizontal dashed red line (the observer being to the left), the galaxy is represented by the dashed inclined black line at the bottom and the outflowing cone by the increasing black lines. 

The right column of Figure~\ref{fig:J0014m0028_0834_1} shows, from top to bottom, the MUSE host galaxy \OII\ map, the \galpak\ model and the model velocity map.
On the top right observed flux map we represent the galaxy PA by the dashed black line and the direction of the quasar with the orange arrow. 

The last two panels of this Figure show the simulated profile of our wind model (middle left) and the UVES \MgI\ absorption lines (bottom left). 
On both panels the galaxy systemic redshift is represented by the vertical yellow dashed line. 
With an outflow velocity \Vout\ = 150~\kms and a cone opening angle of 35$^\circ$, we reproduced the width and asymmetry of the observed \MgI\ absorption. 

Outflow velocities and cone opening angles fit with our model are listed for each wind-pair in Table~\ref{table:megaflow_outflow}. 
Representations of each model are shown in the Appendix.

\subsection{Does the wind material escape?}
\label{escape}

\label{section:samples_does_it_escape}
Here, we will address the question of whether outflows can escape the gravitational potential well of their host galaxy.
To estimate the escape velocity of our galaxies, we use the relation for an isothermal sphere given by equation~\ref{eq:Vesc} from \citet{veilleux_05}:
\begin{equation}
 V_{\rm esc}=V_{\rm vir}\times\sqrt{2\left[1+\ln \left(\frac{R_{\rm vir}}{r}\right)\right]}\label{eq:Vesc}
\end{equation}
where $V_{\rm vir}$ is the virial velocity of the galaxy, $r$ is taken to be the impact parameter $b$ and $R_{\rm vir}$ its virial radius.
The virial radius is defined approximately as $R_{\rm vir} \approx V_{\rm vir}/10H(z)$ where $H(z)$ is the Hubble constant at redshift $z$. 
For our galaxies, we choose to use $1.2\times S_{0.5}$ as a proxy for $V_{\rm vir}$.  Indeed, several groups have shown that $V_{\rm vir}$ is
$V_{\rm max}/ 1.1$--1.3   \citep{Dutton:2010a,CattaneoA_14}, which is a factor similar to $(1.2\times \sqrt{0.5})^{-1}$ in S05.

\begin{figure}
  \centering
  \includegraphics[width=8cm]{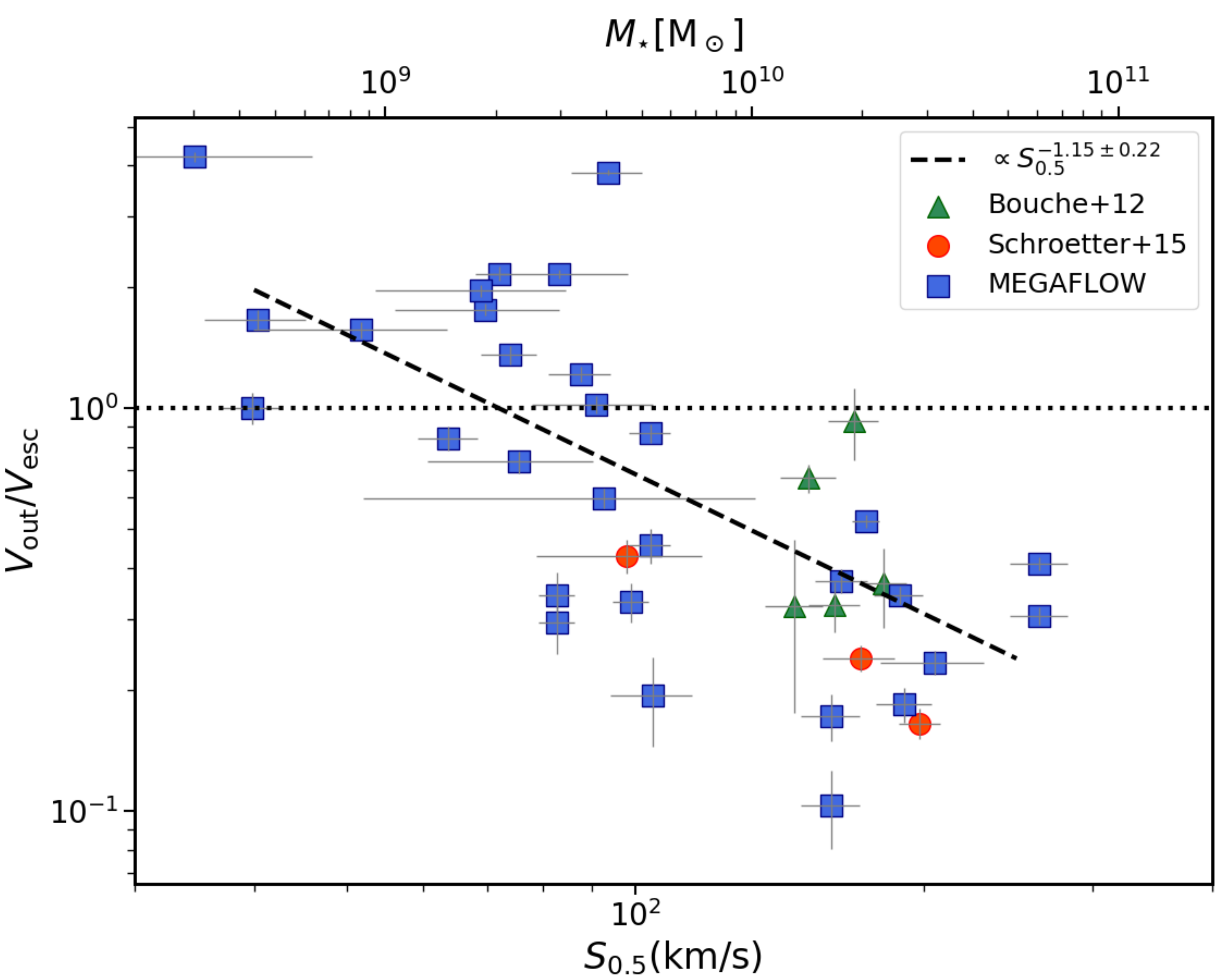}
  \caption{Ratio of best-fit outflow and escape velocity, $V_{\rm out} / V_{\rm esc}$, as a function of dynamical mass indicator $S_{0.5}$ (bottom x-axis) and $M_\star$ (top x-axis). 
  Green triangles are from \citet{bouche_12}, red circles are from \citet{schroetter_15}.
  The blue squares correspond to the MEGAFLOW wind subsample.
  The horizontal line corresponds to $V_{\rm out} = V_{\rm esc}$. 
  The dashed black line corresponds to a fit with coefficients shown in the legend. 
  }
  \label{fig:vesc_mdyn}
\end{figure}

Figure~\ref{fig:vesc_mdyn} presents the ratio between the outflow velocity and the escape velocity ($V_{\rm out} / V_{\rm esc}$) as a function of $S_{0.5}$ (and the galaxy stellar mass along top x-axis). 

Figure~\ref{fig:vesc_mdyn} also shows results from other studies using the background quasar technique. 
In particular, green triangles for \citet{bouche_12} (a combination of LRIS and SDSS data) and red circles for \citet{schroetter_15} (SIMPLE, a combination of SINFONI and UVES) are shown.
The blue squares are the MEGAFLOW wind sub-sample.
We can see that for galaxies with stellar masses lower than $\approx 4\times10^9 \rm M_\odot$, for most of the cases, $V_{\rm out} / V_{\rm esc}>1$.
Those outflows can thus escape the gravitational potential well of their host galaxies.

The ability of the cool wind material (traced by \MgII) to escape the galaxy appears to be limited to low-mass galaxies, with $M_{\star}\lesssim 4\times10^{9}$ M$_\odot$.
For galaxies above this mass, outflows are likely to fall back onto their host and thus fuel future star formation, which is consistent with theoretical expectations \citep[e.g.][]{oppenheimer_08,oppenheimer_10,TorreyP_17,AnglesD_17}. 

\subsection{The mass outflows rate}

For a mass conserving flow, the mass outflow rate    $\dot M_{\rm out}$  is $\rho(R)\,R^2\,V_{\rm out}\,\Omega$, i.e. it  depends critically on four factors, the outflow speed $V_{\rm out}$, the gas mean localization $R$, the column density $N=\rho\,R$ and the wind solid angle  $\Omega$. For a down-the-barrel observations of such a wind,  the
mass outflow rate  reduces to $\propto N_{\rm H}\, R_0\, V_{\rm out}\, \Omega$ \citep{HeckmanT_00a,MartinC_05a,MartinC_12a} where     $R_0$ the launch radius. 
For transverse sight-lines,   $\dot M_{\rm out}$ is proportional to $\propto N_{\rm H}\,b\,V_{\rm out}\,\theta$ where $b$ is the impact parameter and $\theta$ the wind opening angle, as derived in \citet{bouche_12}. This can be understood using the following two observations: (i)  $M_{\rm out}$ is $\propto \rho(b) b^2 \, V_{\rm out}\theta^2$ from mass conservation and (ii) the gas column density $N$ depends linearly on the opening angle $N\propto\rho(b)\,b\, \theta$ for a transverse sight-line.

Hence, for a potentially hollow bi-conical flow,
the mass outflow rate is  \citep[as in][and paper~I]{bouche_12,schroetter_15} :
\begin{equation}
{\dot M_{\rm out} \over \mpy\ } \approx {\mu \over 1.5} \cdot {N_{\rm H}(b)\over 10^{19} \rm cm^{-2}} \cdot {b\over 25 \rm kpc} \cdot 
{V_{\rm out} \over 200 \kms\ } \cdot {\theta_{\rm max}-\theta_{\rm in} \over 30^{\circ}},\label{eq:Mout}
\end{equation}

where $\mu$ is the mean mass per hydrogen particle, $b$ the impact parameter, \thetam\ the cone opening angle\footnote{\thetam\ ($\theta_{\rm in}$) is defined from the central axis, 
and the cone subtends an area $\Sigma$ of $\pi \cdot \theta_{\rm max}^2$.}, $\theta_{\rm in}$ the opening angle of the inner empty cone, \Vout\ the outflow velocity and $N_{\rm H}(b)$ the hydrogen column density at the $b$ distance. 
The numerical factor here includes a factor of $2\times$ to sum the mass flux for both cones. 

The parameters $V_{\rm out}$, $b$ and the cone opening angle can be constrained from our data.
To estimate the last parameter $N_{\rm H}(b)$, we use the empirical relation (Eq.~\ref{eq:menard}) from \citet{menard_09}, re-derived by \citet{LanT_17}, between the neutral gas column density and  $W_r^{\ma}$ : 
\begin{equation}
 N_{\rm HI} (\hbox{cm}^{-2}) = A \left(\frac{W_r^{\lambda 2796}}{1 \rm \AA}\right)^{\alpha}(1+z)^{\beta}\label{eq:menard}.
\end{equation}
Where $A=10^{18.96 \pm 0.10}$, $\alpha=1.69 \pm 0.13$ and $\beta=1.88 \pm 0.29$.

If a region has an \HI\ column density above $\log(N_{\rm HI}/\hbox{cm}^{-2})\approx 19.5$, the ionized gas contribution is negligible.
Thus, one can use the correlation between \MgII\ equivalent width and $N_{\rm HI}$ as a proxy for the hydrogen gas column density \citep[also argued by][]{jenkins_09}.  
Typical errors on our $\log(N_{\rm HI})$ estimates are 0.2-0.3~dex (at 1$\sigma$).
Those errors, together with errors on the other parameters (\Vout, \thetam\ and $b$), allow us to get estimates of mass outflow rates within a factor 2 or 3.
The mass outflow rates are listed in Table~\ref{table:megaflow_outflow}.

\begin{table*}
\centering
\caption{Results on outflow properties for MEGAFLOW galaxies.}
\label{table:megaflow_outflow}
\begin{tabular}{llcrcrcrrrrcc}
\hline
\#&Galaxy & $z_{\rm gal}$ & $b$ & log($N_{\rm H}$($b$)) & $V_{\rm out}$ & $\theta_{\rm max}$ & $\theta_{\rm in}$& $\log(M_\star)$& SFR & $\dot M_{\rm out}$ & $V_{\rm out}/V_{\rm esc}$ & $\eta$ \\
(1)    & (2)           & (3) & (4)                   & (5)           & (6)                & (7) & (8)                & (9)                       & (10)  &(11) &(12)& (13)\\
\hline
 1 & J0014m0028-0834-1-159 & 0.8340 & 9.7 & 20.0$^{+0.2}_{-0.2}$ & 180.0 & 28 &  2 & 8.7 &0.7$^{+0.5}_{-0.3}$ & 3.3$^{+0.3}_{-2.1}$ & 2.07 & 4.6 \\
 2 & J0014m0028-1052-6-268 & 1.0536 & 52.4 & 20.2$^{+0.2}_{-0.2}$ & 360.0 & 15 &  0 & 9.6 &2.7$^{+2.0}_{-1.3}$ & 28.1$^{+5.0}_{-18.0}$ & 2.77 & 10.4 \\
 3 & J0015m0751-0500-4-35 & 0.5073 & 24.1 & 19.6$^{+0.2}_{-0.2}$ & 200.0 & 30 &  0 & 10.7 &3.5$^{+2.6}_{-1.7}$ & 4.2$^{+0.5}_{-2.2}$ & 0.42 & 1.2 \\
 4 & J0015m0751-0731-5-3 & 0.7305 & 35.5 & 20.0$^{+0.2}_{-0.2}$ & 100.0$^\dagger$ & 25 &  0 & 10.7 &3.9$^{+2.9}_{-1.9}$ & 5.3$^{+0.7}_{-3.1}$ & 0.23 & 1.3 \\
 5 & J0015m0751-0810-3-357 & 0.8160 & 20.7 & 20.1$^{+0.2}_{-0.2}$ & 150.0 & 35 &  0 & 10.8 &6.2$^{+4.6}_{-3.0}$ & 9.2$^{+1.9}_{-5.2}$ & 0.29 & 1.5 \\
 6 & J0103p1332-1048-1-136 & 1.0483 & 9.1 & 20.4$^{+0.3}_{-0.3}$ & 170.0 & 30 &  0 & 9.8 &2.9$^{+2.1}_{-1.4}$ & 6.9$^{+1.1}_{-4.6}$ & 0.74 & 2.4 \\
 7 & J0131p1303-1010-3-45 & 1.0103 & 26.4 & 19.6$^{+0.2}_{-0.2}$ & 70.0$^\dagger$ & 40 & 18 & 9.6 &8.6$^{+6.3}_{-4.1}$ & 1.2$^{+0.1}_{-0.7}$ & 0.43 & 0.1 \\
   &                          &        &      &                      & 60.0$^\dagger$ & 40 &  0 &     &                    & 1.9$^{+0.2}_{-1.1}$ & 0.37 & 0.2 \\
 8 & J0131p1303-1104-9-351 & 1.1049 & 75.5 & 19.6$^{+0.2}_{-0.2}$ & 650.0 & 30 &  0 & 9.8 &8.9$^{+6.6}_{-4.3}$ & 41.4$^{+9.5}_{-20.4}$ & 5.05 & 4.6 \\
 9 & J0145p1056-0770-1-93 & 0.7699 & 12.9 & 19.6$^{+0.2}_{-0.2}$ & 160.0 & 30 & 10 & 9.5 &0.5$^{+0.4}_{-0.2}$ & 1.0$^{+0.1}_{-0.6}$ & 0.91 & 2.0 \\
10 & J0800p1849-0843-3-254 & 0.8429 & 20.9 & 19.4$^{+0.2}_{-0.2}$ & 90.0 & 30 &  0 & 9.8 &3.7$^{+2.7}_{-1.8}$ & 1.0$^{+0.1}_{-0.6}$ & 0.41 & 0.3 \\
11 & J0800p1849-0993-9-282 & 0.9936 & 78.0 & 19.5$^{+0.2}_{-0.2}$ & 250.0 & 25 &  0 & 9.4 &8.3$^{+5.9}_{-3.9}$ & 10.1$^{+1.7}_{-5.2}$ & 2.90 & 1.2 \\
12 & J0937p0656-0702-10-197 & 0.7019 & 69.0 & 19.0$^{+0.1}_{-0.1}$ & 100.0$^\dagger$ & 30 &  0 & 9.9 &4.6$^{+3.4}_{-2.2}$ & 1.2$^{+0.1}_{-0.6}$ & 0.58 & 0.3 \\
   &                          &        &      &                       & 190.0$^\dagger$ & 25 &  0 &      &                    & 1.5$^{+0.1}_{-0.7}$ & 1.11 & 0.3 \\
13 & J0937p0656-0702-6-209 & 0.7020 & 38.7 & 19.0$^{+0.1}_{-0.1}$ & 45.0$^\dagger$ & 30 &  0 & 10.5 &14.9$^{+11.0}_{-7.1}$ & 0.3$^{+0.1}_{-0.2}$ & 0.13 & 0.0 \\
   &                          &        &      &                       & 75.0$^\dagger$ & 30 &  0 &      &                     & 0.5$^{+0.1}_{-0.3}$ & 0.21 & 0.0 \\
14 & J0937p0656-0933-5-6 & 0.9337 & 41.4 & 19.8$^{+0.2}_{-0.2}$ & 240.0 & 20 &  0 & 9.7 &5.7$^{+4.2}_{-2.7}$ & 6.7$^{+1.0}_{-3.9}$ & 1.21 & 1.5 \\
15 & J1039p0714-0819-3-124 & 0.8192 & 24.5 & 20.2$^{+0.2}_{-0.2}$ & 270.0 & 30 &  0 & 10.6 &3.2$^{+2.4}_{-1.5}$ & 21.2$^{+5.0}_{-12.1}$ & 0.65 & 6.6 \\
16 & J1039p0714-0949-9-344 & 0.9492 & 72.2 & 18.5$^{+0.1}_{-0.1}$ & 40.0$^\dagger$ & 35 &  0 & 9.9 &2.8$^{+2.1}_{-1.4}$ & 0.2$^{+0.1}_{-0.1}$ & 0.25 & 0.1 \\
17 & J1039p0714-1359-1-123 & 1.3589 & 8.6 & 20.3$^{+0.3}_{-0.3}$ & 220.0 & 27 & 12 & 9.0 &3.2$^{+2.2}_{-1.5}$ & 4.0$^{+1.2}_{-3.0}$ & 1.94 & 1.3 \\
18 & J1107p1021-1015-10-272 & 1.0150 & 80.9 & 20.0$^{+0.2}_{-0.2}$ & 270.0$^\dagger$  & 20 &  0 & 11.1 &11.5$^{+8.2}_{-5.4}$ & 30.1$^{+6.3}_{-17.5}$ & 0.52 & 2.6 \\
   &                          &        &      &                        & 200.0$^\dagger$  & 25 &  0 &      &                     & 27.9$^{+6.4}_{-15.6}$ & 0.38 & 2.4 \\
19 & J1107p1757-1063-3-140 & 1.0637 & 22.1 & 20.6$^{+0.3}_{-0.3}$ & 300.0 & 30 & 20 & 9.4 &2.5$^{+1.8}_{-1.2}$ & 16.7$^{+15.0}_{-12.8}$ & 2.21 & 6.8 \\
20 & J1107p1757-1163-6-166 & 1.1618 & 44.4 & 19.9$^{+0.2}_{-0.2}$ & 200.0 & 20 &  5 & 9.7 &8.8$^{+6.5}_{-4.2}$ & 7.6$^{+1.0}_{-4.9}$ & 1.30 & 0.9 \\
21 & J1236p0725-0639-10-256 & 0.6382 & 66.9 & 18.8$^{+0.1}_{-0.1}$ & 150.0$^\dagger$ & 25 &  0 & 10.5 &5.8$^{+4.3}_{-2.8}$ & 1.0$^{+0.1}_{-0.5}$ & 0.47 & 0.2 \\
22 & J1352p0614-0604-2-260 & 0.6039 & 14.0 & 19.3$^{+0.2}_{-0.2}$ & 360.0 & 30 &  15 & 8.5 &0.3$^{+0.2}_{-0.1}$ & 1.1$^{+0.1}_{-0.7}$ & 4.21 & 3.5 \\
23 & J1358p1145-0809-2-202 & 0.8093 & 12.7 & 19.9$^{+0.2}_{-0.2}$ & 150.0 & 45 &  0 & 9.3 &1.6$^{+1.1}_{-0.8}$ & 4.2$^{+0.6}_{-2.4}$ & 1.05 & 2.6 \\
24 & J1425p1209-0597-1-87 & 0.5968 & 9.6 & 19.6$^{+0.2}_{-0.2}$ & 110.0 & 45 &  21 & 8.7 &1.1$^{+0.8}_{-0.5}$ & 0.7$^{+0.1}_{-0.4}$ & 1.25 & 0.6 \\
25 & J1425p1209-0865-8-353 & 0.8657 & 60.8 & 19.4$^{+0.2}_{-0.2}$ & 190.0 & 35 &  0 & 9.5 &4.2$^{+3.0}_{-2.0}$ & 7.0$^{+1.2}_{-3.3}$ & 1.76 & 1.7 \\
26 & J2152p0625-1319-4-187 & 1.3181 & 32.5 & 19.9$^{+0.2}_{-0.2}$ & 285.0 & 12 &  3 & 9.4 &2.1$^{+1.5}_{-1.0}$ & 4.0$^{+0.1}_{-3.0}$ & 2.50 & 1.9 \\
\hline
\end{tabular}\\
{
(1) Galaxy number;
(2) Galaxy name;
(3) Galaxy redshift;
(4) Impact parameter (kpc);
(5) Gas column density at the impact parameter (cm$^{-2}$);
(6) Wind velocity (\kms );
(7) Cone opening angle (degrees)
(8) Inner empty cone opening angle (degrees)
(9) Galaxy stellar mass $\log(M_\circ)$, errors are 0.14 dex
(10) Star Formation Rate (\mpy) from \OII\  (see text);
(11) Ejected mass rate (\mpy);
(12) Ejection velocity divided by escape velocity;
(13) Mass loading factor: ejected mass rate divided by star formation rate;
$^\dagger$: cases of less convincing wind model (see text) 
}
\end{table*}

 \subsection{Mass loading factors}

\begin{figure*}
  \centering
  \includegraphics[width=18cm]{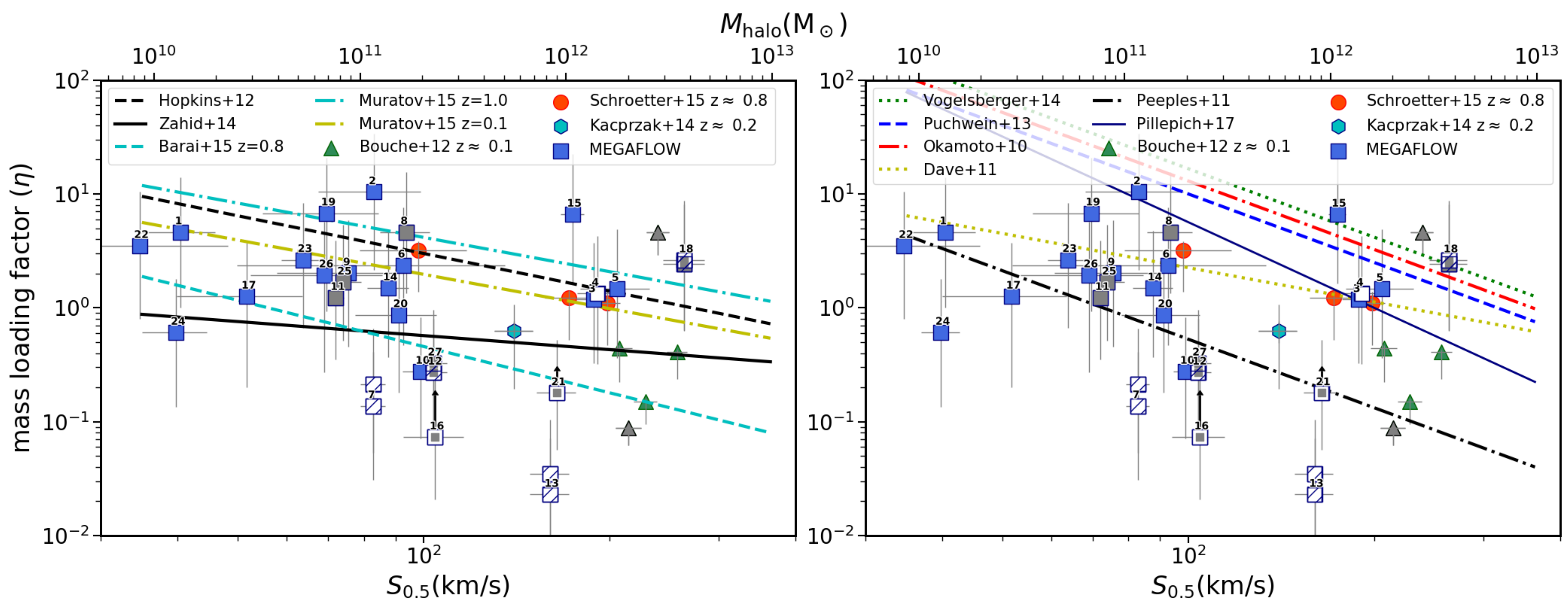}
  \caption{Comparison of mass loading factors (left: measured, right: injected) by theoretical/empirical models (curves) with values derived from background quasar observations 
  (data points) as a function of the maximum rotational velocity. 
  MEGAFLOW results are represented by blue squares. 
 The dashed squares correspond to the 4 cases with multiple possible wind models. 
 The orange circles show the results for galaxies at $z\approx0.8$ from \citet{schroetter_15}. 
 The light blue hexagon shows the mass loading factor for a $z\approx0.2$ galaxy \citep{KacprzakG_14a}. 
 The green triangles show the results for $z\approx0.2$ galaxies from \citet{bouche_12}.
 The gray triangles and squares show the galaxies with quasars located at $b>$60kpc where the mass loading factor is less reliable due to the large travel time 
 needed for the outflow to cross the quasar LOS (several 100 Myr) compared to the short time scale of the derived SFRs ($\sim 10$Myr). 
 The white squares represent the cases where the agreement between the wind model and the UVES data is poor and lower limits indicate systems where ionization correction might be significant.
 The upper halo mass axis is scaled by $V_{\rm max}$ at redshift 0.8 from \citet{mo_02}.
  }
  \label{fig:eta_vs_vmax_megaflow}
\end{figure*}

Figure~\ref{fig:eta_vs_vmax_megaflow} shows the loading factor (defined as $\dot M_{\rm out}/$SFR) as a function of galaxy halo mass (derived from $V_{\rm max}$ and redshift 0.8 from \citet{mo_02} relation). 
The blue squares represent the MEGAFLOW results, and the gray  symbols represent the galaxy-quasar pairs where the quasar is located at an impact parameters $b$ larger than $60$~\kpc\footnote{the $b>$60~\kpc\ is an arbitrary value, see the discussion on this criterion in Paper I and later in this \S.}. 
The mass loading factors were all derived taking into account the empty inner cone (when needed). 
For the 4 cases (IDs 7, 12, 13 and 18) with multiple wind model possibilities, the squares are hatched. 

In addition, we show in white squares the cases for which wind models are found less convincing at reproducing the absorption. 
Those cases are the following numbers: $\# 4, 7, 12, 13, 16, 18$ and $\#21$. 
The main reasons we classify those cases into less convincing are: 
\begin{itemize}
    \item $\#4$ has another absorption component at $\approx 200$~\kms\ which cannot be reproduced by our wind model. 
    \item $\#7$ has two different blended absorptions centered around the systemic redshift.
    It is thus difficult to determine where one absorption begins and the other ends.
    \item $\#12$ and $\#13$ are two different galaxies for the same absorption system. 
    We either fit the two absorption components closer to the systemic redshift or the two others. However, we cannot reproduce the three components simultaneously.
    \item $\#16$ has two absorption components.
    We choose to fit the closest from the systemic redshift as this galaxy is the second detected in Paper II for this system. 
    This galaxy could also contribute to the absorption at $\approx -150$~\kms\ which is identified as an accretion component in Paper II.
    \item $\#18$ also has two absorption components, one blue-shifted and one redshifted with respect to the galaxy systemic redshift. 
    Even if both wind models for this system are similar in outflow velocities (270~\kms\ and 200~\kms for the black and red models respectively), we consider this case as complex and therefore less convincing. 
    \item $\#21$ has a complex absorption system. 
    Giving the geometrical configuration of the system, our wind model can reproduce the closest component to the systemic redshift. 
    We assumed this component to be the signature of the outflowing gas but the other components at $\approx150-200$~\kms\ could also be a part of it.
\end{itemize}

Errors on mass loading factors are described in details in \citet{schroetter_15} and Paper I. 
As a short summary, for the derived parameters (i.e. \Vout\ and \thetam), we assume a Gaussian error distribution and the errors are given by the range of values given by the data. 
Errors on \Vout\ are 10~\kms, which correspond to a step of this parameter while eye-fitting the data. 
Those errors are over-estimated since \Vout$+10$~\kms\ and \Vout$-10$~\kms\ give simulated absorption profiles which does not fit the data at all. 
The same is used for the cone opening angle \thetam. 
The most important source of errors is given by the SFR and the hydrogen column density estimations.

Compared to the plot from Paper I, we separated simulation results in two panels.
On left panel, we show loading factors in which simulations measure them. 
On right panel, we show the injected loading factors (and thus not measured). 

From the two panels on this figure, we can see that the measured loading factors (curves in left panel) tend to be in agreement with the data points whereas injected loading factors on right panel appears to over-estimate them (apart from \citet{dave_11} and \citet{PeeplesM_11a}).
Overall, theoretical and empirical wind models are in agreement with the observational constraints but it seems that simulations in which they measure loading factors are a better estimation to compare with observations.

As already discussed in Paper I, there is a timescale problem concerning the mass loading factor.
Indeed, the SFR measured from \OII\ emission lines has a typical timescale of $\sim 10$~Myr whereas the mass outflow rate $\dot M_{\rm out}$ has a typical timescale of hundreds of Myr (assuming $b>20$~\kpc\ and \Vout$\approx$200~\kms). 
Therefore, both numerator and denominator of $\eta$ are, in most cases, on a different timescale. 
This leads to the conclusion that the mass loading factor may not be physically meaningful, if the SFR changes on short time scales. 

In addition, $\eta$ comparison with simulations may not be the best solution as we do not have the radius dependency for them. 
However, since we can only compare with what has been done so far, we can claim that, even regarding those differences, the mass loading factor does not seem to evolve strongly with the host galaxy mass. 
If we do not take into account the white squares, our results are less scattered and press on the previous statement. 
Systems with low REW or $\log N_{\rm HI}$ lower than $\approx$19.0 might suffer from an unknown ionization correction, and thus their outflow rate should be treated as lower limits.

\section{Summary and conclusions}
\label{conclusions}
Using our MEGAFLOW survey \citep[][Bouché et al. (in prep.)]{schroetter_16, ZablJ_19} which aims to observe galaxies responsible for $\sim80$ strong \MgII\ absorbers ($W_r^{\lambda2796}>0.3$~\AA) seen in quasar spectra at $0.4<z<1.5$ with MUSE and UVES, we investigated the distribution of the gas surrounding those galaxies.
Without any pre-selection on their geometrical configuration, we clearly see a bi-modal distribution of this low-ionized gas (see Figure~\ref{fig:surveys_alpha}). 
This distribution of azimuthal angles suggests a bi-conical outflow geometry and a co-planar extended gas disk.
This in turn supports our geometrical assumption for such phenomena. 

We then selected \NpairsFinal\ galaxy-quasar pairs suitable for wind study (i.e. $\alpha \geq 55^\circ$). 
Outflowing gas properties for 26 of the host galaxies were constrained. 
Those properties were the outflow velocity $V_{\rm out}$, the mass outflow rate $\dot M_{\rm out}$ and the mass loading factor $\eta$ (as shown in Figure~\ref{fig:eta_vs_vmax_megaflow} and Table~\ref{table:megaflow_outflow}). 

A summary of our results is as follows: 
\begin{itemize}
 \item Without morphology or geometry pre-selection (only absorption-selection), we find a bimodal distribution of azimuthal angles (Figure~\ref{fig:surveys_alpha}). 
 This suggests that the geometry of the gas surrounding galaxies is outflow dominated with a cone along the galaxy minor axis and accretion dominated coplanar to the disk, within 100~\kpc. 
 \item Mass loading factors tend to be $\eta \sim 1$, which means that the mass outflow rate is of the same order of magnitude as the galaxy SFR.
 \item The cool gas traced with the low-ionization element \MgII\  is likely to fall back onto the galaxy for galaxies with stellar mass larger than 4$\times$10$^9 \rm M_\odot$   (Figure~\ref{fig:vesc_mdyn}).
\end{itemize}

\section*{Acknowledgments}
We thank the referee for a careful and constructive report, which helped to improve the quality of this manuscript.
This work has been carried out thanks to the support of the ANR FOGHAR (ANR-13-BS05-0010-02), ANR 3DGasFlows (ANR-17-CE31-0017), the OCEVU Labex (ANR-11-LABX-0060), and the A*MIDEX project
(ANR-11-IDEX-0001-02) funded by the ``Investissements d'avenir'' French government program.

\bibliographystyle{apj}
\bibliography{references,megaflow_II_refs}	
\label{lastpage}
\section{appendix}
\setcounter{figure}{0} \renewcommand{\thefigure}{A.\arabic{figure}} 
Here we present the wind models for each wind subsample galaxy-quasar pairs.

\begin{figure*} 
   \centering 
   \includegraphics[width=12cm]{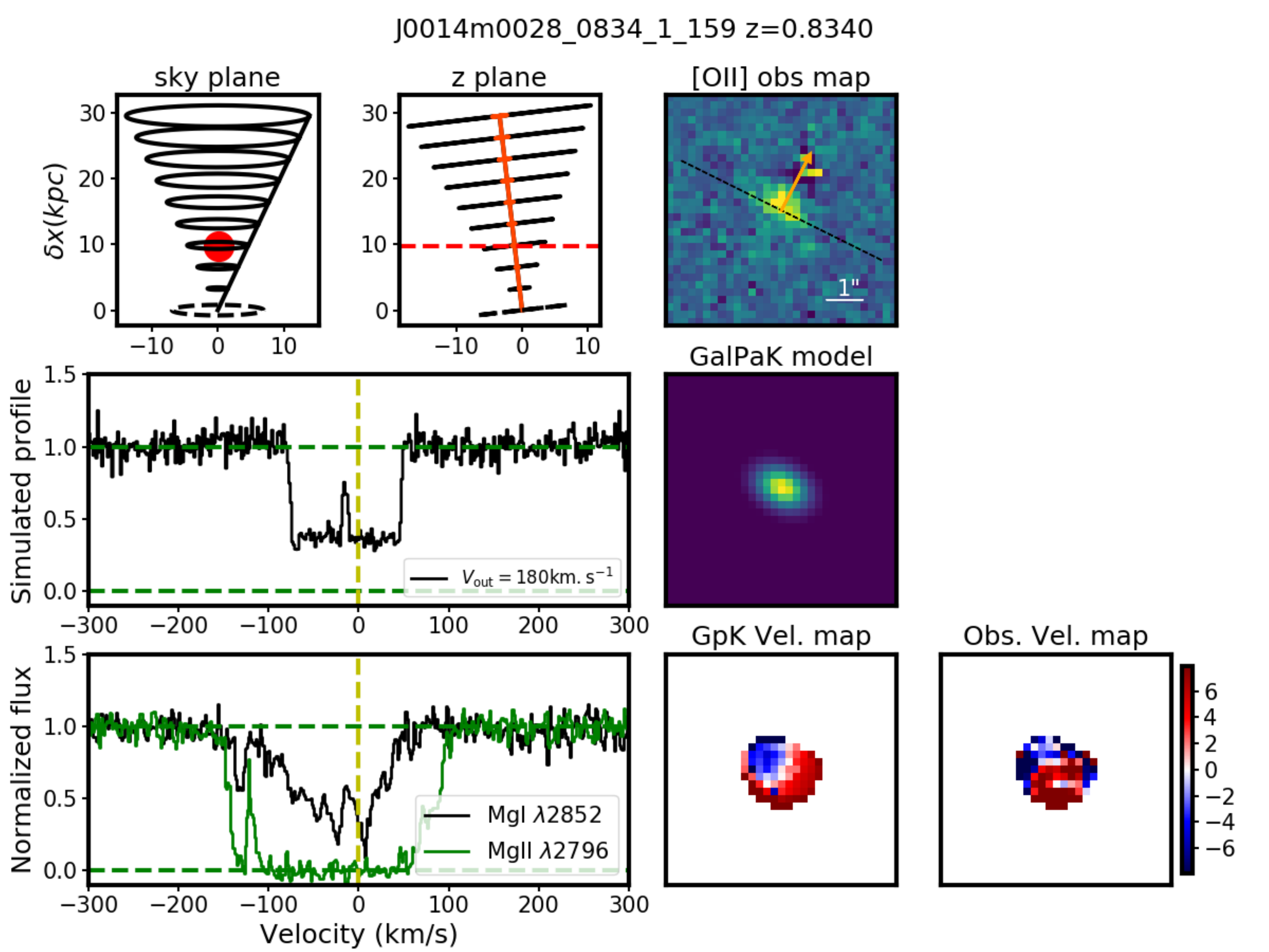} 
   \caption{Same as Figure~\ref{fig:J0014m0028_0834_1} but for the galaxy \#1 at redshift $z=0.8340$.
   This outflow has a \Vout\ of $180\pm10$ \kms, an opening angle \thetam\ of $28\pm2^\circ$ and an inner empty cone $\theta_{\rm in}$ of 2$^\circ$.} 
   \label{fig:J0015m0751_0810_3} 
\end{figure*}

\begin{figure*} 
   \centering 
   \includegraphics[width=12cm]{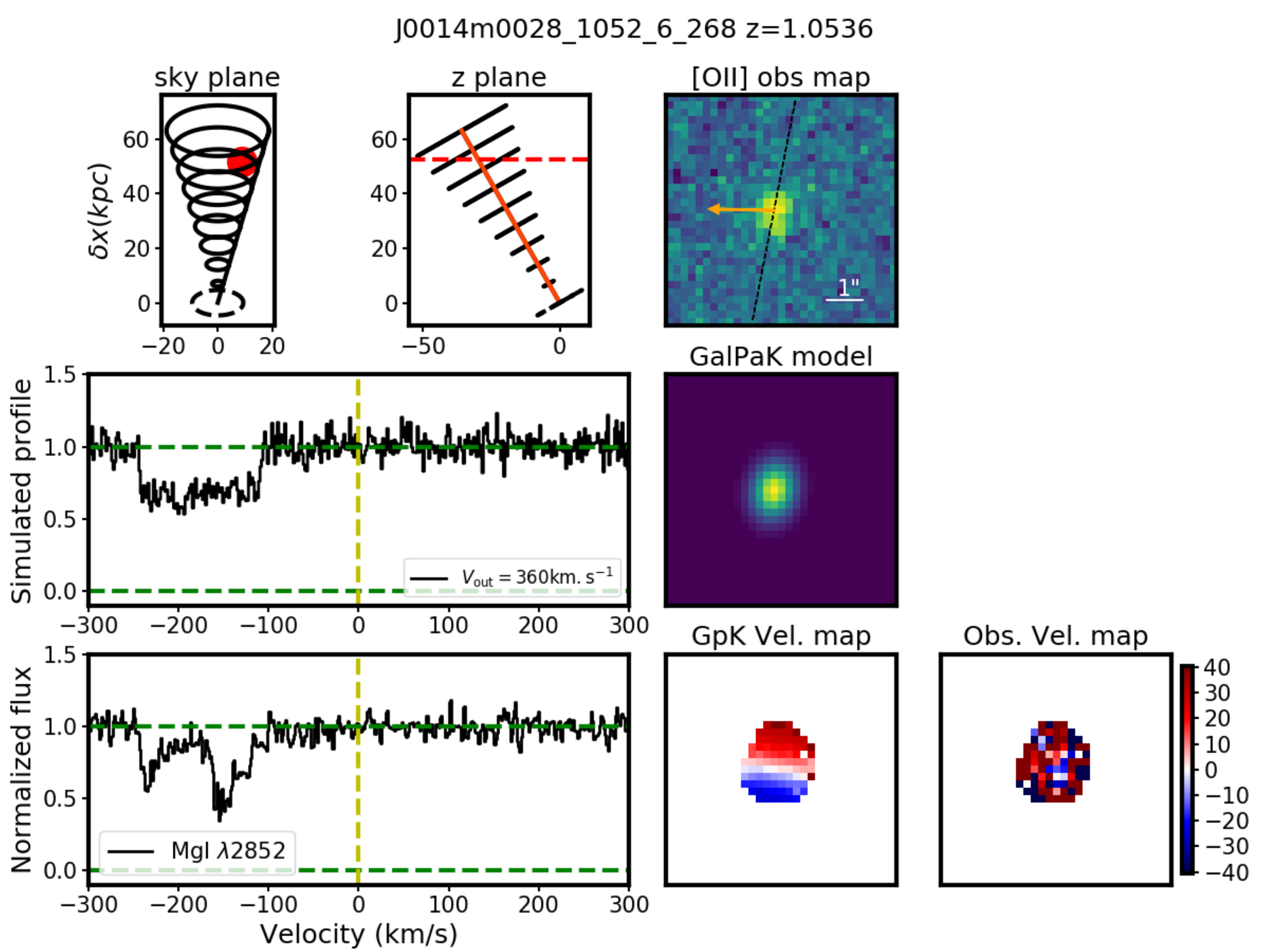} 
   \caption{Same as Figure~\ref{fig:J0014m0028_0834_1} but for the galaxy \#2 at redshift $z=1.0536$.
   This outflow has a \Vout\ of $360\pm10$ \kms\ and an opening angle \thetam\ of $15\pm2^\circ$.} 
   \label{fig:J0014m0028_1052_6} 
\end{figure*}

\begin{figure*} 
   \centering 
   \includegraphics[width=12cm]{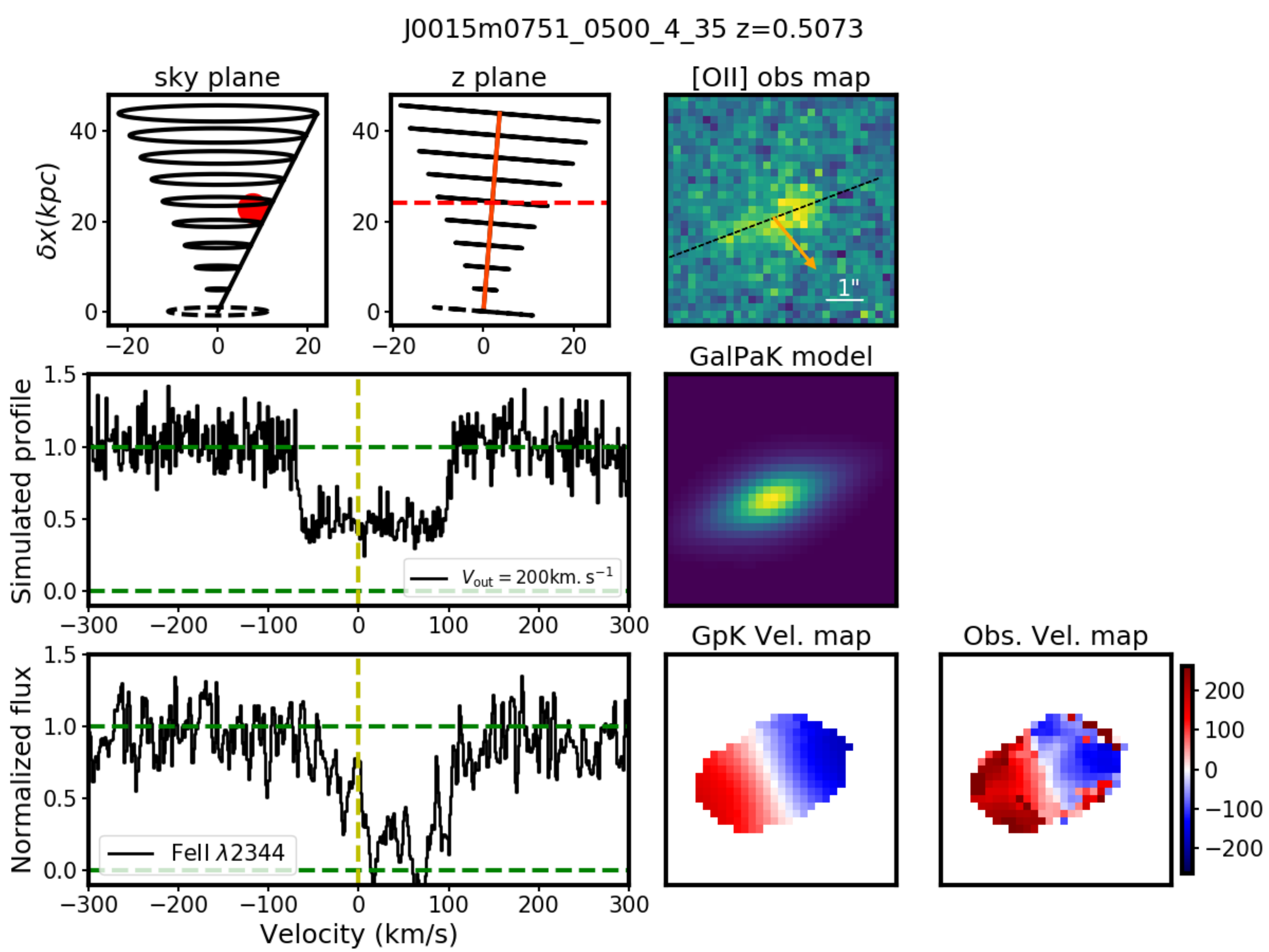} 
   \caption{Same as Figure~\ref{fig:J0014m0028_0834_1} but for the  galaxy \#3 at redshift $z=0.5073$.
   This outflow has a \Vout\ of $200\pm10$ \kms\ and an opening angle \thetam\ of $30\pm2^\circ$.} 
   \label{fig:J0015m0751_0500_4} 
\end{figure*}

\begin{figure*} 
   \centering 
   \includegraphics[width=12.0cm]{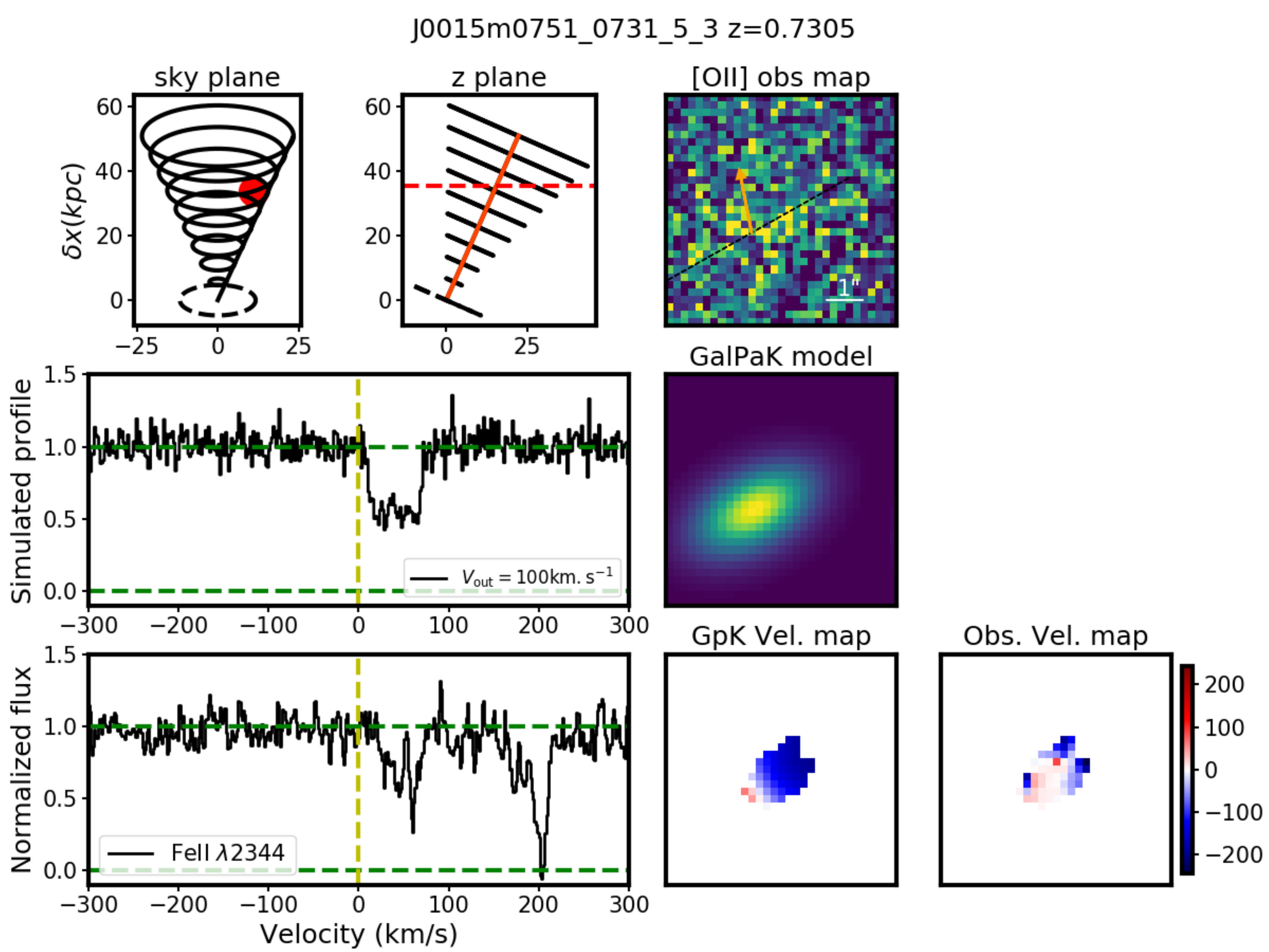} 
   \caption{Same as Figure~\ref{fig:J0014m0028_0834_1} but for the  galaxy \#4 at redshift $z=0.7305$.
   This outflow has a \Vout\ of $100\pm10$ \kms\ and an opening angle \thetam\ of $25\pm2^\circ$.} 
   \label{fig:J0015m0751_0731_5} 
\end{figure*}

\begin{figure*} 
   \centering 
   \includegraphics[width=12cm]{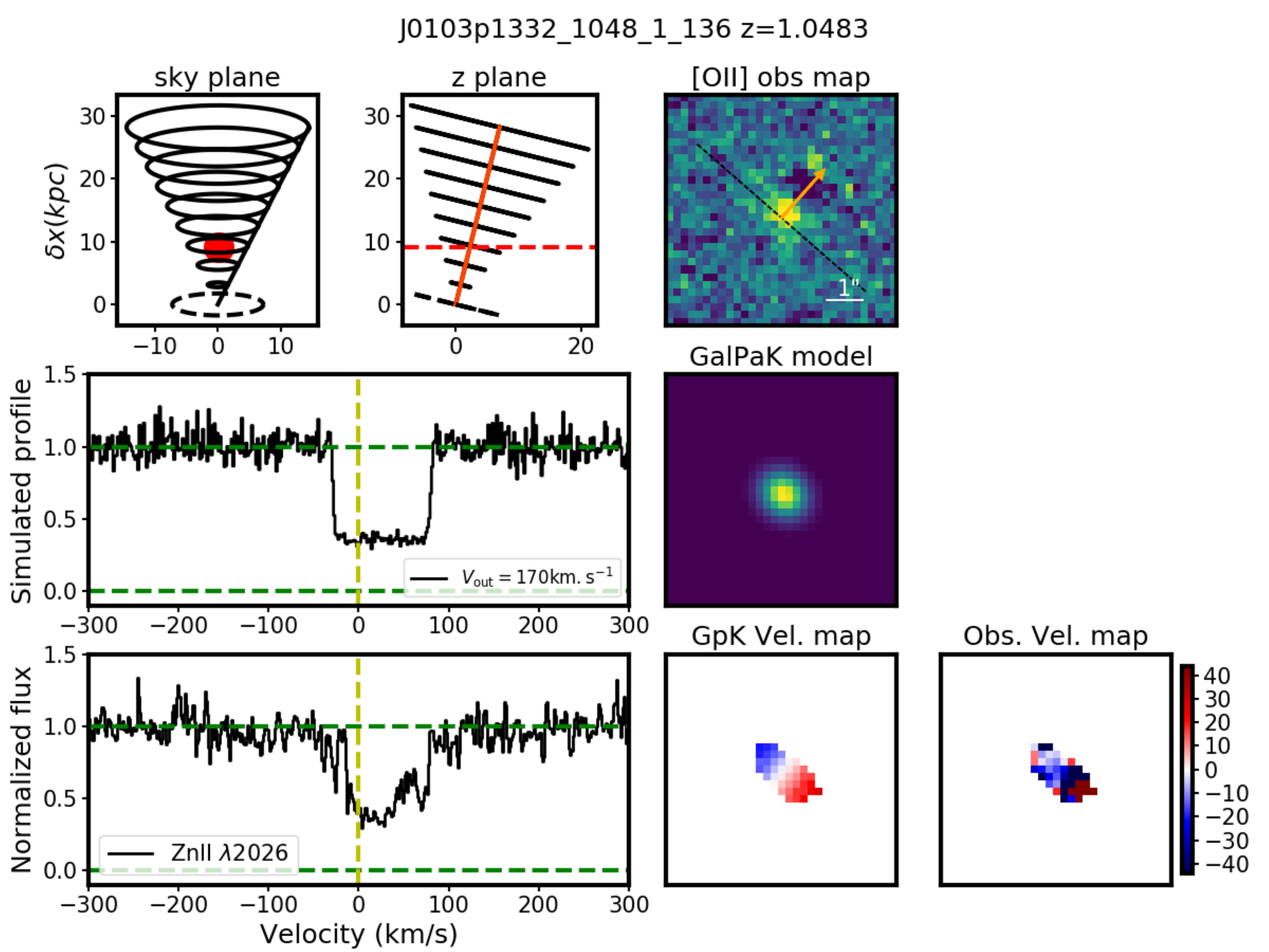} 
   \caption{Same as Figure~\ref{fig:J0014m0028_0834_1} but for the  galaxy \#6 at redshift $z=1.0483$.
   This outflow has a \Vout\ of $170\pm10$ \kms\ and an opening angle \thetam\ of $30\pm2^\circ$.} 
   \label{fig:J0103p1332_1048_1} 
\end{figure*}
\begin{figure*} 
   \centering 
   \includegraphics[width=12cm]{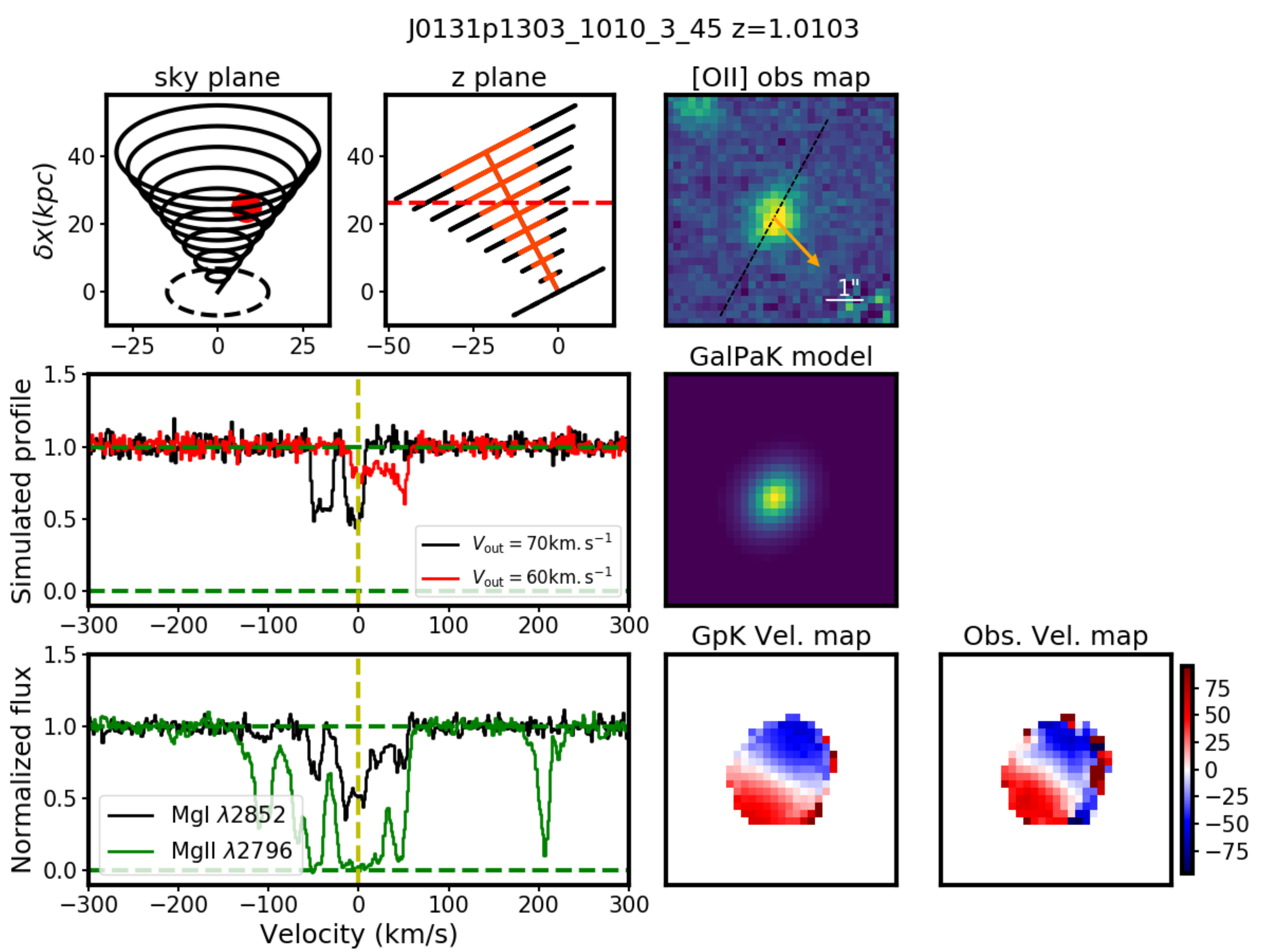} 
   \caption{Same as Figure~\ref{fig:J0014m0028_0834_1} but for the  galaxy \#7 at redshift $z=1.0103$.
   This is one of the "multiple model" ourflows cases.
   The black (red) outflow has a \Vout\ of $70\pm10$ ($60\pm10$) \kms, an opening angle \thetam\ of $40\pm2^\circ$ (same) and an empty inner cone $\theta_{\rm in}$ of 18$^\circ$ (0$^\circ$).} 
   \label{fig:J0131p1303_1010_3} 
\end{figure*}
\begin{figure*} 
   \centering 
   \includegraphics[width=12cm]{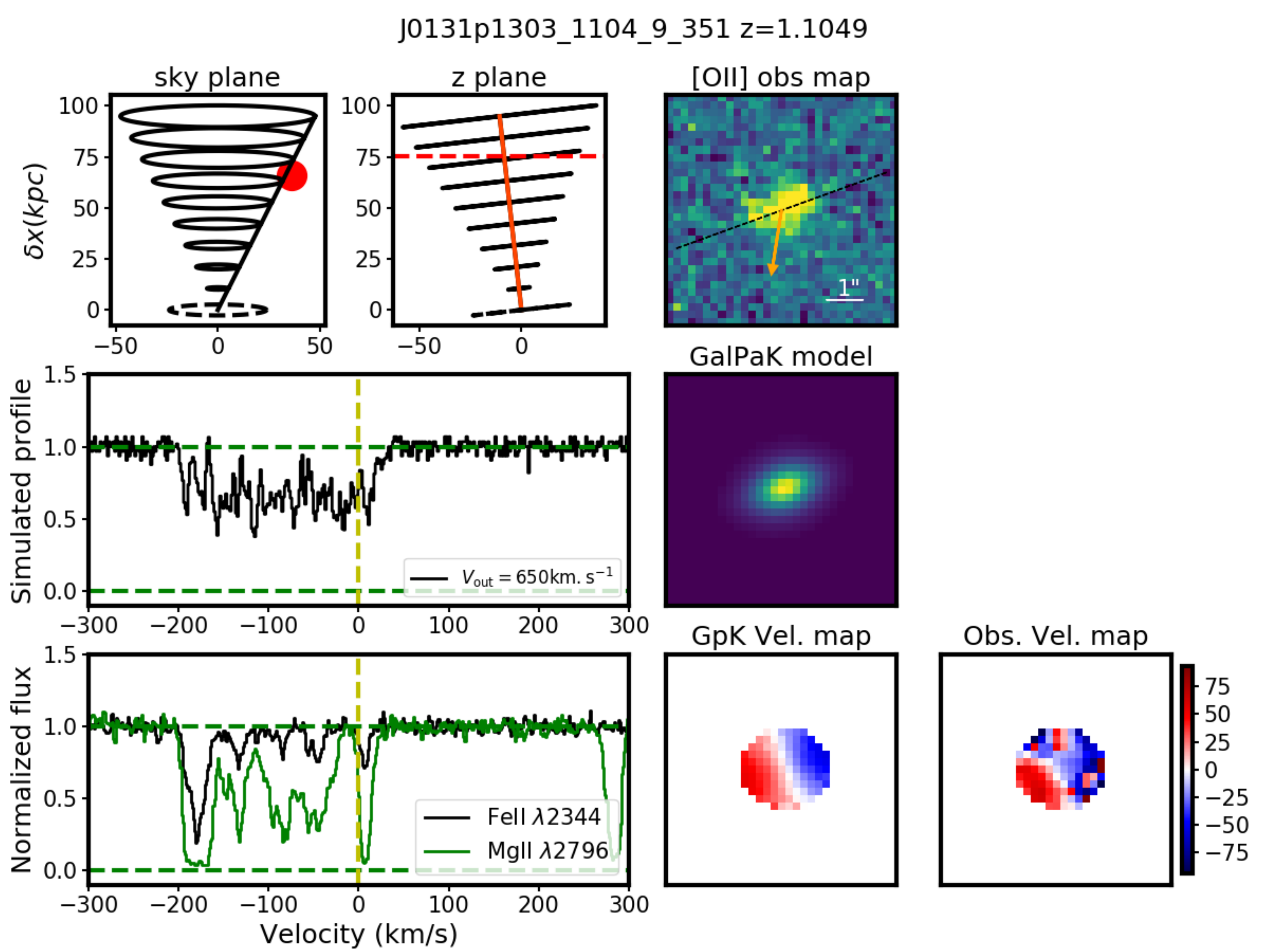} 
   \caption{Same as Figure~\ref{fig:J0014m0028_0834_1} but for the  galaxy \#8 at redshift $z=1.1049$.
   This outflow has a \Vout\ of $650\pm10$ \kms\ and an opening angle \thetam\ of $30\pm2^\circ$.
   We not here that to reproduce the data, only a fraction of the outflow cone is crossed by the QSO LOS, therefore a very high \Vout\ is nedeed.} 
   \label{fig:J0131p1303_1104_9} 
\end{figure*}
\begin{figure*} 
   \centering 
   \includegraphics[width=12cm]{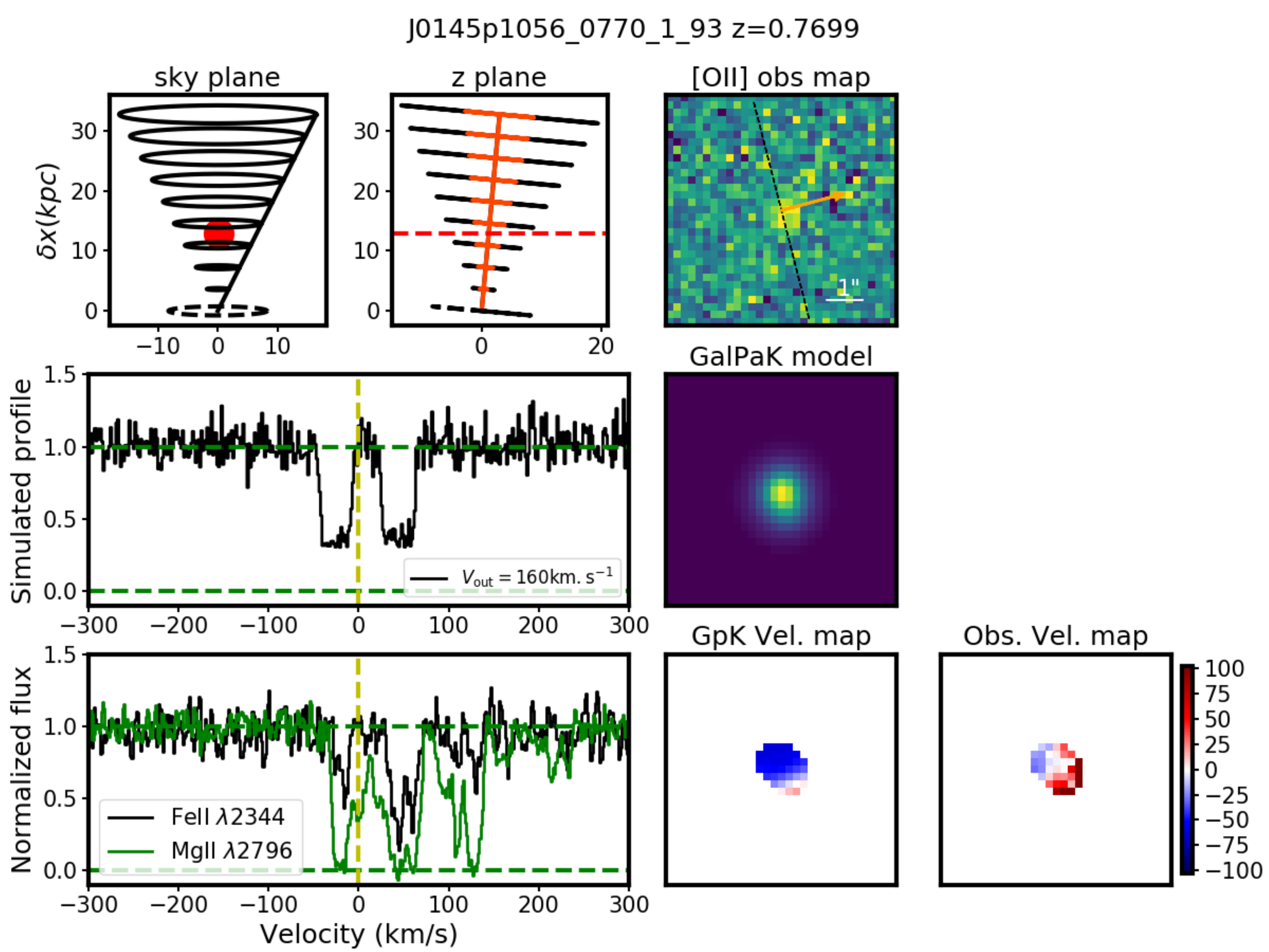} 
   \caption{Same as Figure~\ref{fig:J0014m0028_0834_1} but for the  galaxy \#9 at redshift $z=0.7699$.
   This outflow has a \Vout\ of $160\pm10$ \kms, an opening angle \thetam\ of $15\pm2^\circ$ and an empty inner cone $\theta_{\rm in}$ of 10$^\circ$.
   The component at $\sim 120$~\kms\ could not be reproduced given the geometry of the system.} 
   \label{fig:J0134p0051_1449_7} 
\end{figure*}

\begin{figure*} 
   \centering 
   \includegraphics[width=12cm]{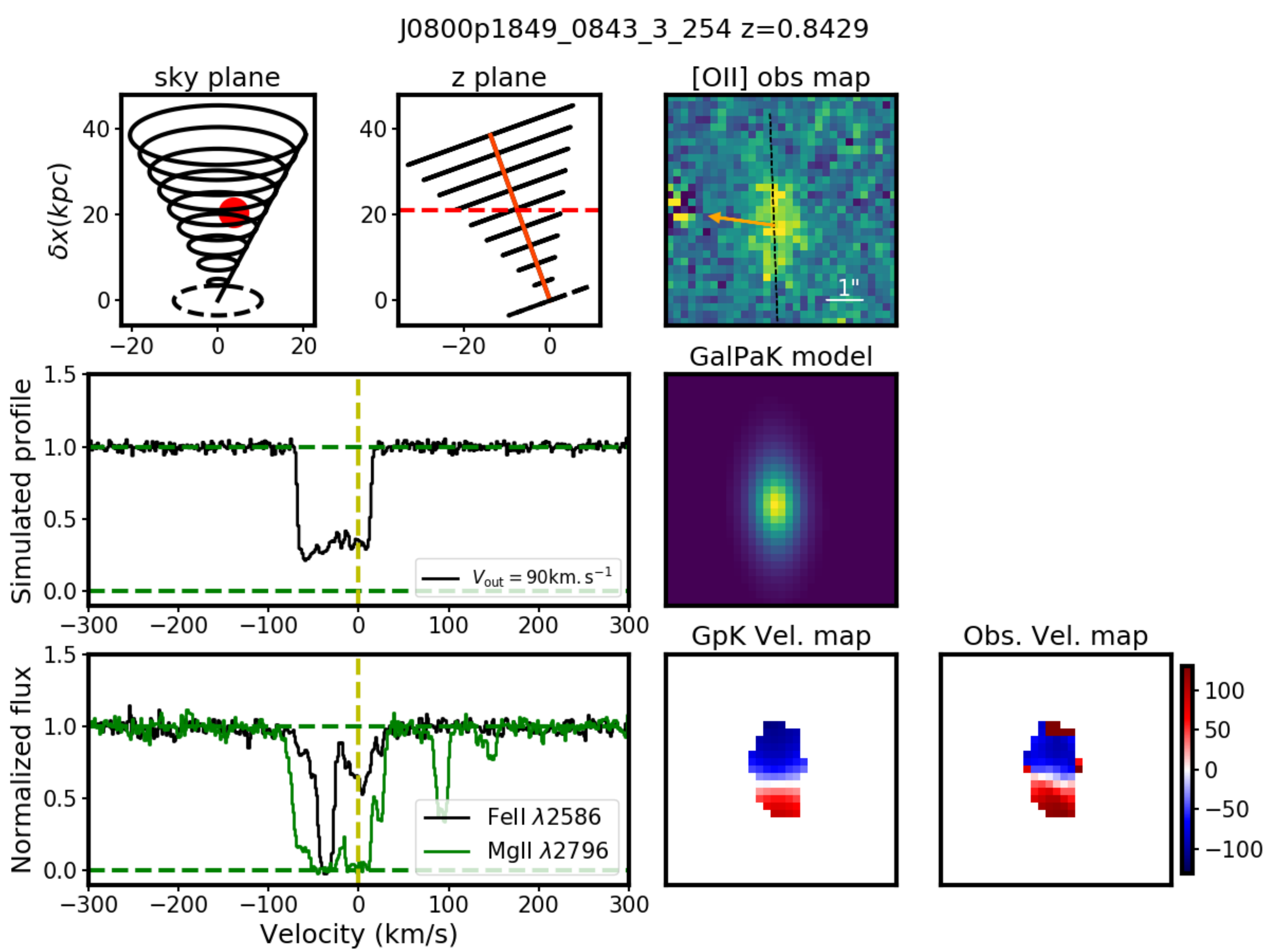} 
   \caption{Same as Figure~\ref{fig:J0014m0028_0834_1} but for the  galaxy \#10 at redshift $z=0.8429$.
   This outflow has a \Vout\ of $90\pm10$ \kms\ and an opening angle \thetam\ of $30\pm2^\circ$.} 
   \label{fig:J0800p1849_0843_3} 
\end{figure*}
\begin{figure*} 
   \centering 
   \includegraphics[width=12.0cm]{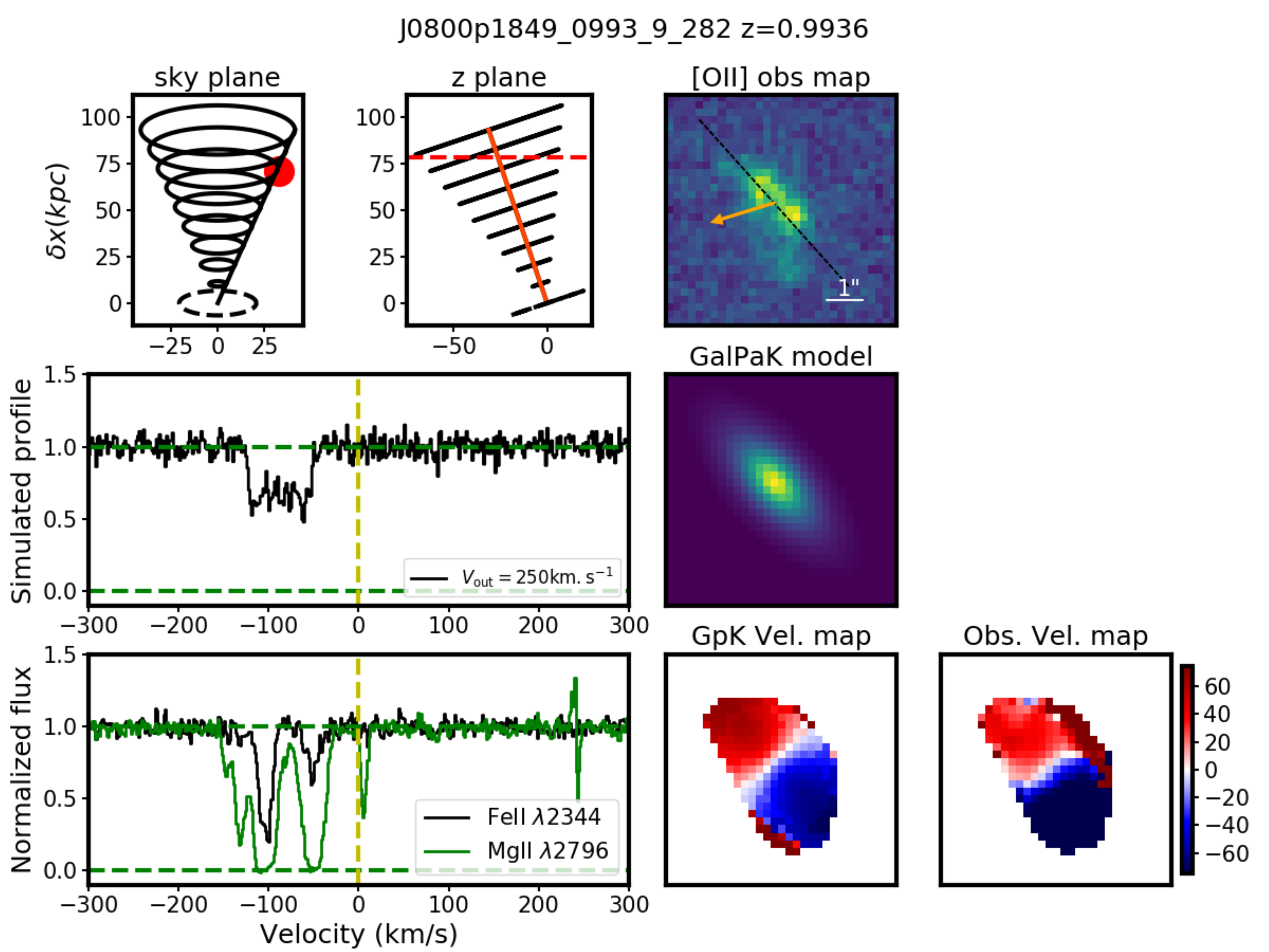} 
   \caption{Same as Figure~\ref{fig:J0014m0028_0834_1} but for the  galaxy \#11 at redshift $z=0.9936$.
   This outflow has a \Vout\ of $250\pm10$ \kms\ and an opening angle \thetam\ of $25\pm2^\circ$.} 
   \label{fig:J0800p1849_0993_9} 
\end{figure*}
\begin{figure*} 
   \centering 
   \includegraphics[width=12.0cm]{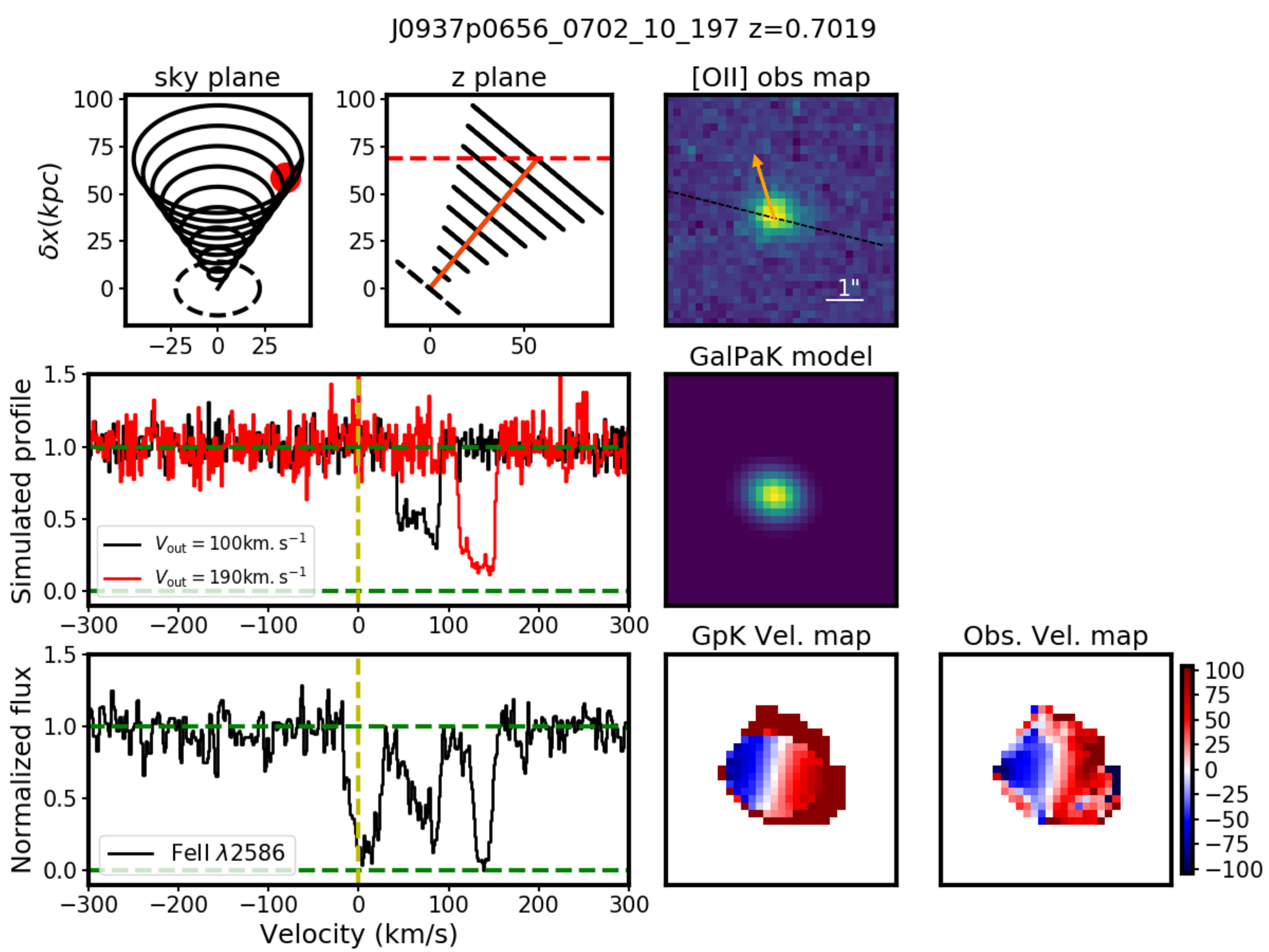} 
   \caption{Same as Figure~\ref{fig:J0014m0028_0834_1} but for the  galaxy \#12 at redshift $z=0.7019$.
   This is one of the "multiple model" ourflows cases.
   The black (red) outflow has a \Vout\ of $100\pm10$ ($190\pm10$) \kms\ and an opening angle \thetam\ of $30\pm2^\circ$ (25$^\circ$).} 
   \label{fig:J0937p0656_0702_1} 
\end{figure*}
\begin{figure*} 
   \centering 
   \includegraphics[width=12.0cm]{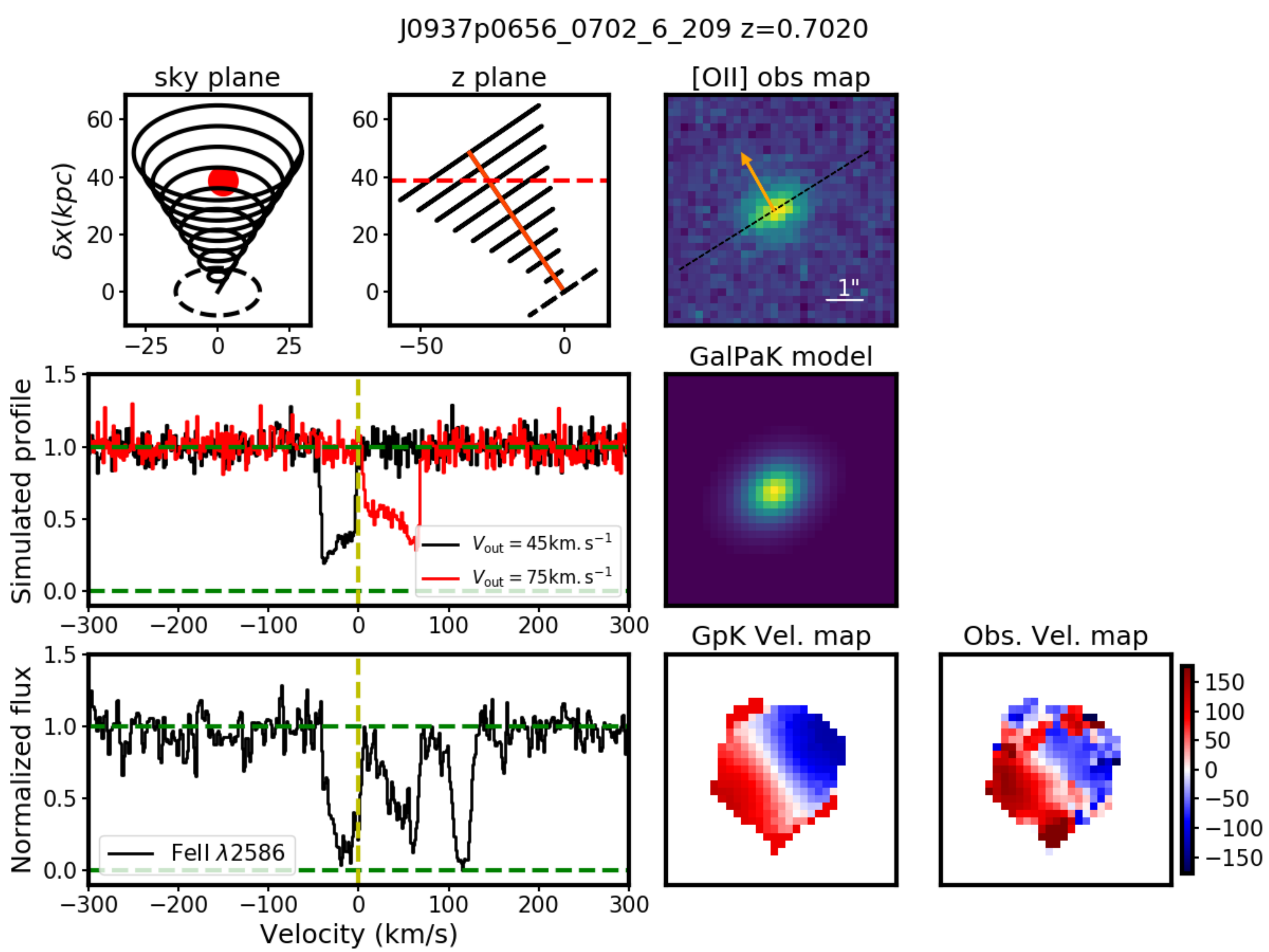} 
   \caption{Same as Figure~\ref{fig:J0014m0028_0834_1} but for the  galaxy \#13 at redshift $z=0.7020$.
   This is one of the "multiple model" ourflows cases.
   The black (red) outflow has a \Vout\ of $45\pm10$ ($75\pm10$) \kms\ and an opening angle \thetam\ of $30\pm2^\circ$ (same).} 
   \label{fig:J0937p0656_0702_6} 
\end{figure*}

\begin{figure*} 
   \centering 
   \includegraphics[width=12.0cm]{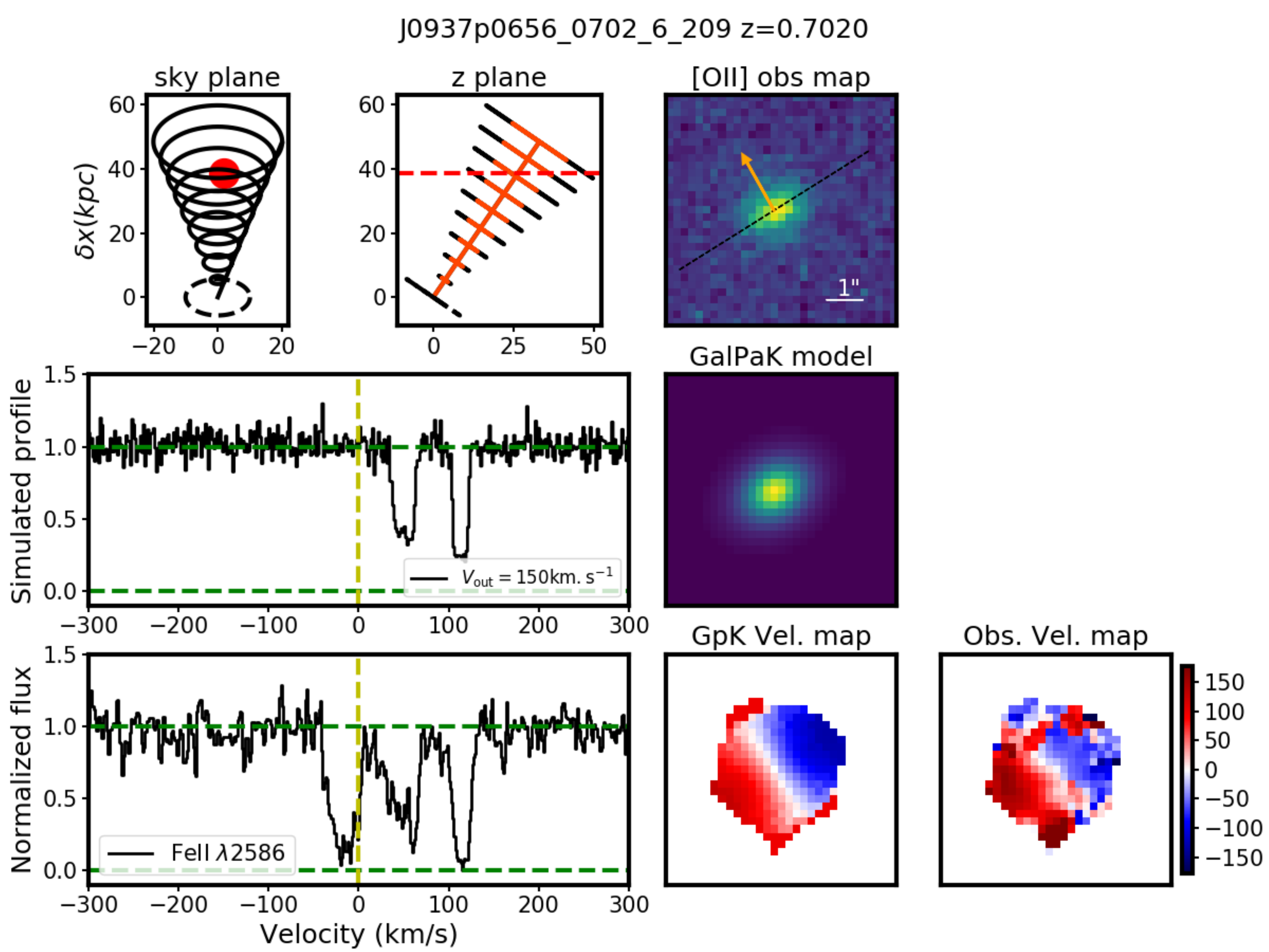} 
   \caption{Same as Figure~\ref{fig:J0014m0028_0834_1} but for the  galaxy \#13 at redshift $z=0.7020$.
   This is one other alternative wind model for this galaxy.
   The black (red) outflow has a \Vout\ of $150\pm10$ \kms\, an opening angle \thetam\ of $30\pm2^\circ$ and an empty inner cone $\theta_{\rm in}$ of 10$^\circ$.} 
   \label{fig:J0937p0656_0702_6} 
\end{figure*}

\begin{figure*} 
   \centering 
   \includegraphics[width=12.0cm]{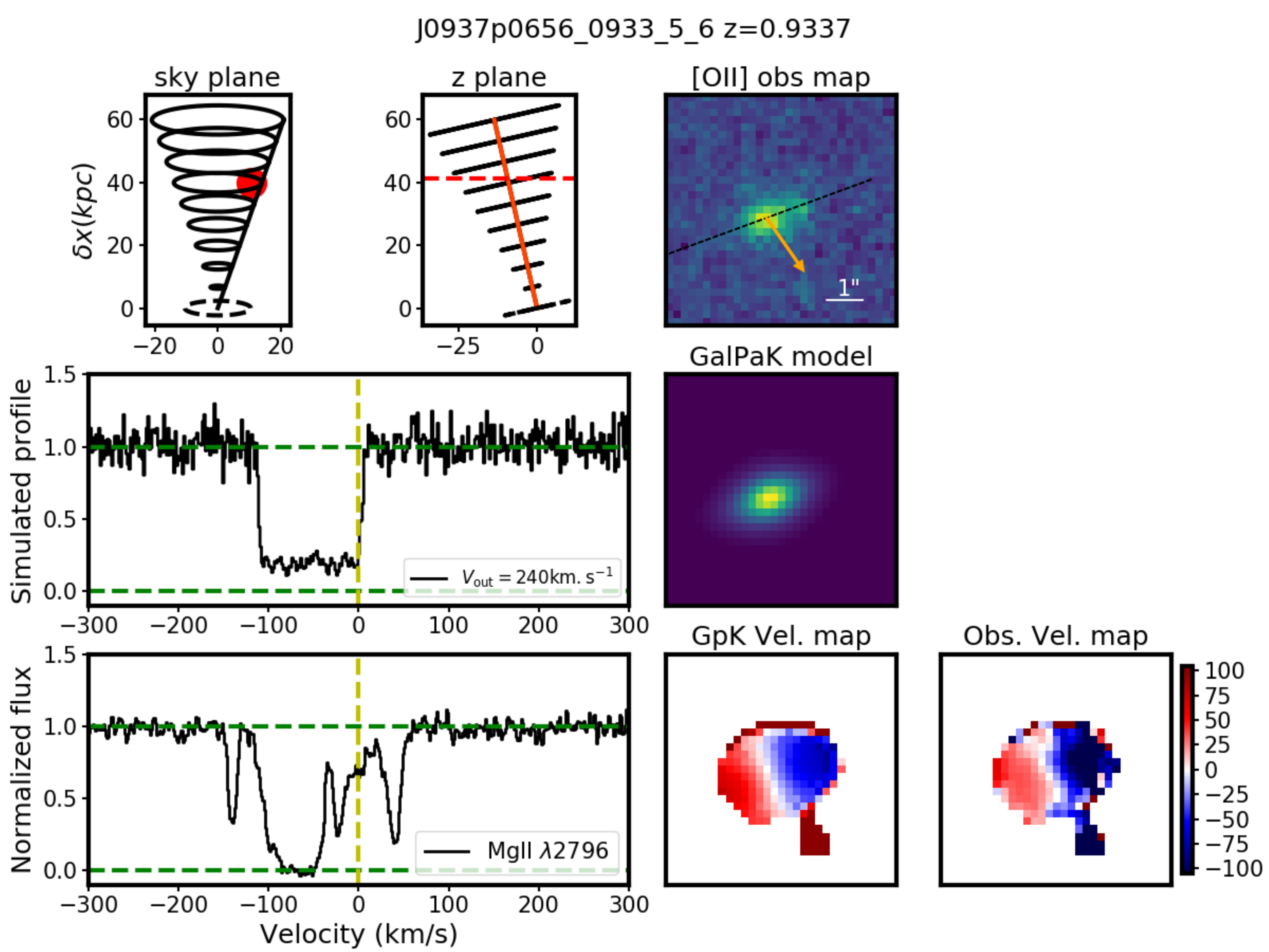} 
   \caption{Same as Figure~\ref{fig:J0014m0028_0834_1} but for the  galaxy \#14 at redshift $z=0.9337$.
   This outflow has a \Vout\ of $240\pm10$ \kms\ and an opening angle \thetam\ of $20\pm2^\circ$} 
   \label{fig:J0937p0656_0933_5} 
\end{figure*}
\begin{figure*} 
   \centering 
   \includegraphics[width=12.0cm]{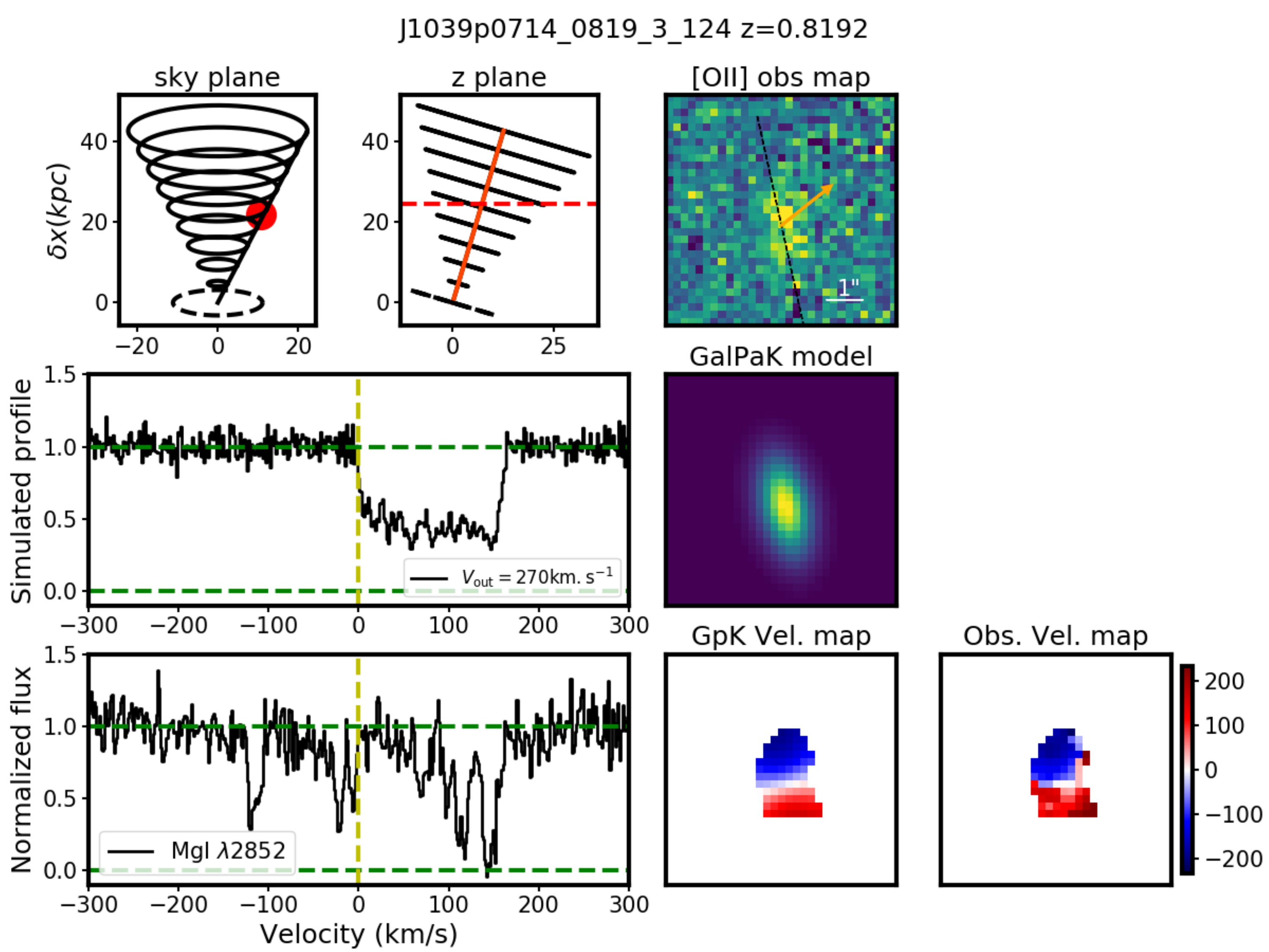} 
   \caption{Same as Figure~\ref{fig:J0014m0028_0834_1} but for the  galaxy \#15 at redshift $z=0.8192$.
   This outflow has a \Vout\ of $270\pm10$ \kms\ and an opening angle \thetam\ of $30\pm2^\circ$.} 
   \label{fig:J1039p0714_0819_3} 
\end{figure*}
\begin{figure*} 
   \centering 
   \includegraphics[width=12.0cm]{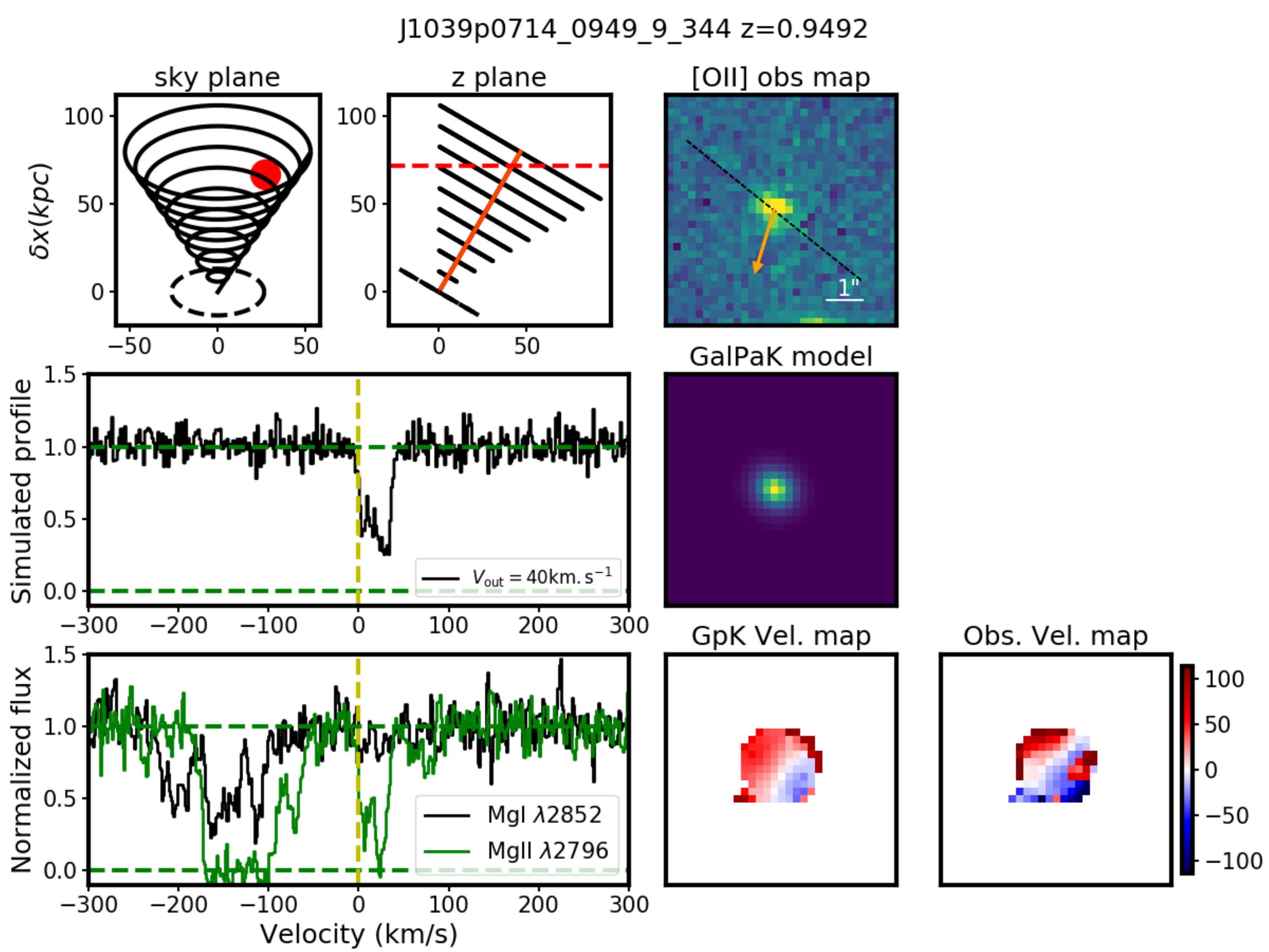} 
   \caption{Same as Figure~\ref{fig:J0014m0028_0834_1} but for the  galaxy \#16 at redshift $z=0.9492$.
   This outflow has a \Vout\ of $40\pm10$ \kms\ and an opening angle \thetam\ of $35\pm2^\circ$. 
   This galaxy is believed to produced the absorption the closest to the systemic redshift as the other absorption appears to come from a closer galaxy described in Paper~II.} 
   \label{fig:J1039p0714_0949_9} 
\end{figure*}
\begin{figure*} 
   \centering 
   \includegraphics[width=12.0cm]{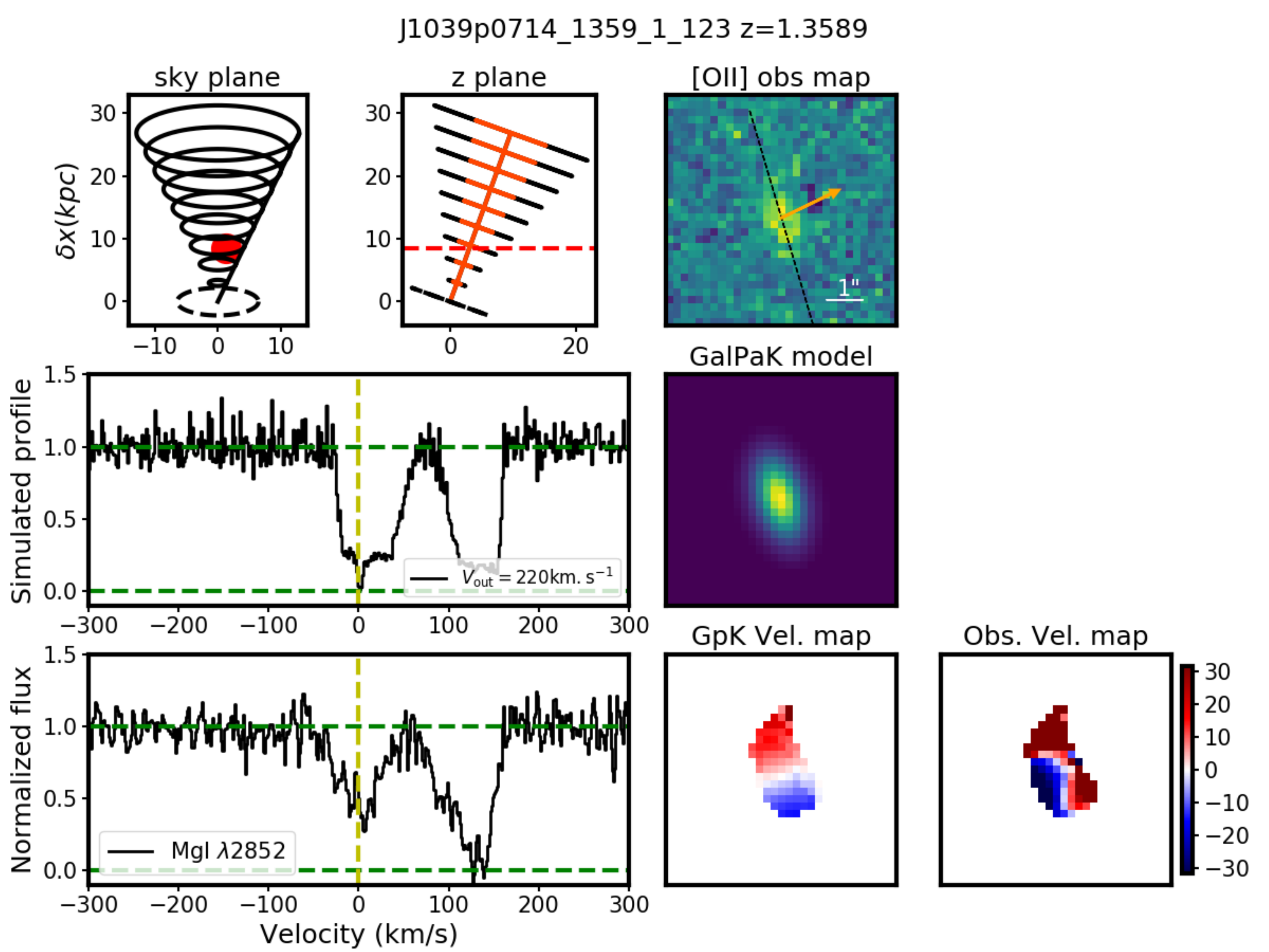} 
   \caption{Same as Figure~\ref{fig:J0014m0028_0834_1} but for the  galaxy \#17 at redshift $z=1.3589$.
   This outflow has a \Vout\ of $220\pm10$ \kms, an opening angle \thetam\ of $27\pm2^\circ$ and an empty inner cone $\theta_{\rm in}$ of 12$^\circ$.
   The observed velocity map appears slightly different due to the quasar subtraction of the subcube.} 
   \label{fig:J1039p0714_1359_1} 
\end{figure*}
\begin{figure*} 
   \centering 
   \includegraphics[width=12.0cm]{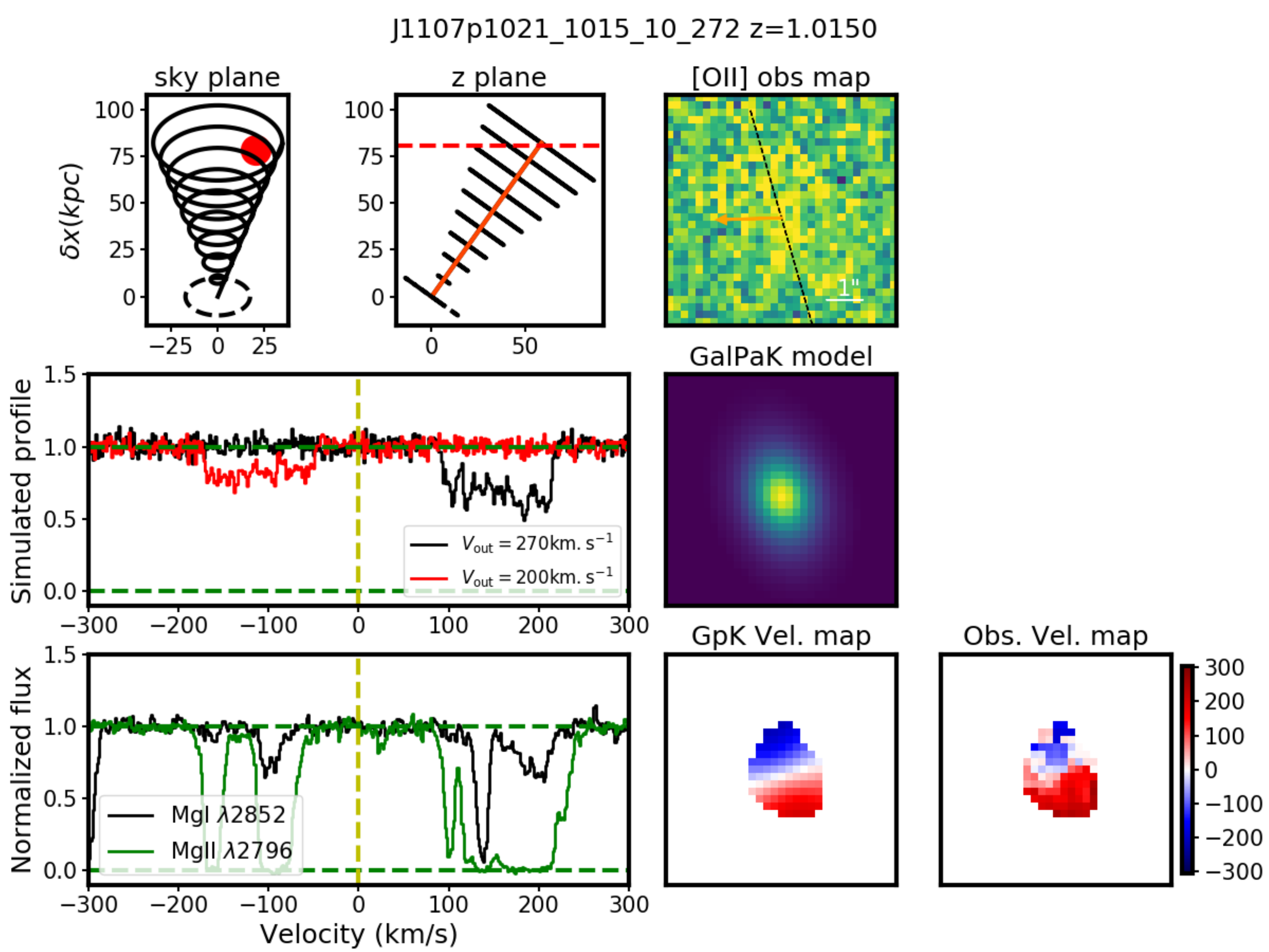} 
   \caption{Same as Figure~\ref{fig:J0014m0028_0834_1} but for the  galaxy \#18 at redshift $z=1.0150$.
   This is one of the "multiple model" ourflows cases.
   The black (red) outflow has a \Vout\ of $270\pm10$ ($200\pm10$) \kms\ and an opening angle \thetam\ of $20\pm2^\circ$ (25$^\circ$).
   } 
   \label{fig:J1107p1021_1015_1} 
\end{figure*}
\begin{figure*} 
   \centering 
   \includegraphics[width=12.0cm]{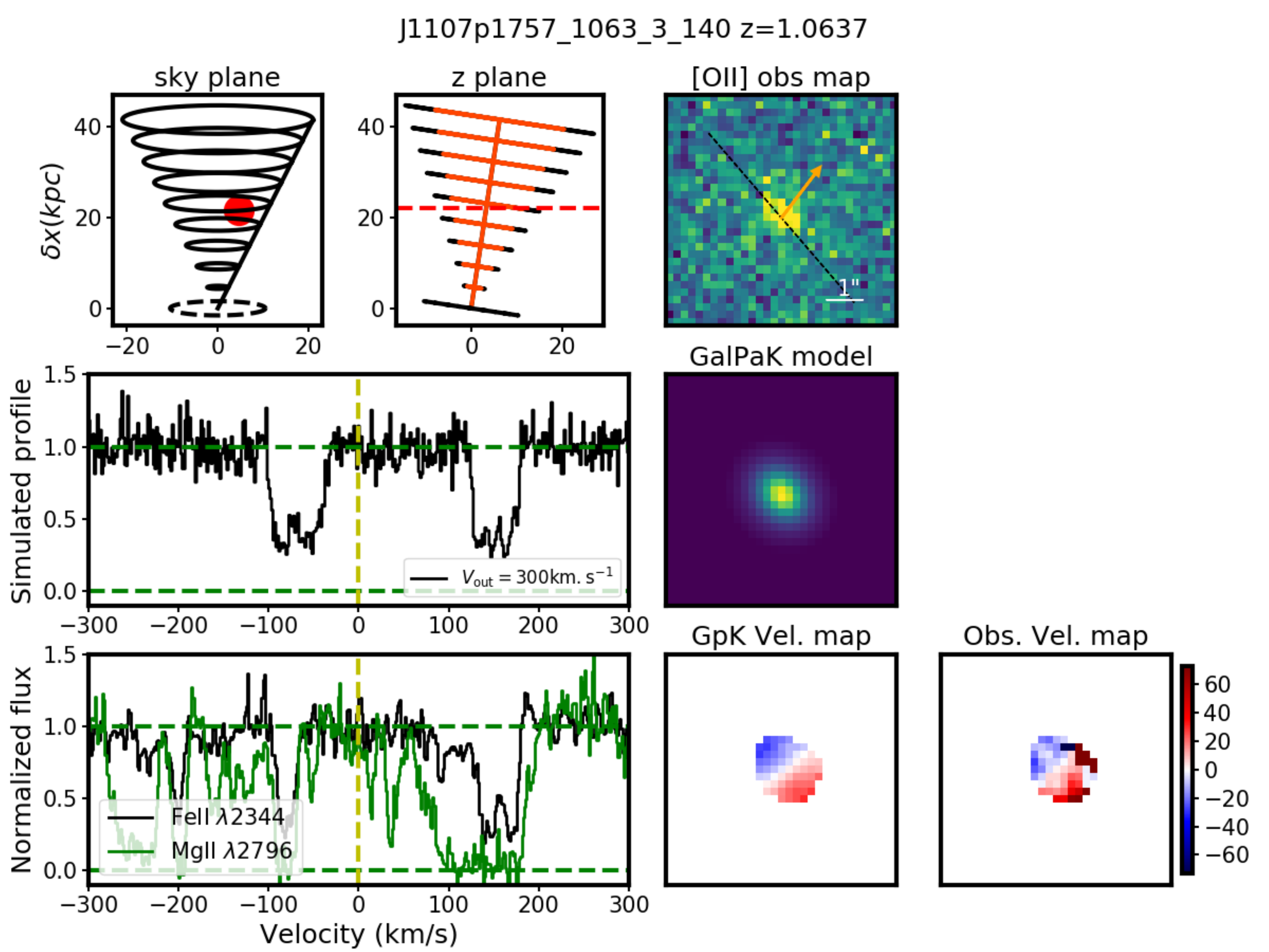} 
   \caption{Same as Figure~\ref{fig:J0014m0028_0834_1} but for the  galaxy \#19 at redshift $z=1.0637$.
   This outflow has a \Vout\ of $300\pm10$ \kms, an opening angle \thetam\ of $30\pm2^\circ$ and an empty inner cone $\theta_{\rm in}$ of 20$^\circ$.} 
   \label{fig:J1107p1757_1063_3} 
\end{figure*}
\begin{figure*} 
   \centering 
   \includegraphics[width=12.0cm]{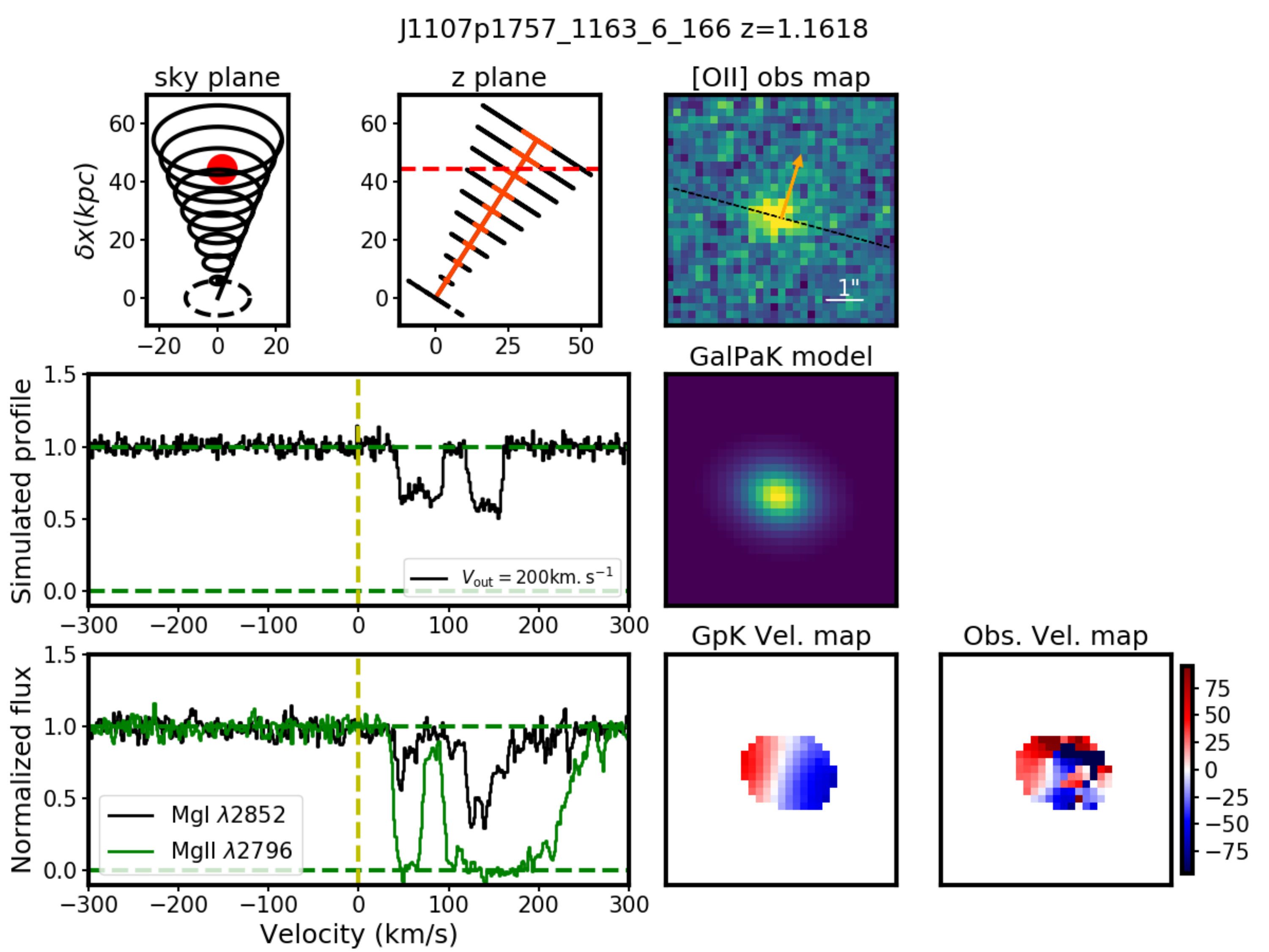} 
   \caption{Same as Figure~\ref{fig:J0014m0028_0834_1} but for the  galaxy \#20 at redshift $z=1.1618$.
   This outflow has a \Vout\ of $200\pm10$ \kms, an opening angle \thetam\ of $20\pm2^\circ$ and an empty inner cone $\theta_{\rm in}$ of 5$^\circ$.} 
   \label{fig:J1236p0725_0639_1} 
\end{figure*}
\begin{figure*} 
   \centering 
   \includegraphics[width=12.0cm]{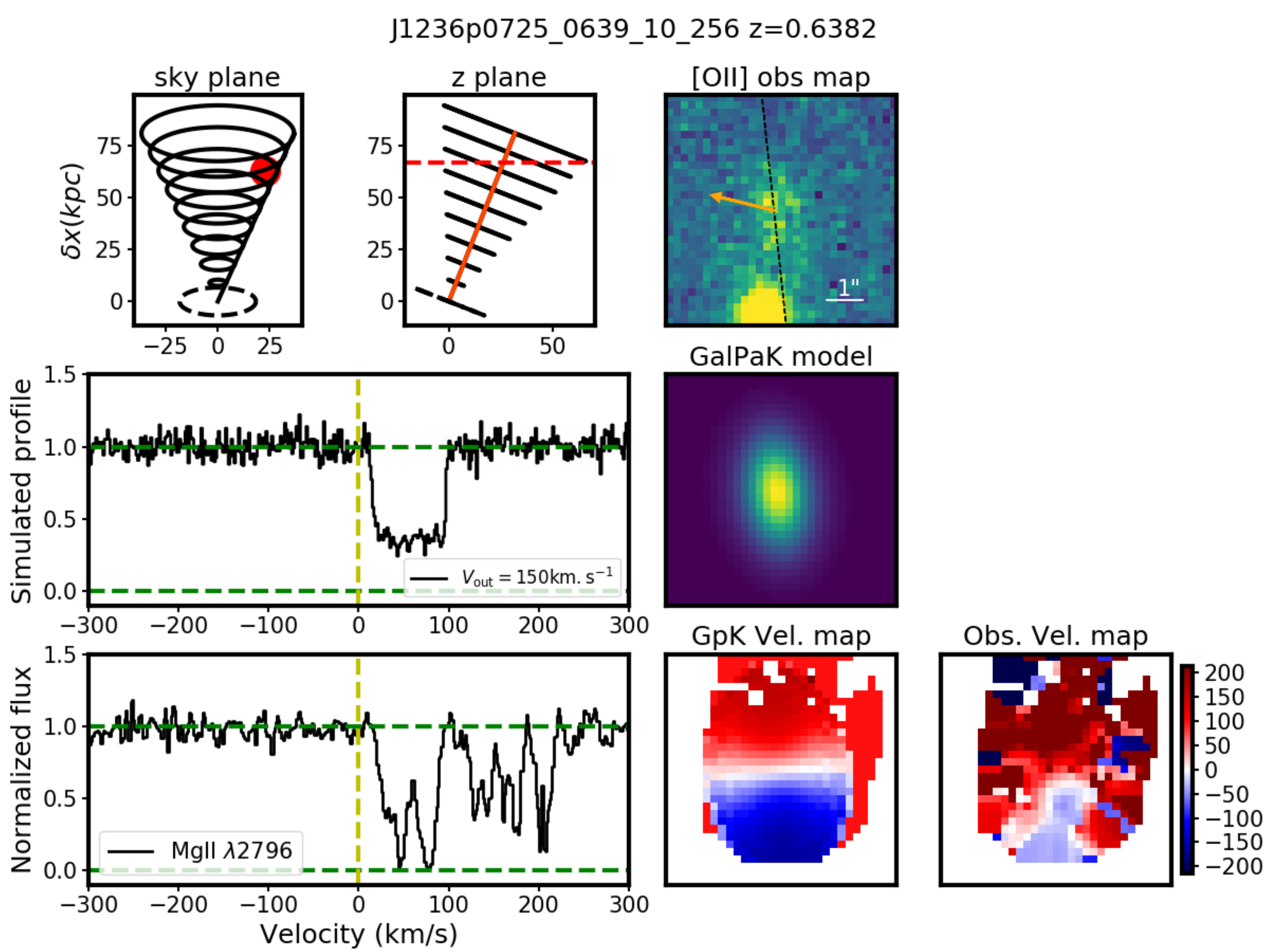} 
   \caption{Same as Figure~\ref{fig:J0014m0028_0834_1} but for the  galaxy \#21 at redshift $z=0.6382$.
   This outflow has a \Vout\ of $150\pm10$ \kms\ and an opening angle \thetam\ of $25\pm2^\circ$.
   The measured velocity map appears to be different at the lower left part since there is a very close galaxy at the same redshift (which we can see on the observed \OII\ map).
   However, the galaxy PA and rotational velocity are in good agreement.}
   \label{fig:J1236p0725_0639_1} 
\end{figure*}
\begin{figure*} 
   \centering 
   \includegraphics[width=12.0cm]{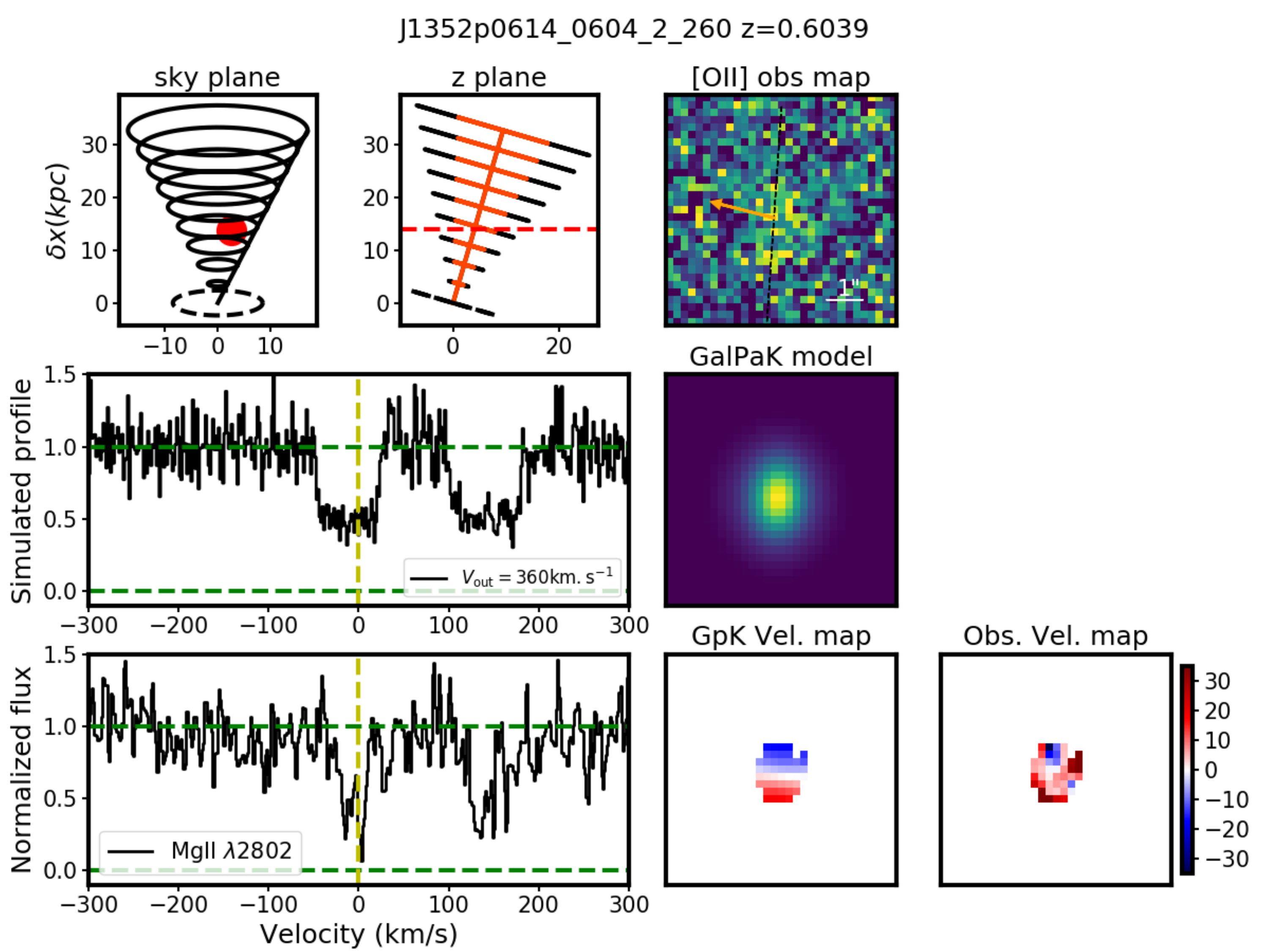} 
   \caption{Same as Figure~\ref{fig:J0014m0028_0834_1} but for the  galaxy \#22 at redshift $z=0.6039$.
   This outflow has a \Vout\ of $80\pm10$ \kms\ and an opening angle \thetam\ of $30\pm2^\circ$.} 
   \label{fig:J1352p0614_0604_2} 
\end{figure*}
\begin{figure*} 
   \centering 
   \includegraphics[width=12.0cm]{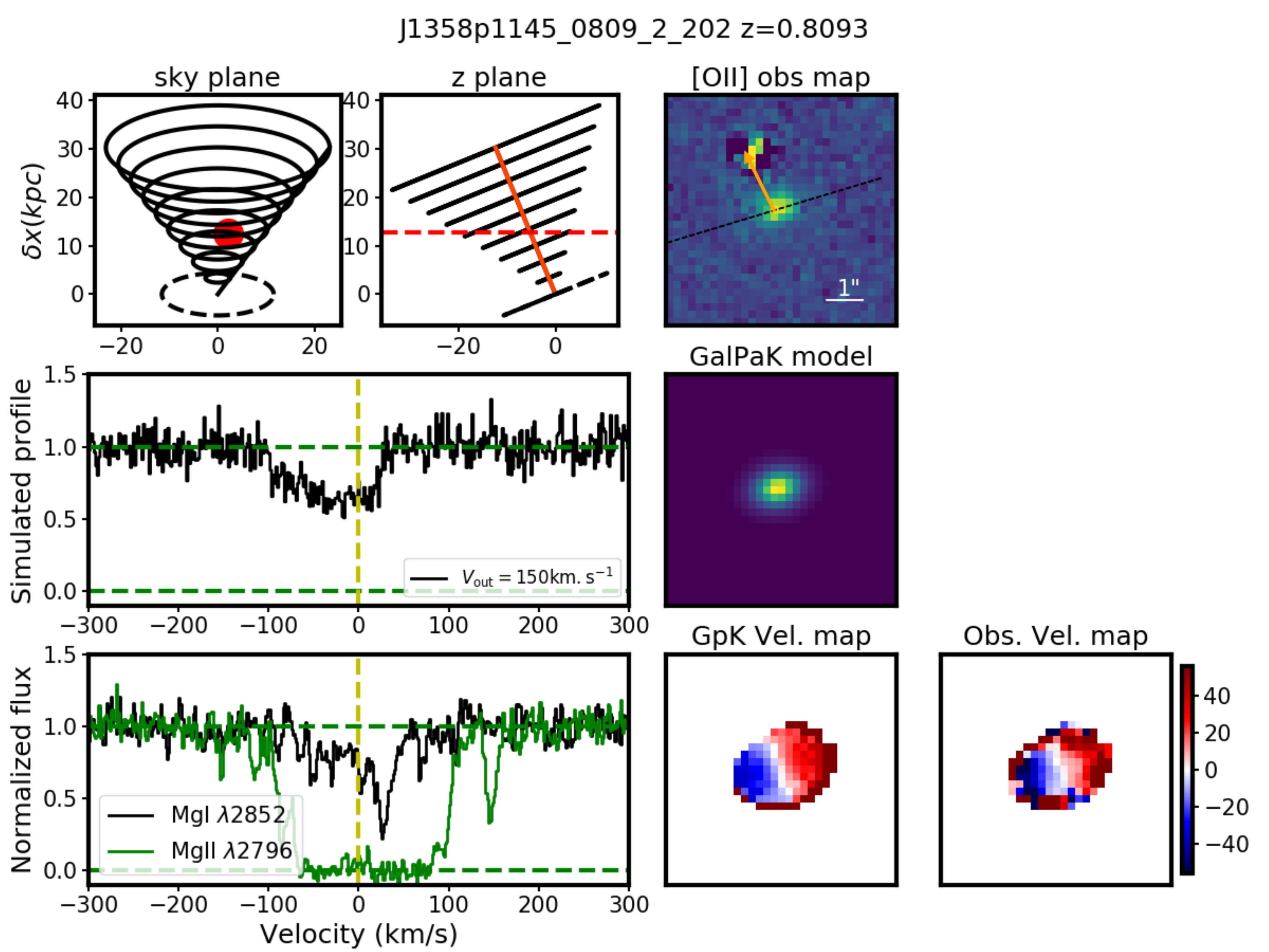} 
   \caption{Same as Figure~\ref{fig:J0014m0028_0834_1} but for the  galaxy \#23 at redshift $z=0.8093$.
   This outflow has a \Vout\ of $150\pm10$ \kms\ and an opening angle \thetam\ of $45\pm2^\circ$.} 
   \label{fig:J1358p1145_0809_2} 
\end{figure*}
\begin{figure*} 
   \centering 
   \includegraphics[width=12.0cm]{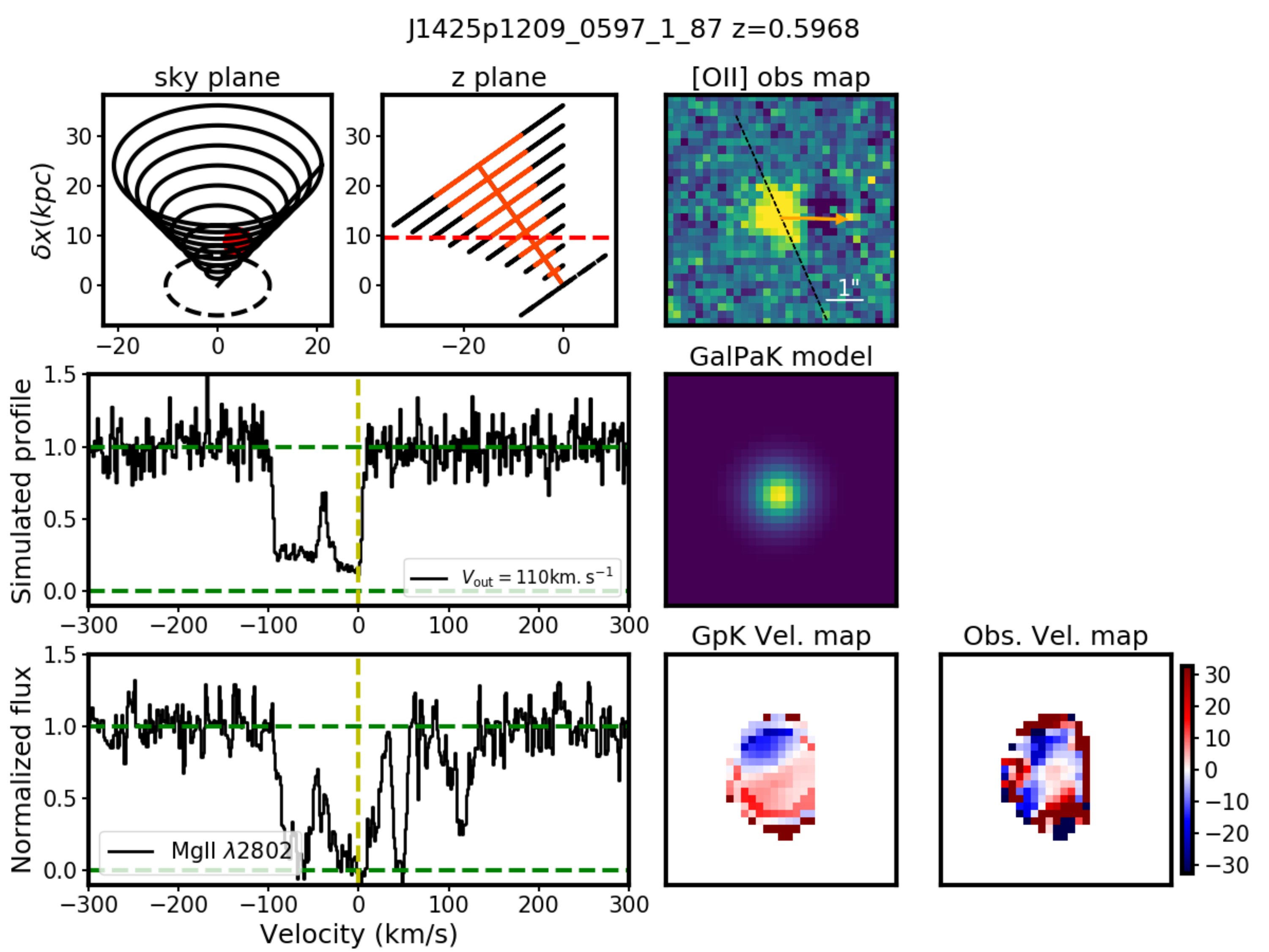} 
   \caption{Same as Figure~\ref{fig:J0014m0028_0834_1} but for the  galaxy \#24 at redshift $z=0.5968$.
   This outflow has a \Vout\ of $110\pm10$ \kms, an opening angle \thetam\ of $45\pm2^\circ$ and an empty inner cone $\theta_{\rm in}$ of 21$^\circ$.} 
   \label{fig:J1425p1209_0597_1} 
\end{figure*}
\begin{figure*} 
   \centering 
   \includegraphics[width=12.0cm]{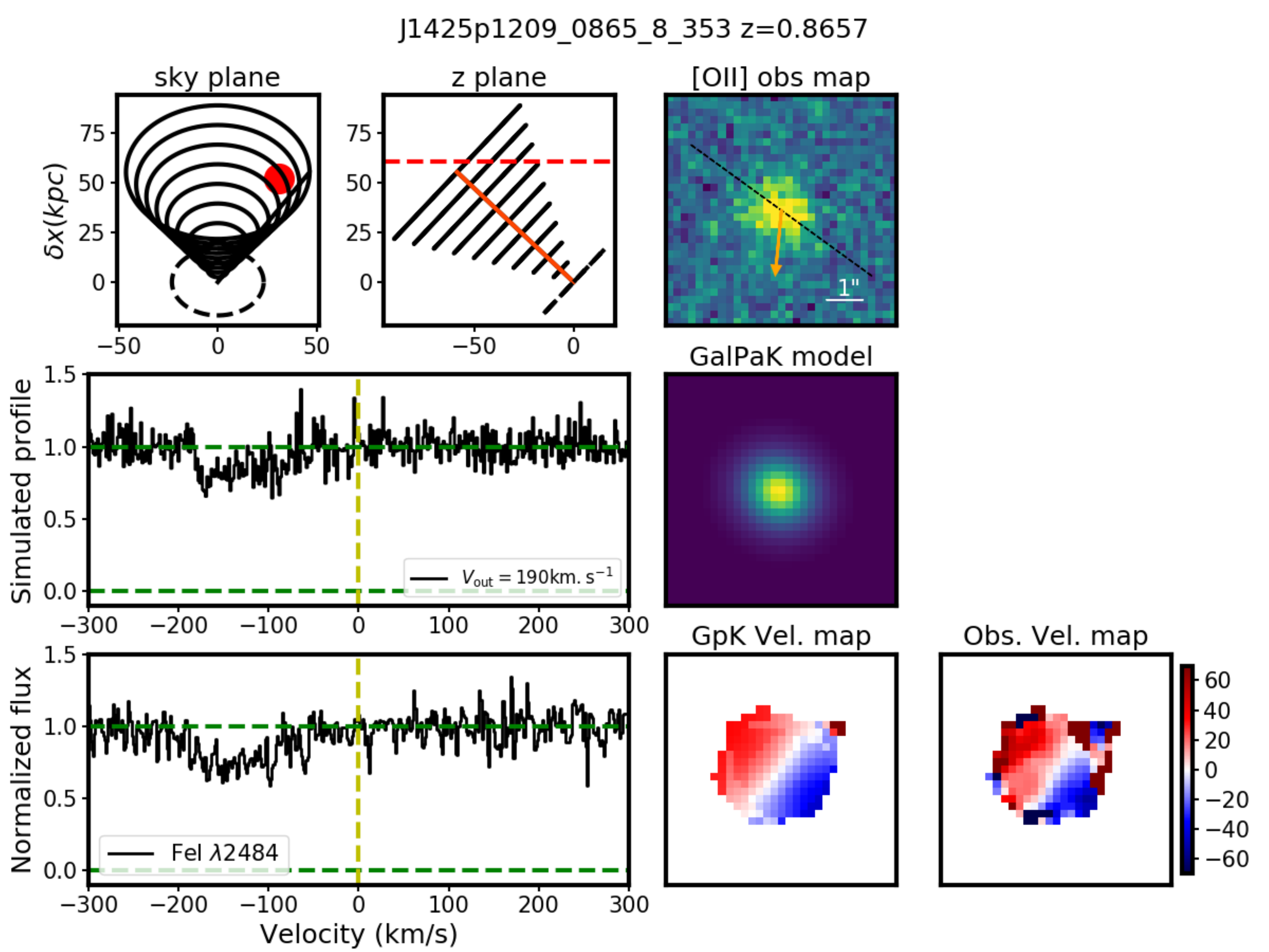} 
   \caption{Same as Figure~\ref{fig:J0014m0028_0834_1} but for the  galaxy \#25 at redshift $z=0.8657$.
   This outflow has a \Vout\ of $190\pm10$ \kms\ and an opening angle \thetam\ of $35\pm2^\circ$.} 
   \label{fig:J1425p1209_0865_8} 
\end{figure*}

\begin{figure*} 
    \centering 
    \includegraphics[width=12.0cm]{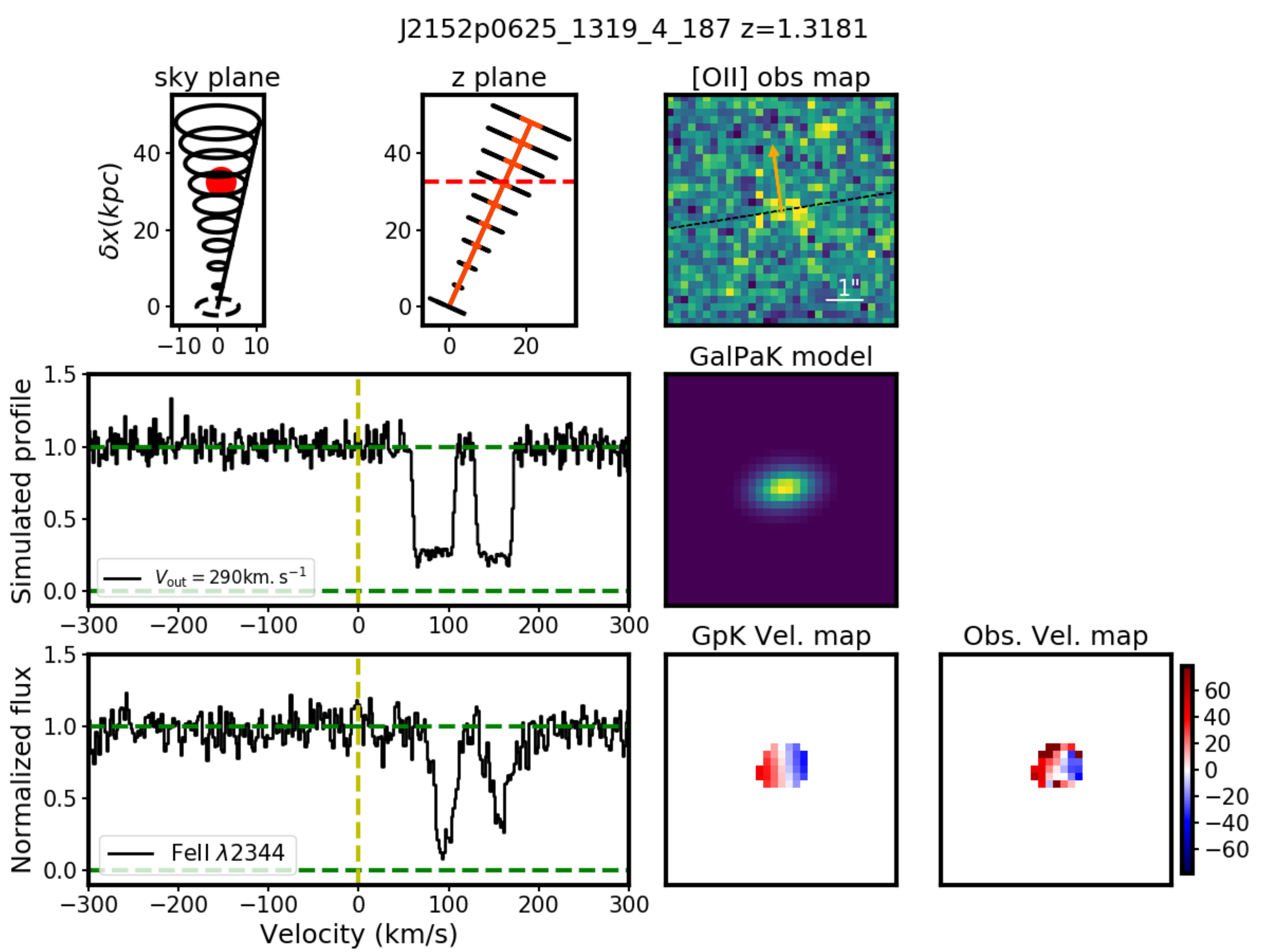} 
    \caption{Same as Figure~\ref{fig:J0014m0028_0834_1} but for the  galaxy \#26 at redshift $z=1.3181$.
    This outflow has a \Vout\ of $290\pm10$ \kms, an opening angle \thetam\ of $12\pm2^\circ$ and an empty inner cone $\theta_{\rm in}$ of 3$^\circ$.} 
    \label{fig:J2152p0625_1319_4} 
 \end{figure*}
 
\end{document}